\documentclass[12pt]{article}
\usepackage[utf8]{inputenc} 
\usepackage[T1]{fontenc}    
\usepackage{hyperref}       
\usepackage{url}            
\usepackage{booktabs}       
\usepackage{amsfonts}       
\usepackage{nicefrac}       
\usepackage{microtype}      
\usepackage{amsmath}
\usepackage{float}
\RequirePackage{graphicx}
\usepackage{caption}
\usepackage{subcaption}
\usepackage{relsize}
\usepackage{mathtools}
\usepackage{multirow}
\newcommand{\diff}{\mathrm{d}}

\newcommand{\difff}{\mathrm{~d}}
\newcommand{\cov}{\text{Cov}}
\newcommand{\thickhline[1]}{%
    \noalign {\ifnum 0=`}\fi \hrule height #1
    \futurelet \reserved@a \@xhline
}

\usepackage[table]{xcolor}
\usepackage{array}

\usepackage{setspace}
\usepackage{enumitem}
\newlength\mylen
\newlist{mycases}{enumerate}{1}
\setlist[mycases,1]{label=\textbf{Scenario~\arabic*.}, 
  labelwidth=\dimexpr-\mylen-\labelsep\relax,leftmargin=40pt,align=right}

\hypersetup{
    colorlinks=true,
    citecolor=black,
    linkcolor=black,     
    urlcolor=blue
    }

\newcommand{\blind}{0}

\begin{document}

\def\spacingset#1{\renewcommand{\baselinestretch}%
{#1}\small\normalsize} \spacingset{1}


\if0\blind
{
  \title{\bf Multivariate disaggregation modeling of air pollutants: a case-study of PM2.5, PM10 and ozone prediction in Portugal and Italy}
  \author{\textbf{Fernando Rodriguez Avellaneda} \\
Computer, Electrical and Mathematical Sciences and Engineering Division,\\
King Abdullah University of Science and Technology (KAUST),\\
Thuwal, Saudi Arabia,\\
    \textbf{Erick A. Chacón-Montalván} \\
Computer, Electrical and Mathematical Sciences and Engineering Division,\\
King Abdullah University of Science and Technology (KAUST),\\
Thuwal, Saudi Arabia,\\
    \textbf{Paula Moraga} \\
Computer, Electrical and Mathematical Sciences and Engineering Division,\\
King Abdullah University of Science and Technology (KAUST),\\
Thuwal, Saudi Arabia}
  \maketitle
} \fi

\if1\blind
{
  \bigskip
  \bigskip
  \bigskip
  \begin{center}
    {\LARGE\bf Multivariate disaggregation modeling of air pollutants: a case-study of PM2.5, PM10 and ozone prediction in Portugal and Italy}
\end{center}
  \medskip
} \fi

\bigskip
\begin{abstract}
Air pollution remains a critical environmental and public health challenge, demanding high-resolution spatial data to better understand its spatial distribution and impacts. This study addresses the challenges of conducting multivariate spatial analysis of air pollutants observed at aggregated levels, particularly when the goal is to model the underlying continuous processes and perform spatial predictions at varying resolutions. To address these issues, we propose a continuous multivariate spatial model based on Gaussian processes (GPs), naturally accommodating the support of aggregated sampling units. Computationally efficient inference is achieved using \texttt{R-INLA}, leveraging the connection between GPs and Gaussian Markov random fields (GMRFs). A custom projection matrix maps the GMRFs defined on the triangulation of the study region and the aggregated GPs at sampling units, ensuring accurate handling of changes in spatial support. This approach integrates shared information among pollutants and incorporates covariates, enhancing interpretability and explanatory power. This approach is used to downscale PM$_{2.5}$, PM$_{10}$ and ozone levels in Portugal and Italy, improving spatial resolution from $0.1^\circ$ (~10 km) to $0.02^\circ$ (~2 km), and revealing dependencies among pollutants. Our framework provides a robust foundation for analyzing complex pollutant interactions, offering valuable insights for decision-makers seeking to address air pollution and its impacts.%
\end{abstract}

\vfill

\newpage
\spacingset{1.45} 
\section{Introduction}
\label{introduction}

Exposure to air pollutants has been consistently associated with an increased risk of all-cause mortality and a variety of diseases, including respiratory infections, chronic obstructive pulmonary disease (COPD), ischemic heart disease (IHD), lung cancer, and stroke \cite{cohen2017}. Ozone (\(O_3\)), a highly reactive oxidative gas, has been linked to all-cause, cardiovascular, and respiratory mortality \cite{gryparis2004acute, nuvolone2018effects}, and it may exacerbate climate impacts on food security in future scenarios \cite{tai2014threat}. Particulate matter with diameters smaller than 2.5 microns (PM\(_{2.5}\)) and 10 microns (PM\(_{10}\)) has also been shown to adversely impact health, even with short-term exposure, contributing to mortality, morbidity, and adverse birth outcomes \cite{pascal2014shortterm, lu2015systematic, feng2016health}. Collectively, PM\(_{2.5}\), PM\(_{10}\), and \(O_3\) are among the most significant air pollutants, with substantial evidence documenting their harmful effects on human health \cite{orellano2020shortterm,kim2020different,hvidtfeldt2019longterm,chen2015effects}.

Understanding the spatial distribution of air pollutants is crucial to assess their effects on human health, the environment, and food security. However, several challenges commonly arise. First, pollutants such as PM\(_{2.5}\), PM\(_{10}\), and \(O_3\) are often interrelated, necessitating careful consideration of their \textit{multivariate spatial structure}. Second, the concentrations of these pollutants are typically measured at sparse locations or aggregated over specific spatial supports. This study addresses the \textit{change of support problem}, where pollutant concentrations are available only at aggregated spatial units, but estimates are required at a continuous level or for alternative supports, such as administrative regions \cite{huang2018multivariatea,cameletti2019bayesian,zhongandmoraga24}. Moreover, predictors for air pollutants frequently exist on spatial supports that differ from those of the pollutant sampling units, a challenge referred to as \textit{spatial misalignment}.


Our study focuses on modeling atmospheric composition data from the Copernicus Atmosphere Monitoring Service (CAMS) \cite{copernicus_data}, which provides multivariate pollutant measurements on a grid resolution of approximately \(10 \times 10\) kilometers across Europe. Specifically, we aim to jointly model PM\(_{2.5}\), PM\(_{10}\), and \(O_3\) concentrations in Italy and Portugal, enabling high-resolution multivariate spatial predictions. We aim to addresses the previously mentioned challenges while maintaining a computationally efficient framework.


A common approach to modeling multivariate spatial data at an aggregated level, such as pollutant concentrations from CAMS, is to use multivariate spatial processes defined on discrete index sets. These methods often focus on constructing valid multivariate processes by extending conditionally autoregressive (CAR) models, which induce spatial association through the neighboring structure of spatial sampling units. For instance, this can be achieved by specifying the multivariate joint density via conditional and marginal densities \cite{jin2005generalized} or by deriving multivariate structures from independent univariate spatial processes \cite{macnab2016linear}. While these approaches are computationally efficient due to the sparsity of their associated precision matrices, they lack a coherent definition of spatial correlation for irregularly shaped areas and are restricted to predictions at the observed sampling units or their unions.


Another approach involves assuming a multivariate spatial process with a continuous index set and linking its aggregation, through change of support, to the observed aggregated values. This is typically achieved using multivariate Gaussian processes (GPs), where the cross-covariance function is specified, and the aggregation of the process is defined as an integral over the sampling units \cite{tanaka2019spatiallya,yousefi2019multitaska}. This method enables inference for the assumed underlying continuous process and facilitates prediction at a continuous level or for arbitrary spatial units. However, full inference and prediction become computationally demanding because the likelihood must be approximated, as the integral definition often lacks a closed form.

While significant progress has been made in spatial modeling of pollutants, many studies address only subsets of the aforementioned challenges. For example, some studies focus on univariate modeling, without explicitly considering the multivariate nature of contaminants \cite{wright2021estimating, cameletti2019bayesian}. Others simplify the multivariate structure by incorporating additional contaminants as predictors or find that a multivariate framework becomes unnecessary after including relevant predictors \cite{gong2021multivariate, huang2018multivariatea}. Similarly, some approaches address the change-of-support problem by relying on centroids of aggregated sampling units as a practical approximation \cite{forlani2020joint}. While these methods provide valuable insights, there remains room to develop more comprehensive approaches that tackle all these challenges simultaneously.


To model multiple pollutants observed at aggregated spatial units, we propose the following approach: (1) We assume an underlying continuous multivariate process, specified as a linear transformation of independent Gaussian processes (GPs), to capture shared spatial variability at the continuous level. This approach is known as the linear model of coregionalization (LMC) \cite{gelfand2004nonstationary}. (2) Predictors are incorporated directly at this continuous level, allowing the model to account for covariate effects. (3) The resulting continuous multivariate system is integrated over the sampling units, yielding a linear coregionalization model defined for aggregated independent processes. The observed pollutant levels are then treated as realizations of the aggregated LMC. This framework effectively addresses the challenges of modeling multivariate spatial data at an aggregated scale, while enabling inference for the underlying continuous process and facilitating predictions at higher spatial resolutions (downscaling).


Unlike traditional methods for Bayesian inference, such as Markov chain Monte Carlo (MCMC), which can be computationally intensive and scale poorly with complex data, our approach exploits the relationship between GPs and Gaussian Markov random fields (GMRFs) via stochastic partial differential equations (SPDEs). This connection facilitates efficient inference using the integrated nested Laplace approximation (INLA) due to the sparse structure of GMRFs. Specifically, we leverage this relationship to represent the aggregated independent GPs required in the resulting LMC under the change of support framework, incorporating a custom projection matrix as proposed in \cite{moraga2017geostatistical}.

The structure of the remainder of the paper is organized as follows. Section \ref{sec:methodology} introduces the theoretical framework including multivariate areal models, continuous models, and the connection between areal and continuous processes. In Section \ref{sec:sim}, we present a simulation study to evaluate the performance of our method in effectively recovering the underlying continuous process from areal data. Section \ref{sec:app} showcases an application of our approach to air pollution data in Europe, focusing on Portugal and Italy. Finally, Section \ref{sec:con} provides conclusions, a discussion of the findings, and directions for future research.

\section{Methodology}
\label{sec:methodology}

\subsection{Multivariate processes based on the linear model of coregionalization}
\label{sec:lmc}


There are various approaches to represent multivariate spatial processes with valid cross-covariance functions \cite{genton2015crosscovariancea}. A widely used method is the linear model of coregionalization (LMC), where a multivariate process is expressed as a linear combination of independent latent processes \cite{grzebyk_et_al_1994}. Consider an $M$-dimensional process $\{\mathbf{W}(\textbf{s})\}$, with locations $\textbf{s}$ belonging to a continuous or discrete index set. Under the LMC framework, $\mathbf{W}(\textbf{s})$ is defined as $\mathbf{W}(\textbf{s}) = \mathbf{\Lambda}\mathbf{{U}}(\textbf{s})$, where $\{\mathbf{U}(\textbf{s})\}$ represents $M$ independent latent spatial processes. The multivariate correlation structure is implicitly defined by the shared latent processes and determined by the matrix $\mathbf{\Lambda}$. This flexible and interpretable approach can be applied to spatial processes with both discrete and continuous index sets \cite{banerjee2004,gelfand2004nonstationary,jin2005generalized,macnab2016linear}.


To ensure model identifiability, it is common to impose a sparse structure on the linear transformation. Specifically, $\mathbf{\Lambda}$ is often assumed to have a lower triangular form with unit values on the diagonal. Under this assumption, the random vector $\mathbf{W}(\mathbf{s}) = [W_1(\mathbf{s}), \dots, W_M(\mathbf{s})]^\top$ can be expressed as:
\begin{equation*}
    \begin{bmatrix}
        W_1(\textbf{s}) \\
        W_2(\textbf{s}) \\
        \vdots \\
        W_M(\textbf{s})
        \end{bmatrix}
        =
        \begin{bmatrix}
        1 & 0 &  \cdots & 0 \\
        \lambda_{21} & 1& \cdots & 0 \\
        \vdots & \vdots & \ddots & \vdots \\
        \lambda_{M1} & \lambda_{M2} & \cdots & 1
        \end{bmatrix}
        \begin{bmatrix}
        U_1(\textbf{s}) \\
        U_2(\textbf{s}) \\
        \vdots \\
        U_M(\textbf{s})
    \end{bmatrix}
\end{equation*}
This leads to the following system of equations:
\begin{align}
\begin{split}
    W_1(\textbf{s}) & = U_1(\textbf{s}), \\[3mm]
    W_2(\textbf{s}) & = \lambda_{21} U_1 (\textbf{s}) + U_2(\textbf{s}), \\[3mm]
    & \hspace{0.2cm} \vdots \\[3mm]
    W_M(\textbf{s}) & =  \lambda_{M1} U_1(\textbf{s}) + \lambda_{M2} (\textbf{s}) + \dots + U_M(\textbf{s}).
\end{split}
\label{equ:lmc1}
\end{align}

Notice that under this specification, $\{U_i(\mathbf{s})\}$ represents the latent process shared among the response processes $\{W_j(\mathbf{s}): j \in i, \dots, M\}$, with $\lambda_{ji}$ denoting the weight of the independent process $\{U_i(\mathbf{s})\}$ for the response process $\{W_j(\mathbf{s})\}$. Let us consider $\mathbf{\Sigma}$, an $M \times M$ diagonal matrix containing the variances of the independent latent processes $U_i(\mathbf{s})$, such that $\sigma_{ii} = \mathrm{Var}(U_i(\mathbf{s}))$. It follows that the covariance matrix of $\mathbf{W}(\mathbf{s})$ at the same location $\mathbf{s}$ is given by $\mathbf{T} = \mathbf{\Lambda} \mathbf{\Sigma} \mathbf{\Lambda}^\top$. Consequently, the correlation between the responses $W_i(\mathbf{s})$ and $W_j(\mathbf{s})$ can be expressed as:
\begin{equation}
    \mathrm{Cor}(W_i(\mathbf{s}), W_j(\mathbf{s})) = \frac{\mathbf{T}_{ij}}{\sqrt{\mathbf{T}_{ii}} \sqrt{\mathbf{T}_{jj}}}.
    \label{equ:corrlmc}
\end{equation}
Thus, the correlation is determined by the linear transformation $\mathbf{\Lambda}$ and the marginal variances $\sigma_{ii}$.


\subsection{Multivariate discrete areal model}
\label{sec:multivariate-areal}


As mentioned in Section \ref{introduction}, our objective is to model multiple pollutant
concentrations observed at aggregated levels. Consider an $M$-dimensional response vector
$\mathbf{Y}_i$ for regions $i = 1, \dots, N$, representing the pollutant levels. Using the
predictors $\boldsymbol{z}_{i}$ for region $i$, a multivariate spatial process
$\{\mathbf{W}_i\}$ and a multivariate term $\mathbf{V}_i$, a common model is $\mathbf{Y}_i = \mathbf{B} \boldsymbol{x}_{i} + \mathbf{W}_i + \mathbf{V}_i$. Using the concept of the LMC framework (Equation \ref{equ:lmc1}), this model can be redefined as follows:
\begin{equation}
    \mathbf{Y}_i = \mathbf{B} \boldsymbol{x}_{i} + \mathbf{\Lambda}\mathbf{U}_i + \mathbf{V}_i.
    \label{equ:mod-area}
\end{equation}
In this model, the matrix $\boldsymbol{B}$ represents the  $M \times (p+1)$ fixed effects associated with the predictors $\boldsymbol{x}_{i}$, which include the intercept and $p$ covariates for each region $i$. This structure allows each covariate to have a distinct effect on each response variable, providing flexibility in capturing pollutant-specific relationships. Furthermore, $\{\mathbf{U}_i\}$ denotes an $M$-dimensional set of independent spatial processes, while $\mathbf{V}_i$ represents an $M$-dimensional set of independent processes to capture any additional variation.

In this formulation, $\mathbf{Y}_i$ is observed over a discrete set of regions, with $\mathbf{U}_i$ and $\mathbf{V}_i$ sharing the same index set. This model extends the univariate Besag-York-Mollié (BYM) model which uses two random effects to account for spatial dependence and uncorrelated residual noise. Similar to the univariate BYM model, the $j$-th spatial random effect $U_{i,j}$ for region $i$ captures the influence of neighboring regions, ensuring that nearby areas are assigned similar values. This is achieved using an intrinsic conditional autoregressive (CAR) model, which smooths data across neighboring regions by borrowing strength from adjacent areas. The conditional distribution of $U_{i,j}$ is given by:
\begin{equation*}
    U_{i,j} \mid U_{-i,j} \sim \mathcal{N} \left( 
\frac{\sum_{k \in \mathcal{N}_i} A_{ik} U_{k,j}}{\sum_{k \in \mathcal{N}_i} A_{ik}}, 
\frac{\sigma^2_j}{\sum_{k \in \mathcal{N}_i} A_{ik}}
\right),
\end{equation*}
where $\mathcal{N}_i$ is the set of neighbors for region $i$, $U_{-i,j}$ denotes the $j$-th random effects for all regions except $i$, and $\mathbf{A}$ is the adjacency matrix defining the neighborhood structure. In contrast, the $j$-th uncorrelated noise $V_{i,j}$ is modeled as an independent Gaussian random variable with zero mean and variance $\sigma_{v_j}^2$, expressed as $V_{i,j} \sim \mathcal{N}(0, \sigma_{v_j}^2)$.

Although this multivariate model, defined over a discrete index set, is computationally efficient due to the sparsity of the precision matrices induced by the CAR model, it has limitations. Specifically, it lacks a clear definition of spatial correlation for irregularly shaped areas and is restricted to predictions at observed sampling units or their unions. For example, while this model is suitable for modeling multivariate pollutant data, it cannot support predictions at a higher spatial resolution. Nevertheless, we use this model as a baseline for comparison with other approaches.

\subsection{Multivariate disaggregation model}

One of the main objectives of this paper is to achieve finer spatial resolution from areal datasets. To accomplish this, we assume that the observed data in regions are underpinned by a continuous $M$-dimensional latent spatial process, $\mathbf{Z}(\mathbf{s})$. This process can be represented using spatial predictors $\boldsymbol{x}(\mathbf{s})$ and Gaussian random fields (GRFs), which offer a flexible framework for capturing spatial dependence. As discussed in the previous sections, this multivariate latent process can be defined using the LMC framework as follows:
\begin{equation}
    \mathbf{Z}(\mathbf{s}) = \mathbf{B} \boldsymbol{x}(\mathbf{s}) + \mathbf{\Lambda}\mathbf{U}(\mathbf{s}),
    \label{equ:mod-con}
\end{equation}
where $\mathbf{B} \boldsymbol{x}(\mathbf{s})$ represents the multivariate fixed effects, and $\mathbf{U}(\mathbf{s})$ denotes the multivariate set of independent spatial processes ${U_j(\mathbf{s})}$. Each $U_j(\mathbf{s})$ is modeled as a Gaussian random field with spatial dependence characterized by the Matérn covariance function \cite{matern_1960}, which is defined as:
\begin{equation}
\label{equ:cov1}
\cov(U_j(\textbf{s}_i),U_j(\textbf{s}_k)) = \frac{\sigma^2_j}{2^{\nu-1} \Gamma(\nu)}\left(\kappa_j\left\|\boldsymbol{s}_i-\boldsymbol{s}_k\right\|\right)^{\nu_j} K_{\nu_j}\left(\kappa_j\left\|\boldsymbol{s}_i-\boldsymbol{s}_k\right\|\right)
\end{equation}

Here, $\sigma^2_j$ represents the marginal variance of the $j$-th spatial field, while $K_{\nu_j}(\cdot)$ denotes the modified Bessel function of the second kind and order $\nu_j > 0$. The parameter $\nu_j$, which controls the smoothness of the field, is typically fixed in practical applications due to challenges in its reliable estimation. The parameter $\kappa_j > 0$ is associated with the spatial range $\rho_j$ for the $j$-th field, defined as the distance at which the correlation between two points effectively diminishes to zero.


The aggregation of the continuous latent process, $\mathbf{Z}(\mathbf{s})$, over an arbitrary region $R$ can be expressed as $\mathbf{Z}(R) = \frac{1}{|R|} \int_{R} \mathbf{Z}(\mathbf{s}) , \mathrm{d} \mathbf{s}$. We assume that the observed response values for region $R_i$ result from the aggregation of the continuous latent process and a measurement error. Using Equation \eqref{equ:mod-con}, this leads to the disaggregation model:
\begin{equation} \mathbf{Y}(R_i) = \mathbf{B} \boldsymbol{x}(R_i) + \mathbf{\Lambda} \mathbf{U}(R_i) + \mathbf{V}_i, \label{equ:mod-con-ag} \end{equation}
where $\boldsymbol{x}(R_i)$ represents the averaged covariates over region $R_i$, $\mathbf{U}(R_i)$ denotes the aggregated multivariate independent process with components defined as $\mathcal{U}j(R_i) = \frac{1}{|R_i|} \int_{R_i} \mathcal{U}_j(\mathbf{s}) \mathrm{d} \mathbf{s}$, and $\mathbf{V}_i$ accounts for the uncorrelated process at the aggregated level. Similar to the previous section, $\mathbf{V}_i$ is modeled such that $V_{i,j} \sim \mathcal{N}(0, \sigma_{v_j}^2)$.

\subsection{Inference using SPDE and INLA}

A GRF with a Matérn covariance structure is also the solution of a continuous stochastic partial differential equation (SPDE) \cite{whittle1963stochastic} defined as follows:
\begin{equation}
    \left(\kappa^2-\Delta\right)^{\alpha / 2}(\tau \cdot U(\mathbf{s}))=\vartheta(\mathbf{s}).
    \label{equ:spde}
\end{equation}

Here, $U(\mathbf{s})$ represents a Gaussian random field (GRF), while $\vartheta(\mathbf{s})$ denotes a Gaussian spatial white noise process. The smoothness of the GRF is governed by the parameter $\alpha$, which is related to the smoothness parameter $\nu$ through the relationship $\nu = \alpha - d/2$, where $d$ is the spatial dimension. The parameter $\tau$ controls the variance of the process, while $\kappa > 0$ acts as a scale parameter, influencing the range of spatial dependence. Finally, the operator $\Delta$ refers to the Laplacian, which is defined in the spatial dimension $d$.

To perform inference on the system of GRFs defined in Equation \ref{equ:mod-con}, we utilized the \texttt{R-INLA} package, which provides a convenient and computationally efficient approach for modeling geostatistical data \cite{rue_et_al_r_inla_2009}. This package employs the finite element method to approximate the solution of the SPDE described in Equation \ref{equ:spde}. Specifically, it constructs a triangulated mesh consisting of $G$ nodes and $G$ basis functions. The basis functions, denoted as $\psi_k$, are piecewise linear functions that take a value of 1 at vertex $k$ and 0 at all other vertices, enabling a localized representation of the spatial process. Then, we can obtain a discrete representation of each continuous GRF defined as the sum of basis functions defined in the triangulated mesh as:
\begin{equation*}
    U_j(\textbf{s}) = \sum_{k=1}^{G} \psi_k(\textbf{s}) \mathcal{U}_{j,k},
\end{equation*}
where $G$ is the number of vertices of the triangulation, and $\{ \mathcal{U}_{j,k}\}$ are zero-mean Gaussian distributed weights that approximate the solution $U_j(\textbf{s})$ of the SPDE on the mesh nodes.


Let consider a set of regions $R_i$ for $i=1,...,N$, using the ideas for disaggregation models presented in \cite{moraga2017geostatistical}, we can link the values of the aggregated fields $U_j(R_i)$ required in Equation \ref{equ:mod-con-ag} with point estimation in the mesh vertices as follows:
\begin{align}
\begin{split}
    U_j(R_i) & = \frac{1}{|R_i|}\int_{R_i} U_j(\mathbf{s}) \text{d} \textbf{s} \\
    & = \frac{1}{|R_i|} \left( \sum_{k=1}^{G} \int_{R_{ik}} U_j(\mathbf{s}) \text{d} \textbf{s}\right) \\
    & \approx \frac{1}{|R_i|} \left( \sum_{k=1}^{G} |R_{ik}| U_{j,k} \right) \\
    & = \sum_{k=1}^{G} A_{ik} U_{j,k} \quad \text{with} \quad A_{ik} = \frac{|R_{ik}|}{|R_{i}|}.
\end{split}
\label{equ:des}
\end{align}
Here, the operation $|\cdot|$ represents the area of a region, and $R_{ik}$ with $k=1,...,G$ refers to the areas of the triangulated mesh that fall within the set $R_i$. The parameter $G$ denotes the number of mesh vertices, while $A$ is an $N \times G$ sparse matrix that maps the GRF values from the $N$ regions to the $G$ triangulation nodes. This matrix, commonly referred to as the projection matrix, has several notable properties: its entries are positive ($A_{ik} > 0$), and the rows of the matrix sum to 1, i.e., $\sum_{k} A_{ik} = 1$.

Additionally, if the triangulated mesh over the study region is uniformly spaced, a simplification can be applied to the matrix $A$. In this case, each entry is defined as $A_{ij} = 1 / n_i$, where $n_i$ is the number of mesh vertices within region $R_i$. This simplification reduces computational complexity while maintaining the mapping's integrity.

\section{Simulation study}
\label{sec:sim}

To assess the performance of our model, we performed a simulation study that tested how well our model can capture multivariate latent processes observed at a aggregated level. We simulated three-dimensional multivariate Gaussian processes with a specific covariance structure on the unit square $[0,1] \times [0,1] = [0,1]^2$. Then, we created three regular and three irregular scenarios with regions of different shapes and sizes and aggregated the values of the latent processes within these regions.
Then, we used the multivariate disaggregation model and the multivariate areal model to predict the original latent processes using as input the aggregated data.
Finally, we evaluated the performance of each model comparing the simulated and predicted surfaces using the Root Mean Squared Error (RMSE).
In each scenario, we simulated 100 realizations to have consistent results.

\subsection{Latent processes generation}

We generated three different continuous random variables defined in the unit square $[0,1]^2$ determined by the following equations:
\begin{align}
\begin{split}
    & W_1(\mathbf{s})=\alpha_1+z_{1}(\mathbf{s}) \\ 
    & W_2(\mathbf{s})=\alpha_2+\lambda_1 z_{1}(\mathbf{s})+z_2(\mathbf{s}) \\ 
    & W_3(\mathbf{s})=\alpha_3+\lambda_2 z_{1}(\mathbf{s})+\lambda_3 z_2(\mathbf{s})+z_3(\mathbf{s})
\end{split}
\label{equ:sim1}
\end{align} 
Here, $\alpha_k$ denotes the intercepts, $z_k(\mathbf{s})$ represents the spatial random effects, $\lambda_k$ are the weights associated with the shared spatial effects, where $k=1,2,3$. In this setup, the spatial effect $z_1(\mathbf{s})$ is shared across all three variables, while $z_2(\mathbf{s})$ is shared between the second and third variables. 

\begin{figure}[!htp]
    \centering
    \begin{minipage}{0.32\textwidth}
        \centering
        \includegraphics[width=\linewidth]{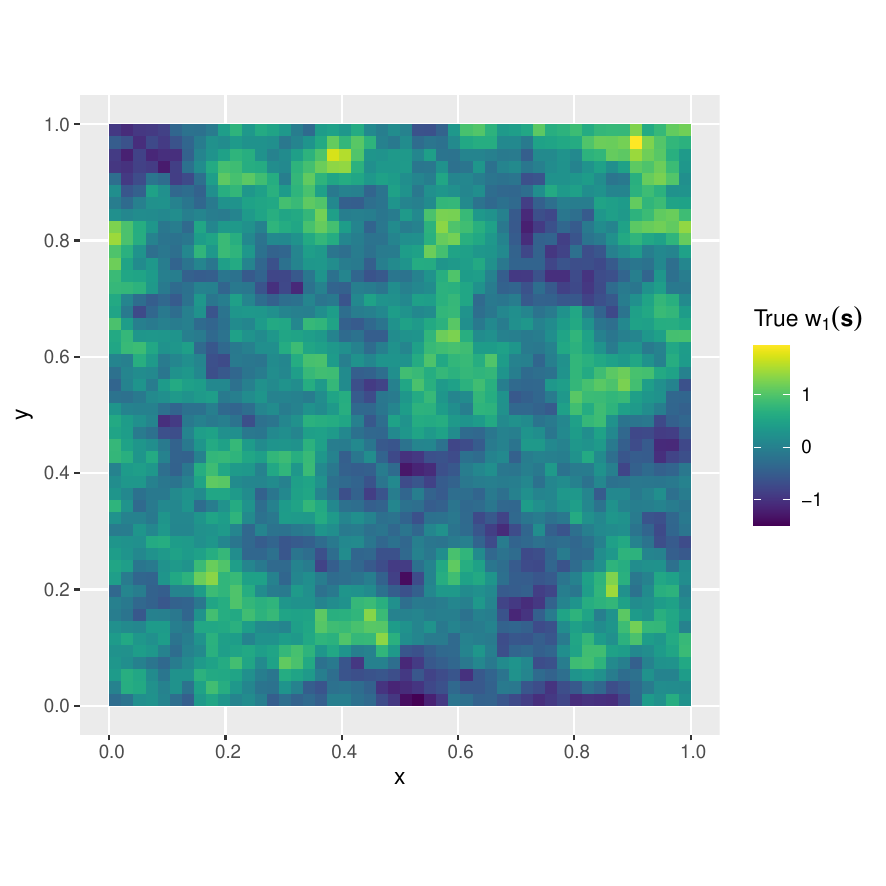}
    \end{minipage}
    \begin{minipage}{0.32\textwidth}
        \centering
        \includegraphics[width=\linewidth]{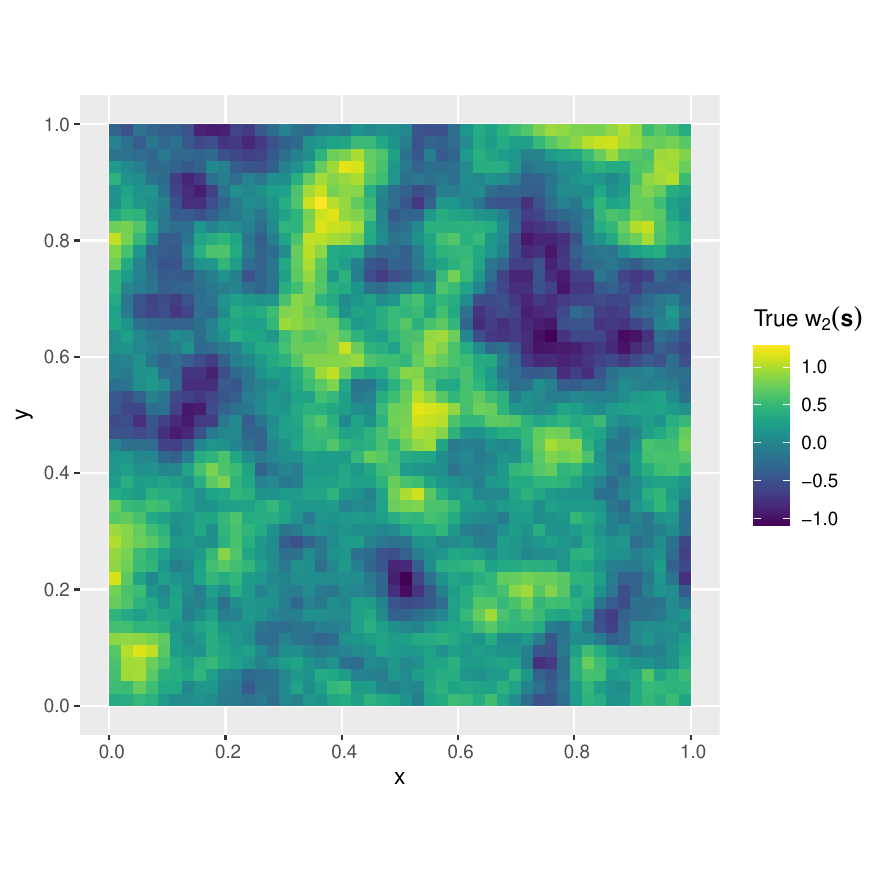}
    \end{minipage}
    \begin{minipage}{0.32\textwidth}
        \centering
        \includegraphics[width=\linewidth]{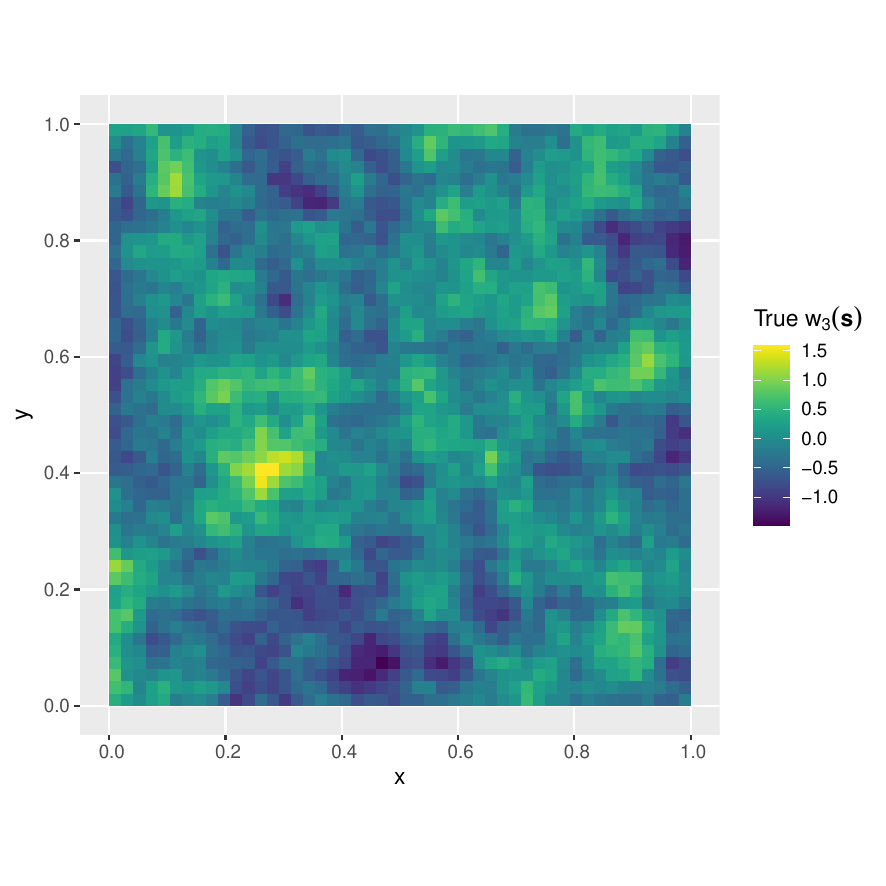}
    \end{minipage}
    \caption{Realization of the three random variables}
    \label{fig:sim1}
\end{figure}

This simulation was conducted within a latent space consisting of $48 \times 48 = 2304$ equally spaced points over the unit square. The random variables $z_k(\mathbf{s})$ were modeled as zero-mean Gaussian fields with a Matérn covariance structure. The parameters used in the simulation are detailed in Table \ref{tab:param}.
In Figure \ref{fig:sim1}, we present a realization of the three processes described by Equation \ref{equ:sim1}.

\begin{table}[h!]
\centering
\begin{tabular}{|c|c|c|c|c|c|}
\hline
& Intercepts & Correlation & Spatial Variance & Spatial Range & Standard Error \\
& ($\alpha_{k}$) & ($\lambda_k$) & ($\sigma_k$) & ($\rho_k$) & ($e_k$) \\ \hline
$k=1$ & 0.1 & 0.3 & 0.25 & 0.1 & 0.1 \\ \hline
$k=2$ & 0.05 & 0.1 & 0.2 & 0.2 & 0.2 \\ \hline
$k=3$ & -0.1 & -0.3 & 0.15 & 0.1 & 0.15 \\ \hline
\end{tabular}
\caption{Simulation parameters.}
\label{tab:param}
\end{table}

\subsection{Regions generation}

To generate the regions of this study, we considered six scenarios, three regular and three irregular, with several number of areas. For the irregular case, first we generate spatial points over the domain $[0,1]^2$ by using an inhomogeneous Poisson process with the following intensity:

\begin{equation}
\label{equ:inten}
    \lambda(x,y) = \gamma \left( \frac{3}{2} \right) \left( x^2 + y^2 \right) \text{\quad with \quad} (x,y) \in [0,1]^2.
\end{equation}

Using this intensity, the expected number of points within the region \([0,1]^2\) is equal to  
\begin{equation*}
\mathbb{E}\left[\mathcal{N}\left([0,1]^2\right)\right] = \int_{0}^{1} \int_{0}^{1} \lambda\left(x,y\right) \diff x \difff y = \int_{0}^{1} \int_{0}^{1} \gamma \left( \frac{3}{2} \right) \left( x^2 + y^2 \right) \diff x \difff y = \gamma.
\end{equation*}

For the simulation, we utilized the R package \texttt{spatstat} \cite{baddeley_2015} to generate various point patterns from the inhomogeneous Poisson process. Once a realization of the process with $m$ points is obtained, we use the Voronoi diagram to partition the unit square into exactly $m$ regions, with the centroids corresponding to the previously generated point pattern. Conversely, for the regular case, the unit square is divided into a fixed number of equal-sized squares. In summary, we simulated the following six scenarios:

\begin{mycases}
    \item Irregular areas randomly generated with a mean value of $\gamma = 36$.
    \item Irregular areas randomly generated with a mean value of $\gamma = 64$.
    \item Irregular areas randomly generated with a mean value of $\gamma = 144$.
    \item Regular areas with $6 \times 6 = 36$ squares.
    \item Regular areas with $8 \times 8 = 64$ squares.
    \item Regular areas with $12 \times 12 = 144$ squares.
\end{mycases}

Once the regions were defined for each scenario, we aggregated the multivariate Gaussian processes from Equation \ref{equ:sim1} by averaging the values within each region. To account for variability, we then added independent and identically distributed zero-mean Gaussian noise, $e_k$, to each variable across the regions $R_1, \dots, R_n$, as detailed below.

\begin{align}
\begin{split}
    & W_1(R_i) = \alpha_1+\bar{z}_{1}(R_i) + e_1(i) \\
    & W_2(R_i) = \alpha_2+\lambda_1 \bar{z}_{1}(R_i)+\bar{z}_2(R_i)+ e_2(i)\\
    & W_3(R_i) = \alpha_3+\lambda_2 \bar{z}_{1}(R_i)+\lambda_3 \bar{z}_2(R_i)+\bar{z}_3(R_i) +e_3(i)
\end{split}
\label{equ:sim2}
\end{align}

Here, $\bar{z}_{k}(R_i)$ represents the average of $z_k$ over the points within region $R_i$, and the error terms $e_k$ follow zero-mean Gaussian distributions with a fixed standard deviation for $k = 1, 2, 3$. Note that regions $R_i$ are generated according to different scenarios as explained in next section.

\begin{figure}[!htp]
    \centering
    \begin{minipage}{0.32\textwidth}
        \centering
        \includegraphics[width=\linewidth]{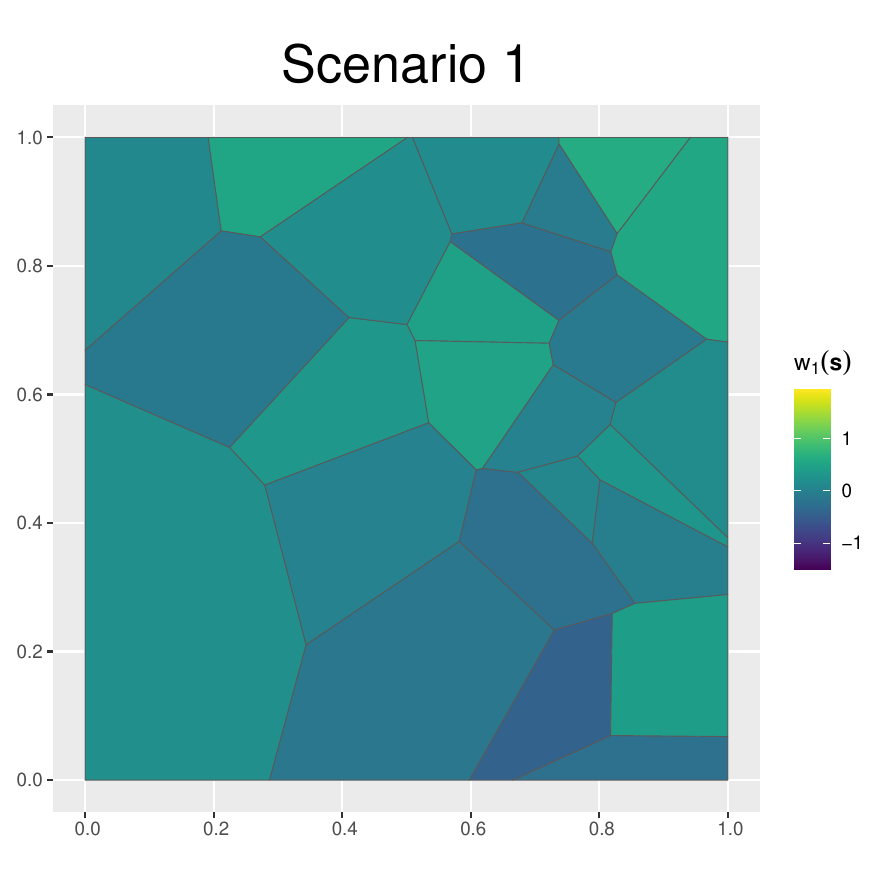}
    \end{minipage}
    \begin{minipage}{0.32\textwidth}
        \centering
        \includegraphics[width=\linewidth]{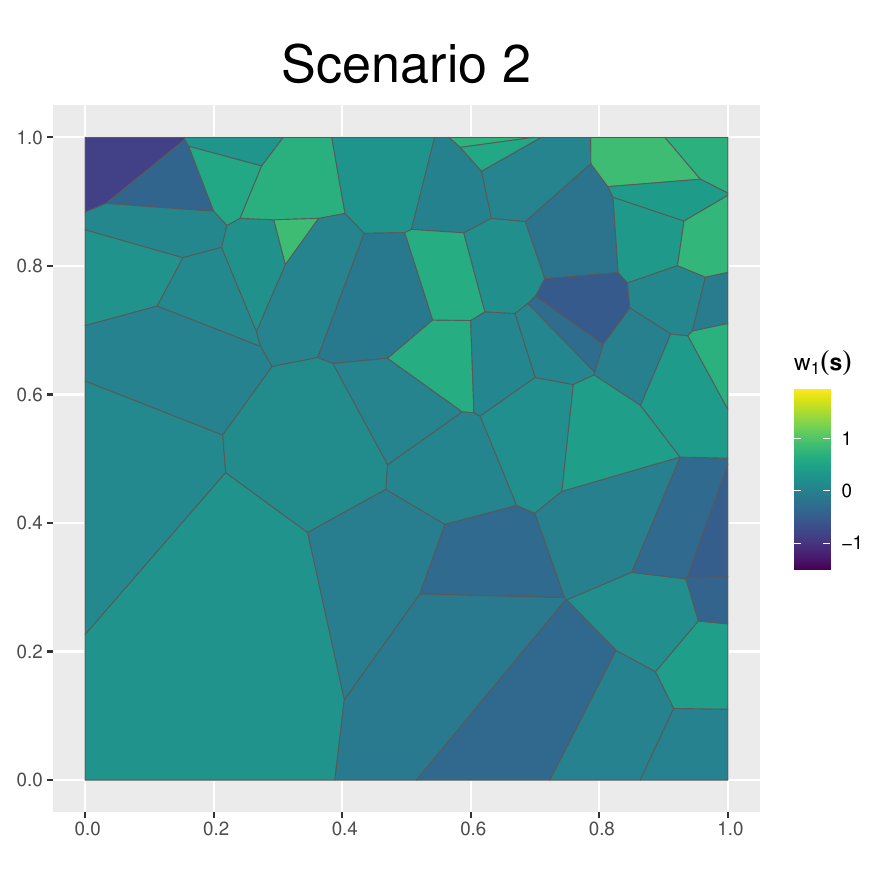}
    \end{minipage}
    \begin{minipage}{0.32\textwidth}
        \centering
        \includegraphics[width=\linewidth]{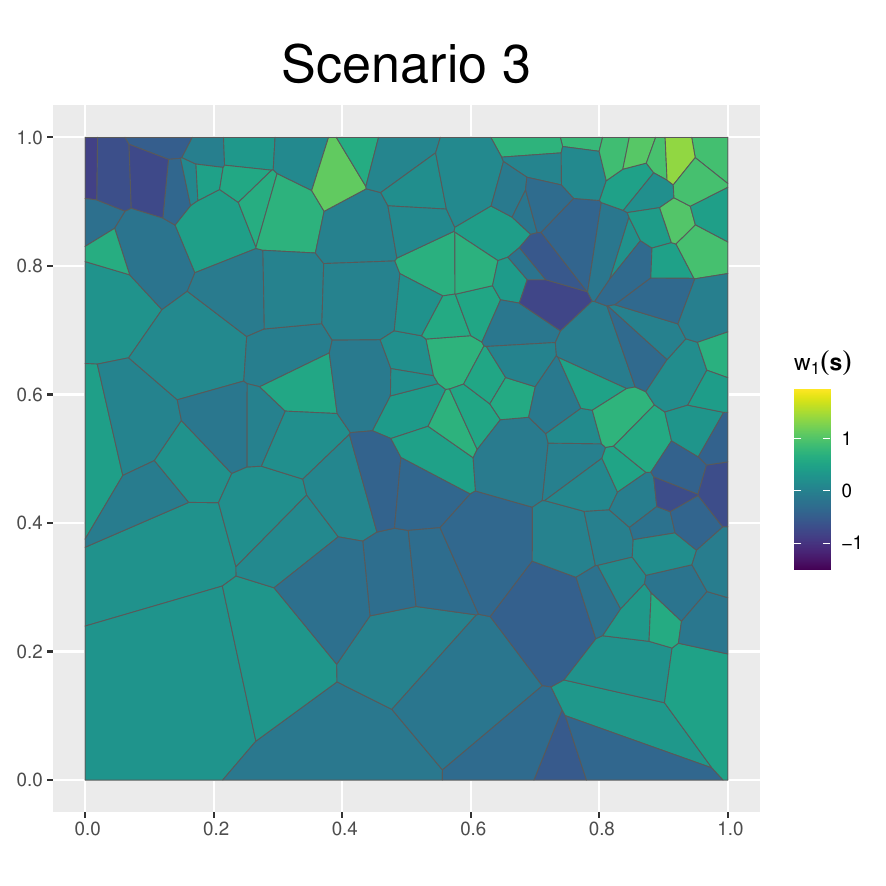}
    \end{minipage}
    \begin{minipage}{0.32\textwidth}
        \centering
        \includegraphics[width=\linewidth]{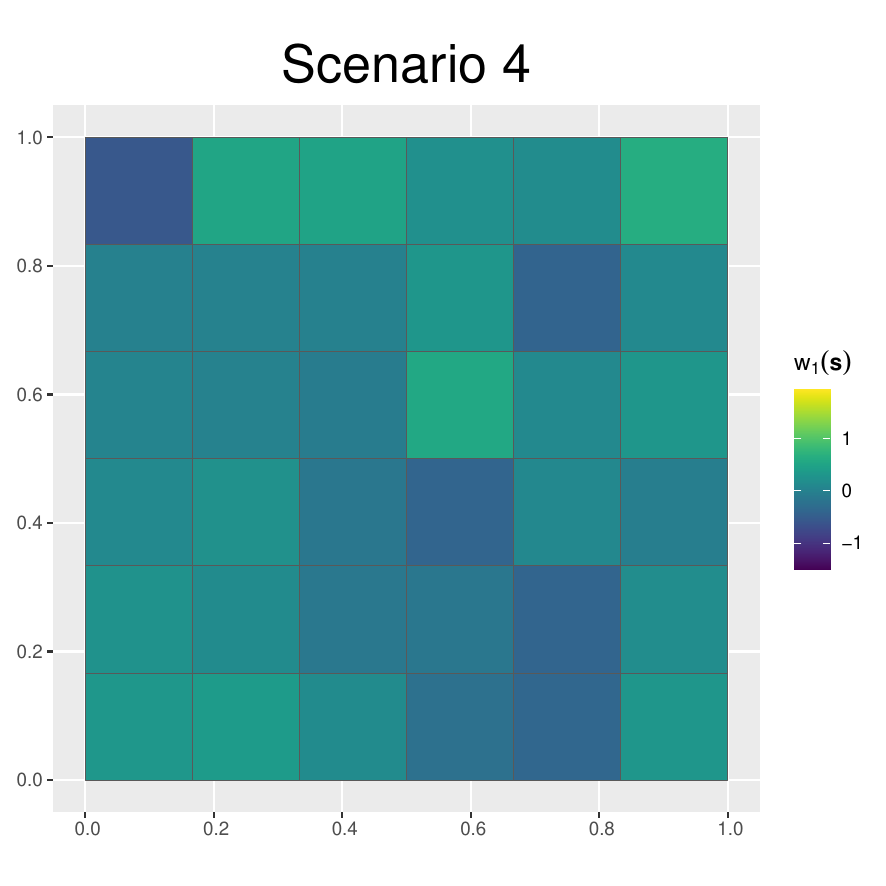}
    \end{minipage}
    \begin{minipage}{0.32\textwidth}
        \centering
        \includegraphics[width=\linewidth]{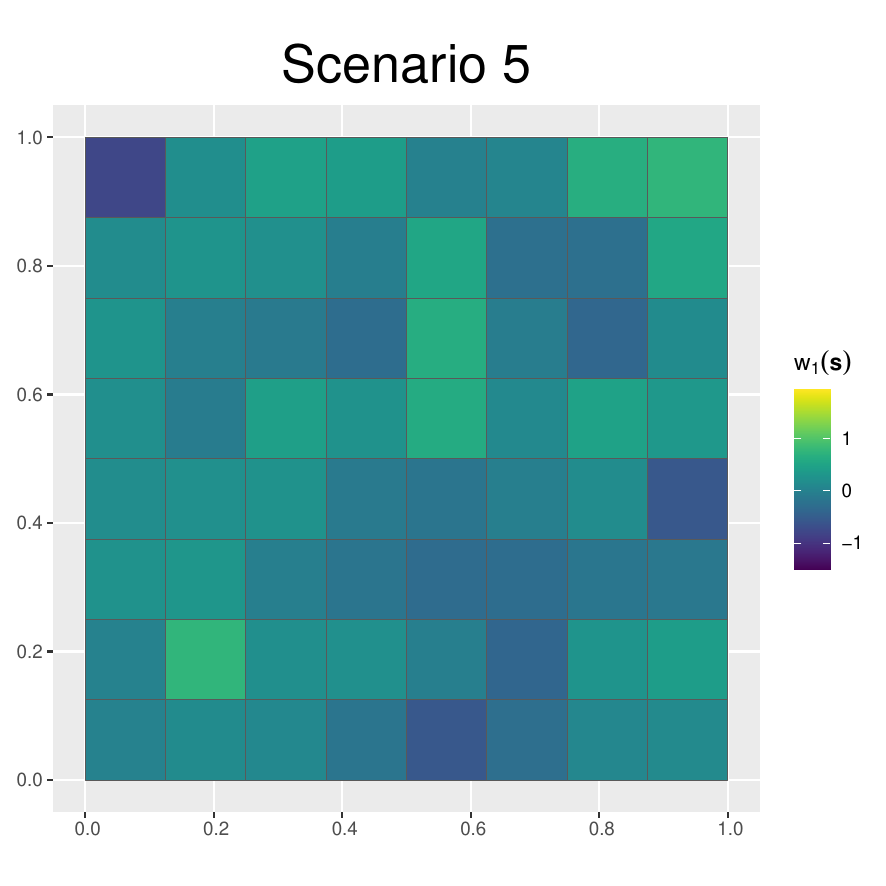}
    \end{minipage}
    \begin{minipage}{0.32\textwidth}
        \centering
        \includegraphics[width=\linewidth]{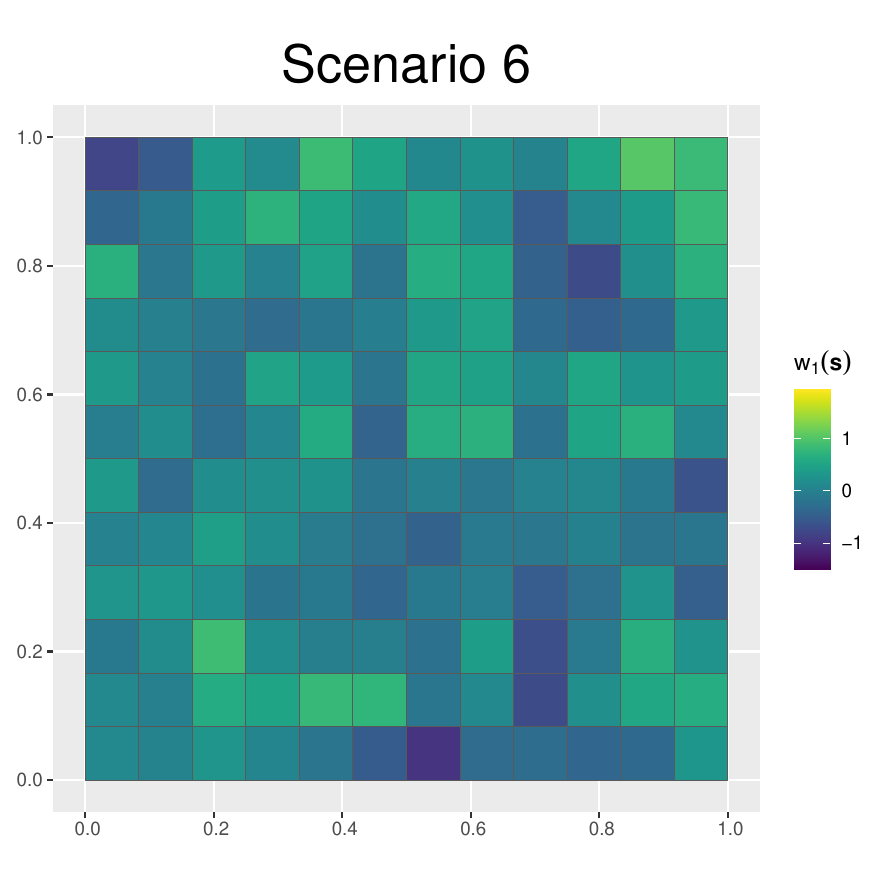}
    \end{minipage}
    \caption{Realization of the first variable, $W_1(\mathbf{s})$, averaged over areas for different scenarios.}
    \label{fig:sim2}
\end{figure}

Figure \ref{fig:sim2} shows a realization of the first process, \( W_1(\mathbf{s}) \), from Figure \ref{fig:sim1}, but observed at the aggregated level in one of the six scenarios. In the first three scenarios, the regions in the lower-left corner are notably larger than those in the upper-right corner. This variation arises because the intensity function of the inhomogeneous Poisson process that generates the irregular regions, defined in Equation \ref{equ:inten}, increases as \((x, y)\) approaches \((1, 1)\).

\subsection{Simulation Results}

\subsubsection{Parameter Estimation}

We present the estimation results for the simulation parameters outlined in Table \ref{tab:param}. For each parameter, we calculated the double-sided $95 \%$ credible intervals and evaluated the proportion of scenarios in which the true parameter value falls within these intervals. Table \ref{tab:sim_parameter_est} summarizes these results, displaying the average coverage rates of the credible intervals across scenarios. Values are color-coded for clarity: green indicates coverage rates equal to or above $95 \%$, yellow corresponds to rates between $80 \%$ and $95 \%$, and red represents rates below $80 \%$.

Table \ref{tab:sim_parameter_est} shows that the intercept parameters ($\alpha_k$) and the correlation parameters ($\lambda_k$) are generally well captured by the credible intervals, with only a few exceptions, possibly due to sampling variability. The spatial variances ($\sigma_k$) for all three processes consistently exhibit excellent coverage, exceeding $95 \%$ across all scenarios. In contrast, the spatial range and standard error display similar patterns, where the average coverage tends to improve as the number of regions increases, regardless of whether the scenario is regular or irregular. Notably, Scenarios 3 and 6 represent the best-performing cases for regular and irregular settings, respectively, with Scenario 6 performing as the overall best scenario in the simulation.

\begin{table}[h!]
\centering
\renewcommand{\arraystretch}{1.5}
\newcolumntype{?}[1]{!{\vrule width #1}}
\begin{tabular}{|c|c|c|c?{0.4mm}c|c|c|}
\hline
\multirow{2}{*}{\textbf{Variable}} & \multicolumn{6}{c|}{\textbf{Scenarios}} \\ \cline{2-7}
                          & \multicolumn{3}{c?{0.4mm}}{\textbf{Irregular}} & \multicolumn{3}{c|}{\textbf{Regular}} \\ \cline{2-7}
                          & \textbf{1} & \textbf{2} & \textbf{3} & \textbf{4} & \textbf{5} & \textbf{6} \\ \hline
Intercept 1 ($\alpha_1$)  & \cellcolor{green!25}95\% & \cellcolor{green!25}96\% & \cellcolor{green!25}98\% & \cellcolor{yellow!25}91\% & \cellcolor{green!25}95\% & \cellcolor{green!25}97\% \\ \hline
Intercept 2 ($\alpha_2$)  & \cellcolor{green!25}97\% & \cellcolor{yellow!25}92\% & \cellcolor{green!25}96\% & \cellcolor{yellow!25}92\% & \cellcolor{green!25}95\% & \cellcolor{green!25}100\% \\ \hline
Intercept 3 ($\alpha_3$)  & \cellcolor{green!25}96\% & \cellcolor{green!25}97\% & \cellcolor{green!25}99\% & \cellcolor{green!25}96\% & \cellcolor{green!25}95\% & \cellcolor{green!25}97\% \\ \hline
Lambda 1 ($\lambda_1$)     & \cellcolor{green!25}98\% & \cellcolor{yellow!25}92\% & \cellcolor{green!25}96\% & \cellcolor{green!25}100\% & \cellcolor{green!25}97\% & \cellcolor{green!25}96\% \\ \hline
Lambda 2 ($\lambda_2$)   & \cellcolor{green!25}96\% & \cellcolor{green!25}96\% & \cellcolor{yellow!25}93\% & \cellcolor{green!25}99\% & \cellcolor{green!25}98\% & \cellcolor{green!25}96\% \\ \hline
Lambda 3 ($\lambda_3$)  & \cellcolor{green!25}95\% & \cellcolor{yellow!25}92\% & \cellcolor{yellow!25}92\% & \cellcolor{green!25}98\% & \cellcolor{green!25}96\% & \cellcolor{green!25}96\% \\ \hline
Spatial Range 1 ($\rho_1$) & \cellcolor{red!25}71\% & \cellcolor{yellow!25}93\% & \cellcolor{yellow!25}94\% & \cellcolor{red!25}40\% & \cellcolor{yellow!25}83\% & \cellcolor{green!25}97\% \\ \hline
Spatial Variance 1 ($\sigma_1$) & \cellcolor{green!25}99\% & \cellcolor{green!25}98\% & \cellcolor{yellow!25}91\% & \cellcolor{green!25}96\% & \cellcolor{green!25}97\% & \cellcolor{green!25}98\% \\ \hline
Spatial Range 2 ($\rho_2$) & \cellcolor{red!25}75\% & \cellcolor{yellow!25}92\% & \cellcolor{green!25}95\% & \cellcolor{red!25}75\% & \cellcolor{yellow!25}93\% & \cellcolor{yellow!25}94\% \\ \hline
Spatial Variance 2 ($\sigma_2$) & \cellcolor{green!25}95\% & \cellcolor{green!25}96\% & \cellcolor{green!25}96\% & \cellcolor{green!25}99\% & \cellcolor{green!25}97\% & \cellcolor{green!25}97\% \\ \hline
Spatial Range 3 ($\rho_3$) & \cellcolor{red!25}40\% & \cellcolor{red!25}67\% & \cellcolor{yellow!25}93\% & \cellcolor{red!25}15\% & \cellcolor{red!25}46\% & \cellcolor{yellow!25}92\% \\ \hline
Spatial Variance 3 ($\sigma_3$) & \cellcolor{green!25}98\% & \cellcolor{green!25}98\% & \cellcolor{green!25}97\% & \cellcolor{green!25}98\% & \cellcolor{green!25}97\% & \cellcolor{green!25}97\% \\ \hline
Standard Error 1 ($e_1$) & \cellcolor{red!25}65\% & \cellcolor{yellow!25}86\% & \cellcolor{yellow!25}91\% & \cellcolor{red!25}40\% & \cellcolor{red!25}78\% & \cellcolor{yellow!25}91\% \\ \hline
Standard Error 2 ($e_2$) & \cellcolor{red!25}79\% & \cellcolor{green!25}97\% & \cellcolor{green!25}98\% & \cellcolor{red!25}77\% & \cellcolor{yellow!25}94\% & \cellcolor{yellow!25}94\% \\ \hline
Standard Error 3 ($e_3$) & \cellcolor{red!25}42\% & \cellcolor{red!25}73\% & \cellcolor{yellow!25}92\% & \cellcolor{red!25}20\% & \cellcolor{red!25}48\% & \cellcolor{yellow!25}92\% \\ \hline
\end{tabular}
\vspace{0.2cm}
\caption{Summary of parameters estimation: Coverage probabilities for the intercept parameters ($\alpha_k$), correlation parameters ($\lambda_k$), spatial variances ($\sigma_k$), spatial ranges ($\rho_k$), and standard errors ($e_k$) across six scenarios. Green indicates coverage $\geq 95\%$, yellow indicates $80\% \leq$ coverage $< 95\%$, and red indicates coverage $< 80\%$. Regular scenarios correspond to Scenarios 1–3, while irregular scenarios correspond to Scenarios 4–6.}
\label{tab:sim_parameter_est}
\end{table}

\subsubsection{Comparison between scenarios}

\begin{figure}[h]
\centering
\includegraphics[scale=0.5]{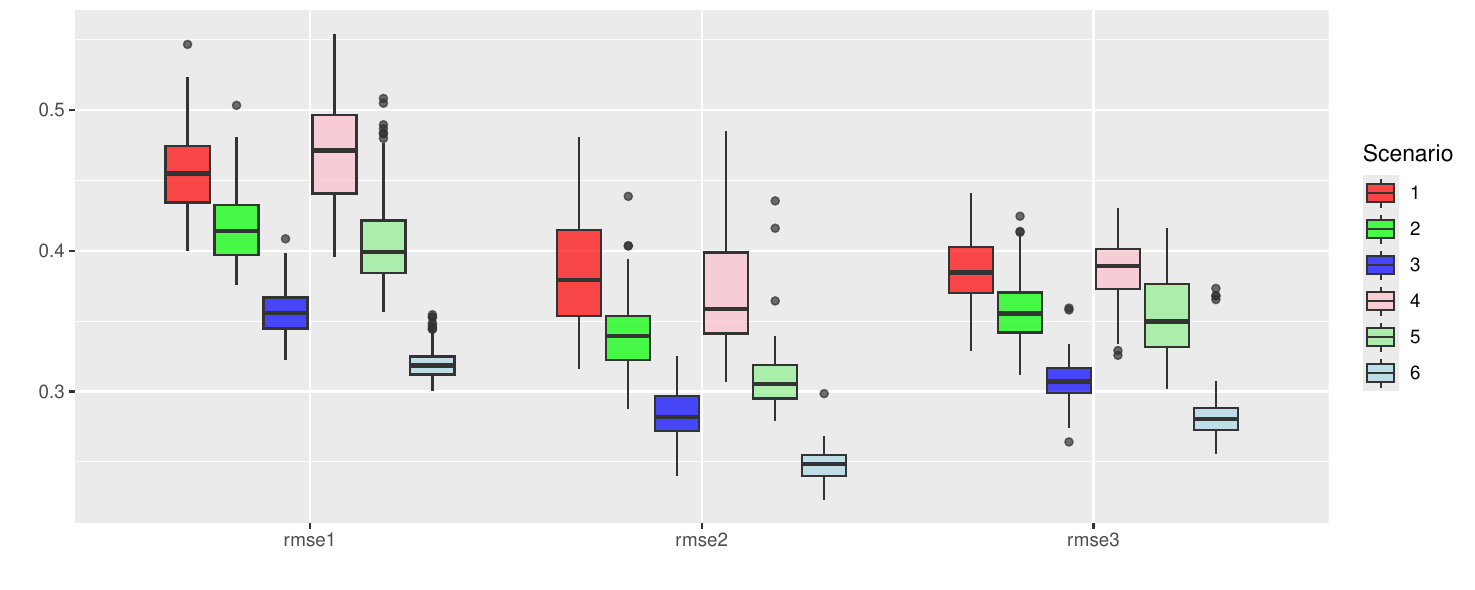}
\caption{Root mean square error between the prediction mean and true values.
}
\label{fig:sim-rmse}
\end{figure}

Figure \ref{fig:sim-rmse} describes the distribution of the root mean squared error (RMSE) for the three processes described in Equation \ref{equ:sim1} and the posterior mean in each of the scenarios. First, we can see that increasing the number of areas improves the approximation in terms of reducing the RMSE in the regular and irregular cases. On the other hand, we can see that the regular case (light colors) has, on average, most of the cases, a better approximation than the irregular case (vivid colors), showing this property in the mean value of the RMSE of the three variables. Ultimately, we can notice that the best approximation was the regular case with the most regions. Nevertheless, the approximation with the irregular areas with the biggest number (Scenario 3) is, on average, better than the regular areas with many regions (Scenario 5). So, we can conclude that increasing the number of regions positively and directly impacts the approximation.

\subsubsection{Comparison between models}

To highlight the main advantages of our model, we compare it with an areal model as specified in Section \ref{sec:multivariate-areal}. For the inference of this areal model, we use the values generated from Equation \ref{equ:sim2} and apply the same structure to model a multivariate areal process for the six previously defined scenarios. The marginal $\text{RMSE}_k$ for the continuous model is computed as the difference between the true and predicted values at each grid point. In contrast, for the areal model, predictions are generated for each region, assigning the same predicted value to all grid points within a region. The marginal $\text{RMSE}_k$ for the areal model is then calculated as the difference between these regional predictions and the true values at each grid point, for $k = 1, 2, 3$.

\begin{figure}[h]
\centering
\includegraphics[scale=0.7]{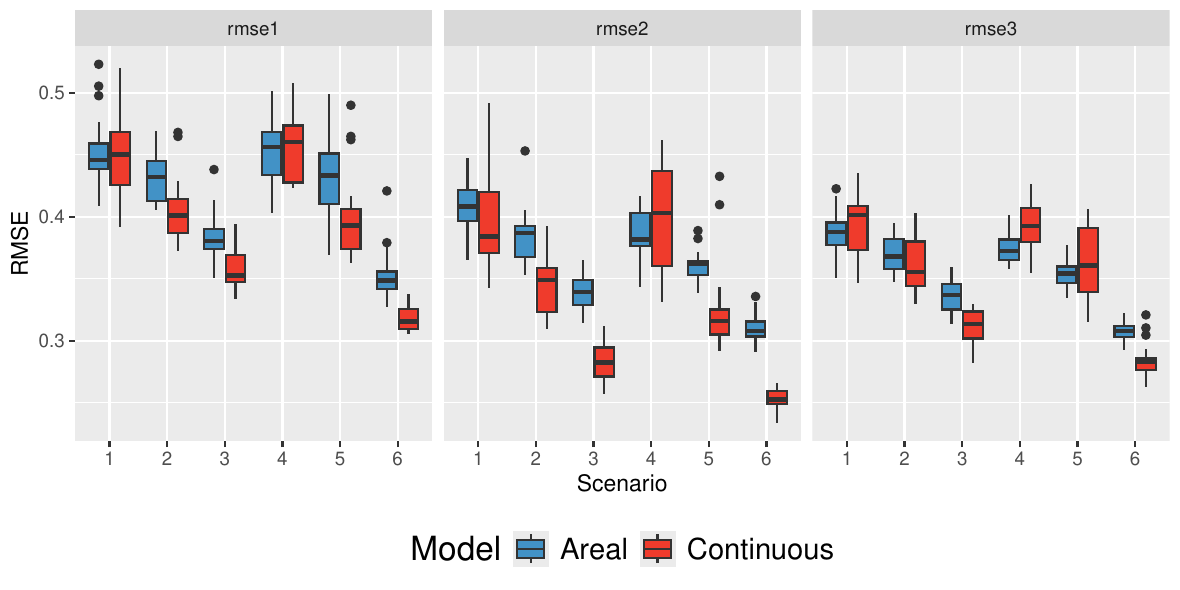}
\caption{Root mean square error between the prediction mean and true values for the areal model (blue) and our continuous disaggregation model (red).}
\label{fig:sim-area-rmse}
\end{figure}

Figure \ref{fig:sim-area-rmse} displays the distribution of $\text{RMSE}_k$ across different realizations for the six scenarios. The results demonstrate that the $\text{RMSE}_k$, is, on average, lower for the continuous model (red) compared to the areal model (blue) across most scenarios. Additionally, the figure highlights that increasing the number of areas, whether consisting of irregular or regular regions, enhances the approximation accuracy of the continuous model more effectively than the areal model. These findings suggest that the continuous model consistently outperforms the areal model in providing a more accurate representation of the underlying spatial process.

Finally, in the Appendix we present a realization of the simulated processes, and the predictions from the disaggregation and areal models, corresponding to the simulation that has the lowest global RMSE in each of the sixth scenarios.

\section{Multivariate modeling of air contaminants in Portugal and Italy}
\label{sec:app}

In this study, we apply our method to atmospheric data, focusing specifically on air pollution measurements. The primary objective is to achieve finer estimations of key air quality indicators, such as PM2.5, PM10, and ozone concentrations.
We jointly modeled these three components to leverage the potential relationships between them. For instance, studies in China demonstrated that specific chemical processes can cause PM2.5 to decrease ozone concentrations \cite{ozoneandpm25}. Moreover, research analyzing particulate matter in Shijiazhuang, northern China, highlighted significant correlations between PM2.5 and PM10 concentrations, with a PM2.5/PM10 mass ratio of 0.7. These findings illustrate the interconnected dynamics of air pollutants and their measurable relationships that we want to extrapolate to European countries.

We utilize PM2.5, PM10, and ozone levels in Portugal and Italy from the reanalysis dataset of atmospheric composition provided by the Copernicus Atmosphere Monitoring Service (CAMS), a high-resolution product developed by the European Centre for Medium-Range Weather Forecasts (ECMWF) \cite{copernicus_data}.
This dataset offers annual air quality reanalyses for Europe based on both unvalidated (interim) and validated observations, making it a suitable resource for testing the effectiveness of our modeling approach in refining spatial estimations.

Portugal and Italy were chosen due to their moderate size, geographical diversity, and proximity to the sea. This choice ensures that the study encompasses a range of environmental and geographical conditions while maintaining computational efficiency and methodological robustness. Additionally, both countries provide valuable case studies for investigating spatial patterns of emissions, given their varying topographies and coastal influences.
The data corresponds to January 2019 and consists of hourly observations with a spatial resolution of 0.1 degrees, approximately equivalent to a 10 km scale. To facilitate the analysis, the data was aggregated to compute a monthly average of hourly emissions for the three variables under consideration.

A primary objective of this model is to achieve spatial downscaling, reducing the resolution from 0.1 degrees to 0.02 degrees, equivalent to increasing the spatial detail from approximately 10 km to 2 km. This approach allows for greater granularity, generating $5 \times 5 = 25$ predictions within each original observation's spatial extent. Such enhanced resolution facilitates a more detailed and nuanced analysis of spatial patterns.
The modeling structure applied in this study follows a framework similar to that defined in Equations \ref{equ:sim1} and \ref{equ:sim2}. However, altitude is incorporated as an additional covariate by using the following model:

\begin{align}
\begin{split}
    \text{pm}_{2.5}(\mathbf{s}) & = \alpha_1+\beta_1 \cdot \text{altitude} + z_{1}(\mathbf{s}) + e_1(\mathbf{s}) \\
    \text{pm}_{10}(\mathbf{s}) & = \alpha_2 +\beta_2 \cdot \text{altitude} + \lambda_1 z_{1}(\mathbf{s})+z_2(\mathbf{s})+ e_2(\mathbf{s})\\
    \text{ozone}(\mathbf{s}) & = \alpha_3 + \beta_3 \cdot \text{altitude} + \lambda_2 z_{1}(\mathbf{s})+\lambda_3 z_2(\mathbf{s})+z_3(\mathbf{s}) +e_3(\mathbf{s})
\end{split}
\label{equ:ap1}
\end{align}

\begin{figure}[!htp]
    \centering
    \begin{minipage}{0.4\textwidth}
        \centering
        \includegraphics[width=\linewidth]{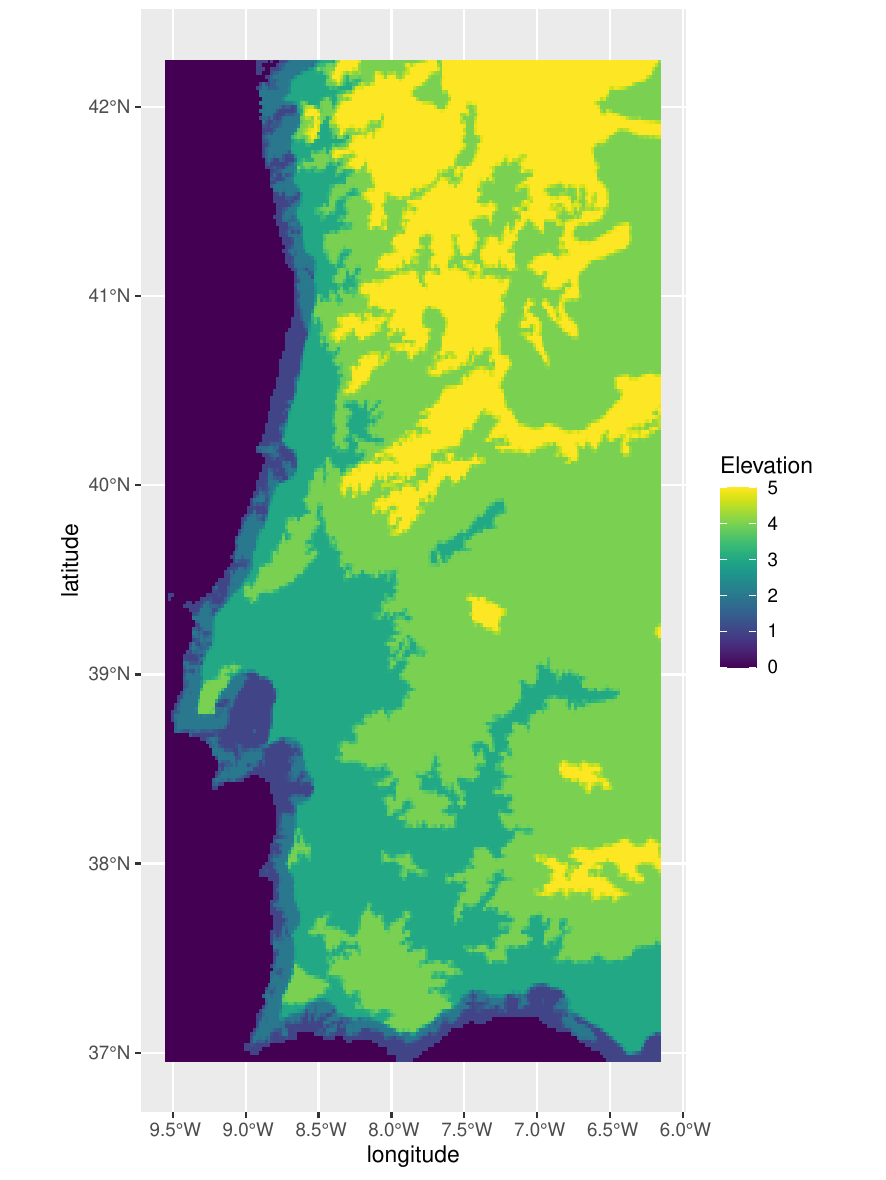}
    \end{minipage}
    \begin{minipage}{0.58\textwidth}
        \centering
        \includegraphics[width=\linewidth]{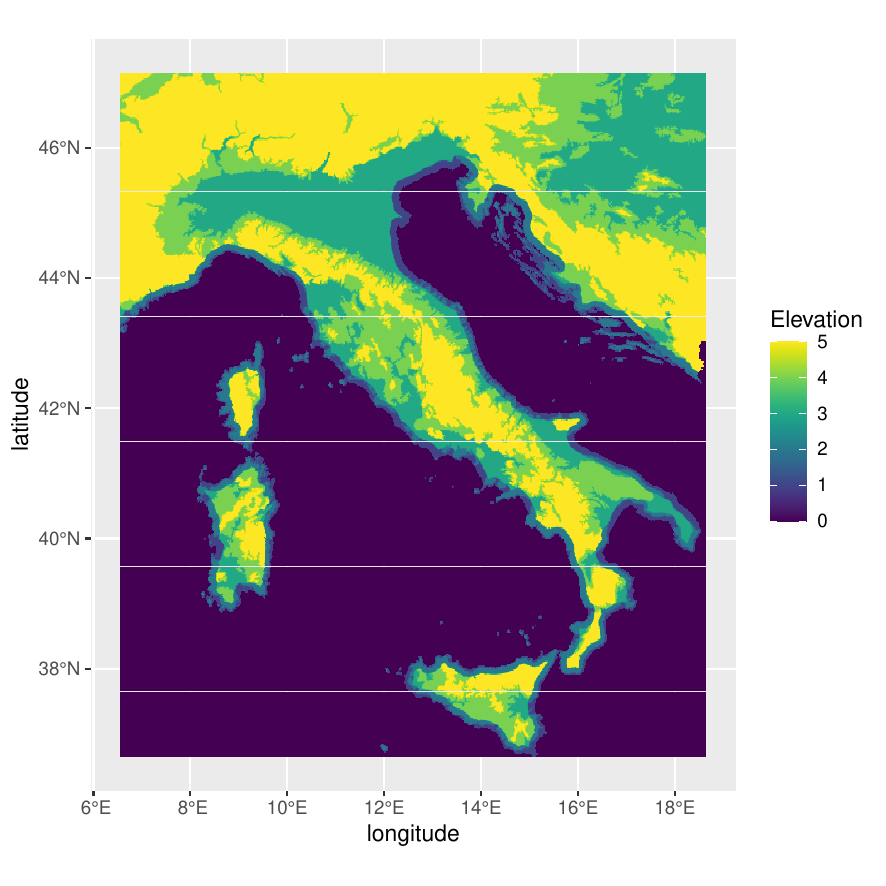}
    \end{minipage}
    \caption{Elevation in $km$ for Portugal (left panel) and Italy (right panel), respectively ($0.01^\circ \times 0.01^\circ \approx 1$km $\times 1$km resolution).}
    \label{fig:elev}
\end{figure}

Figure \ref{fig:elev} presents the elevation data for Portugal and Italy.
The data is provided at a fine spatial resolution of 0.01 degrees, corresponding to approximately 1 km in latitude and longitude. This high-resolution dataset enables the capture of nuanced topographical variations, which is essential for accurately modeling spatial processes influenced by elevation. The dataset was sourced from the European Environment Agency \cite{eea_elevation_2024}, ensuring its reliability and consistency across the study regions. By incorporating elevation as a covariate, we aim to better understand its role in driving spatial variability within the model, particularly in the context of environmental and geographical gradients across these countries.

\subsection{Portugal}
\label{sec:portugal}

\begin{figure}[htp]
\begin{subfigure}{1\textwidth}
    \begin{minipage}{0.32\textwidth}
        \includegraphics[width=\linewidth]{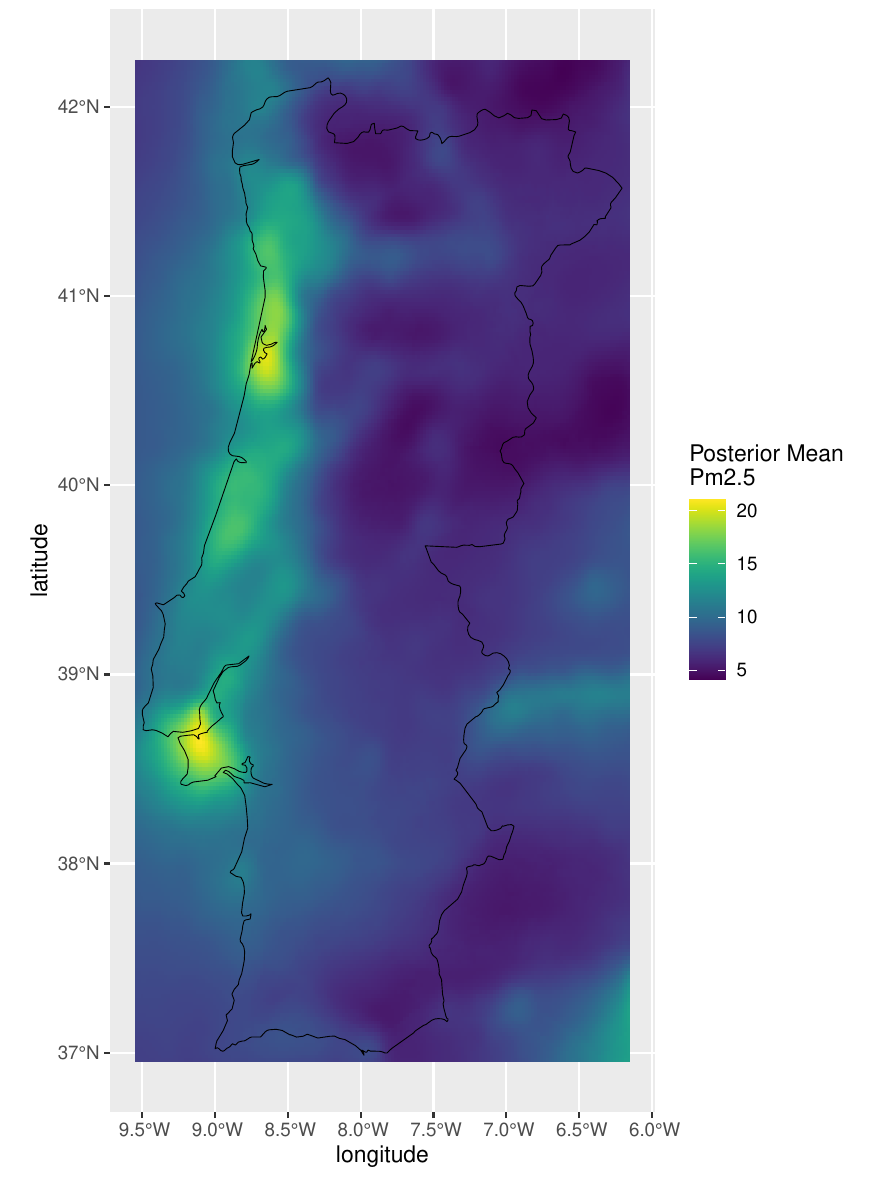}
    \end{minipage}
    \begin{minipage}{0.32\textwidth}
        \includegraphics[width=\linewidth]{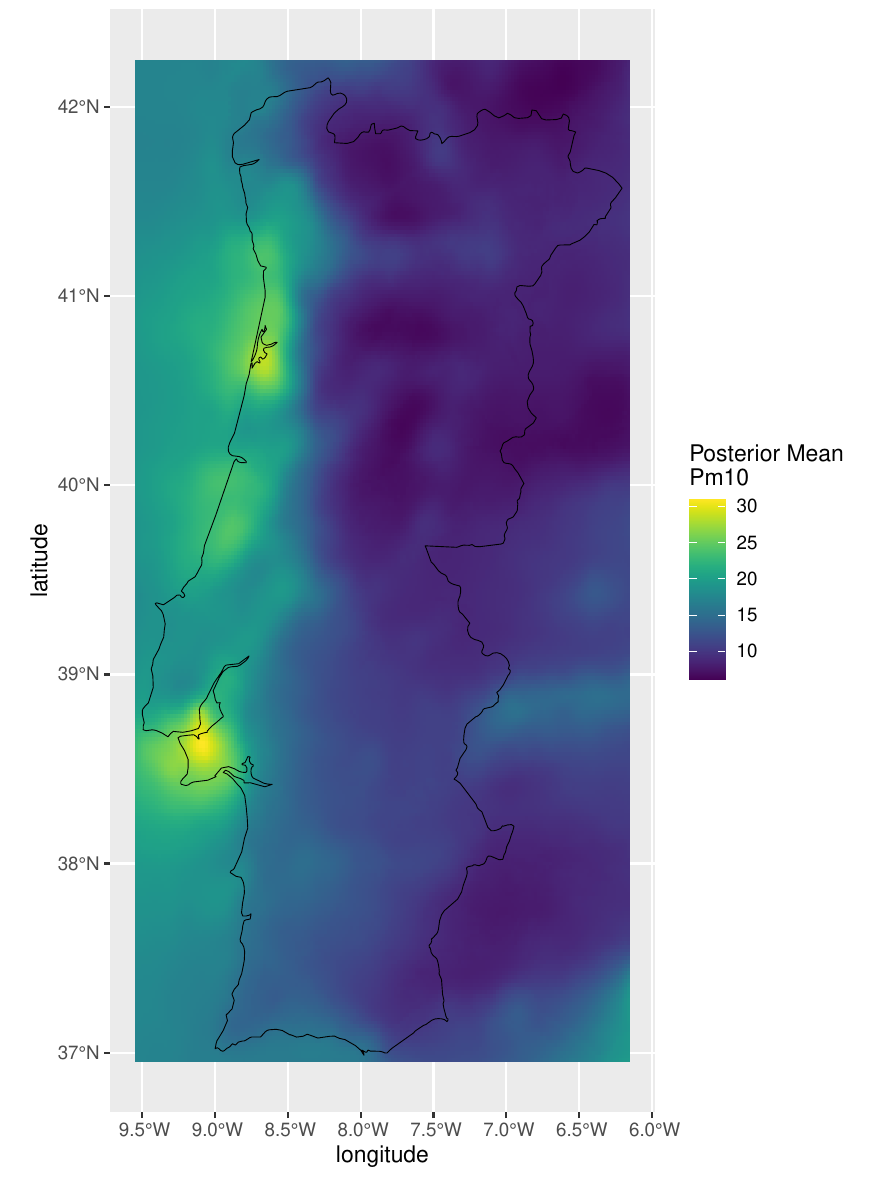}
    \end{minipage}
    \begin{minipage}{0.32\textwidth}
        \includegraphics[width=\linewidth]{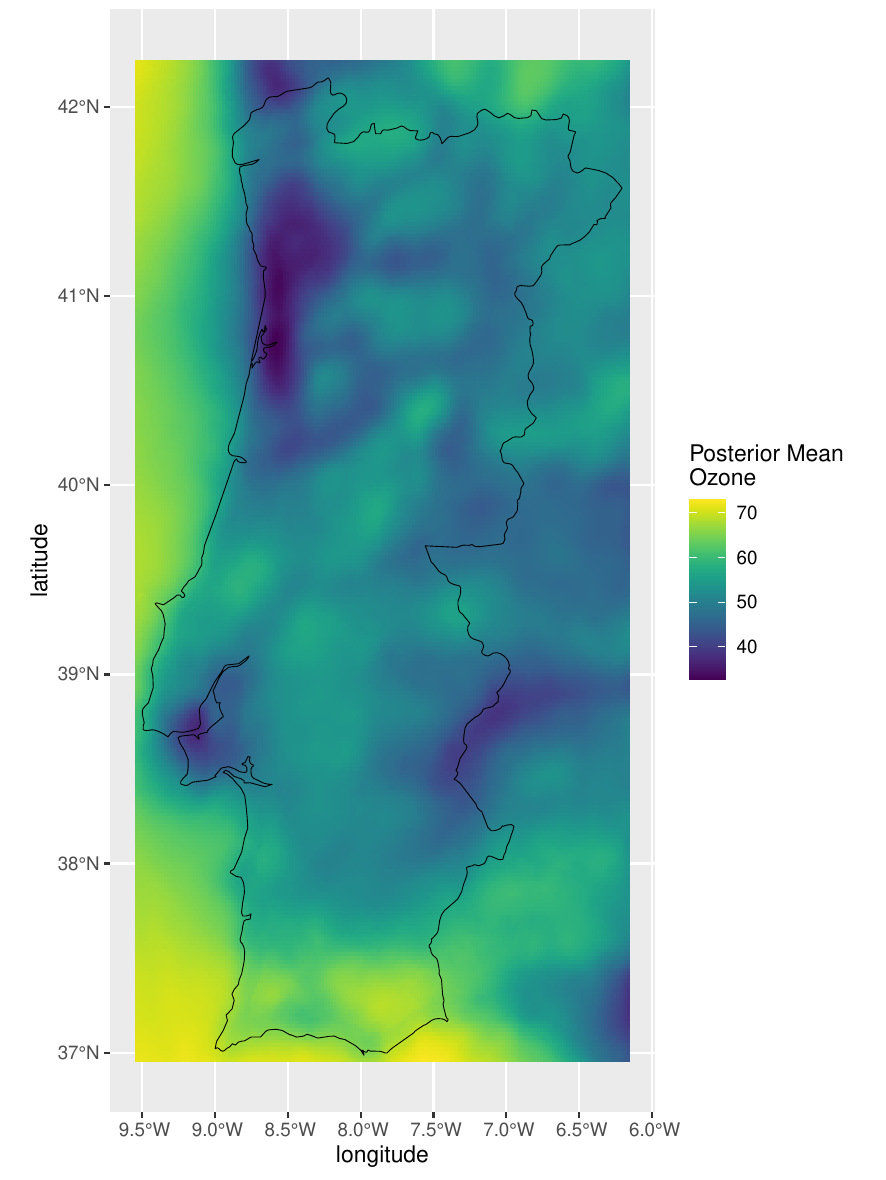}
    \end{minipage}
    \caption{Posterior mean of our model ($0.02^{\circ} \times 0.02^{\circ} \approx 2$km $\times 2$km resolution)}
\end{subfigure}
\begin{subfigure}{1\textwidth}
    \begin{minipage}{0.32\textwidth}
        \includegraphics[width=\linewidth]{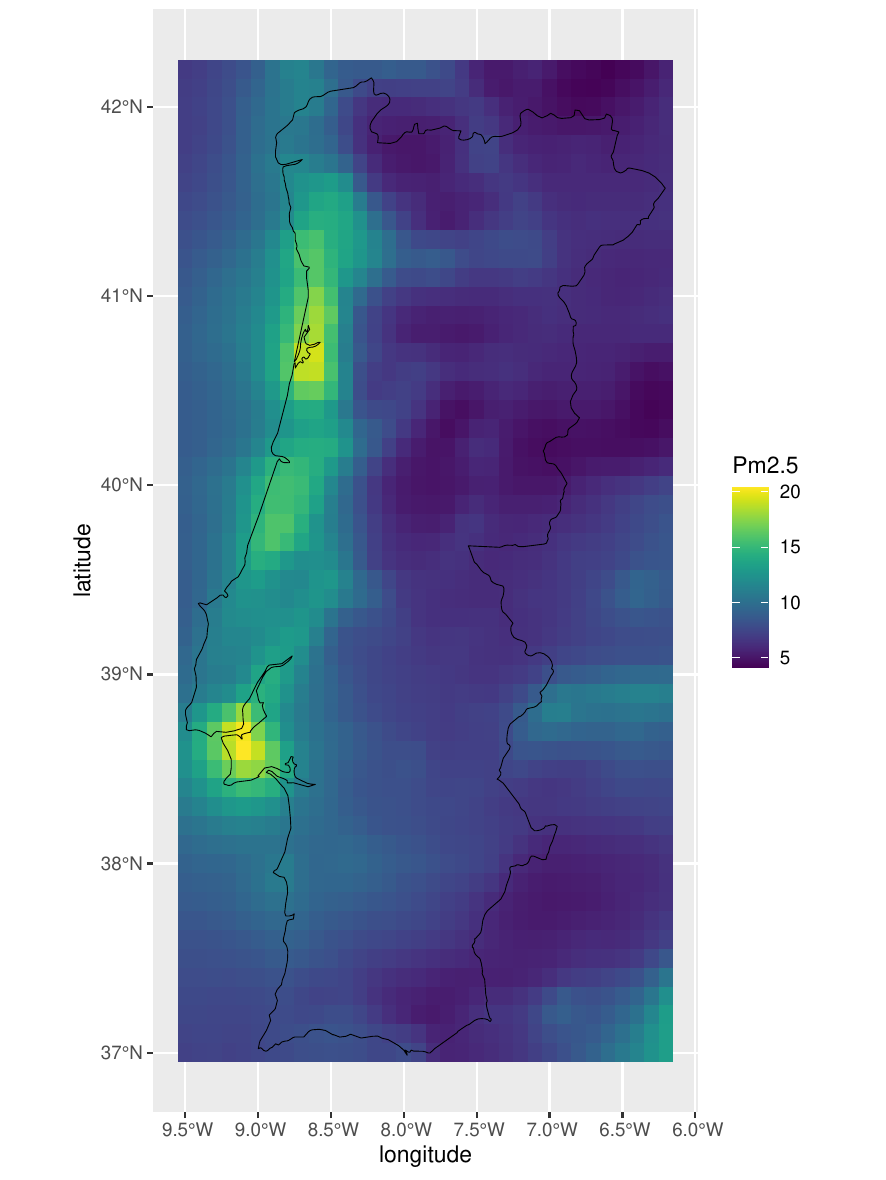}
    \end{minipage}
    \begin{minipage}{0.32\textwidth}
        \includegraphics[width=\linewidth]{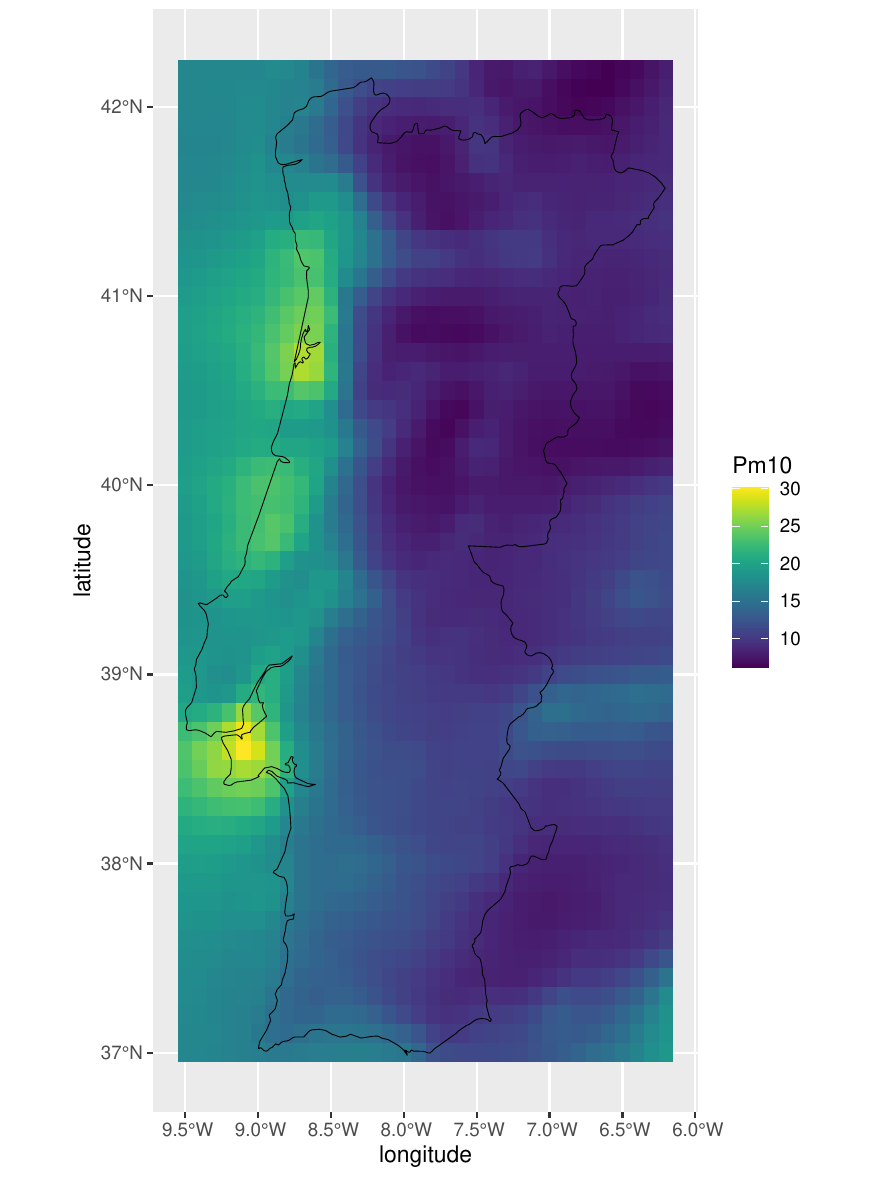}
    \end{minipage}
    \begin{minipage}{0.32\textwidth}
        \includegraphics[width=\linewidth]{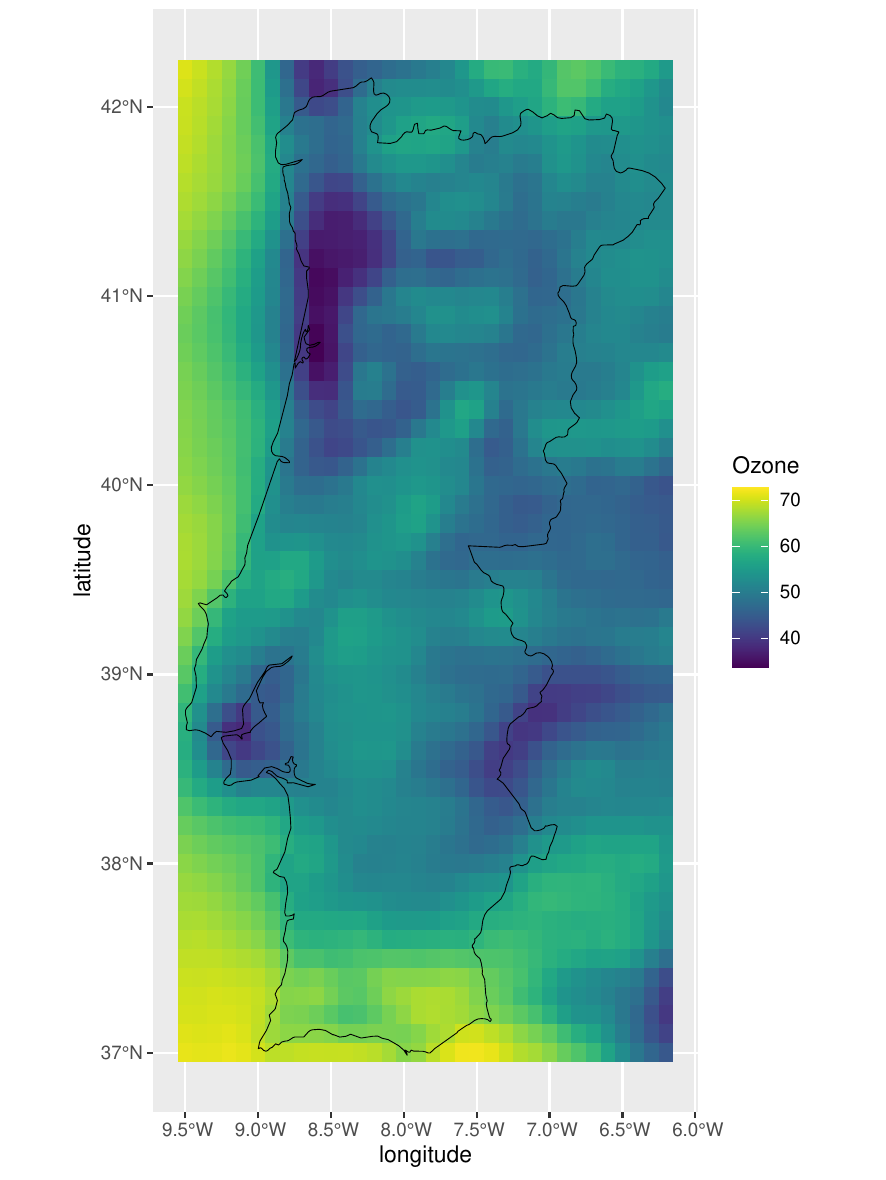}
    \end{minipage}
    \caption{True values ($0.1^{\circ} \times 0.1^{\circ} \approx 10$km $\times 10$km resolution)}
\end{subfigure}
\caption{Posterior mean of our model and true values for the mainland of Portugal.}
\label{fig:port-norm}
\end{figure}

\begin{figure}[htp]
\begin{subfigure}{1\textwidth}
    \begin{minipage}{0.32\textwidth}
        \includegraphics[width=\linewidth]{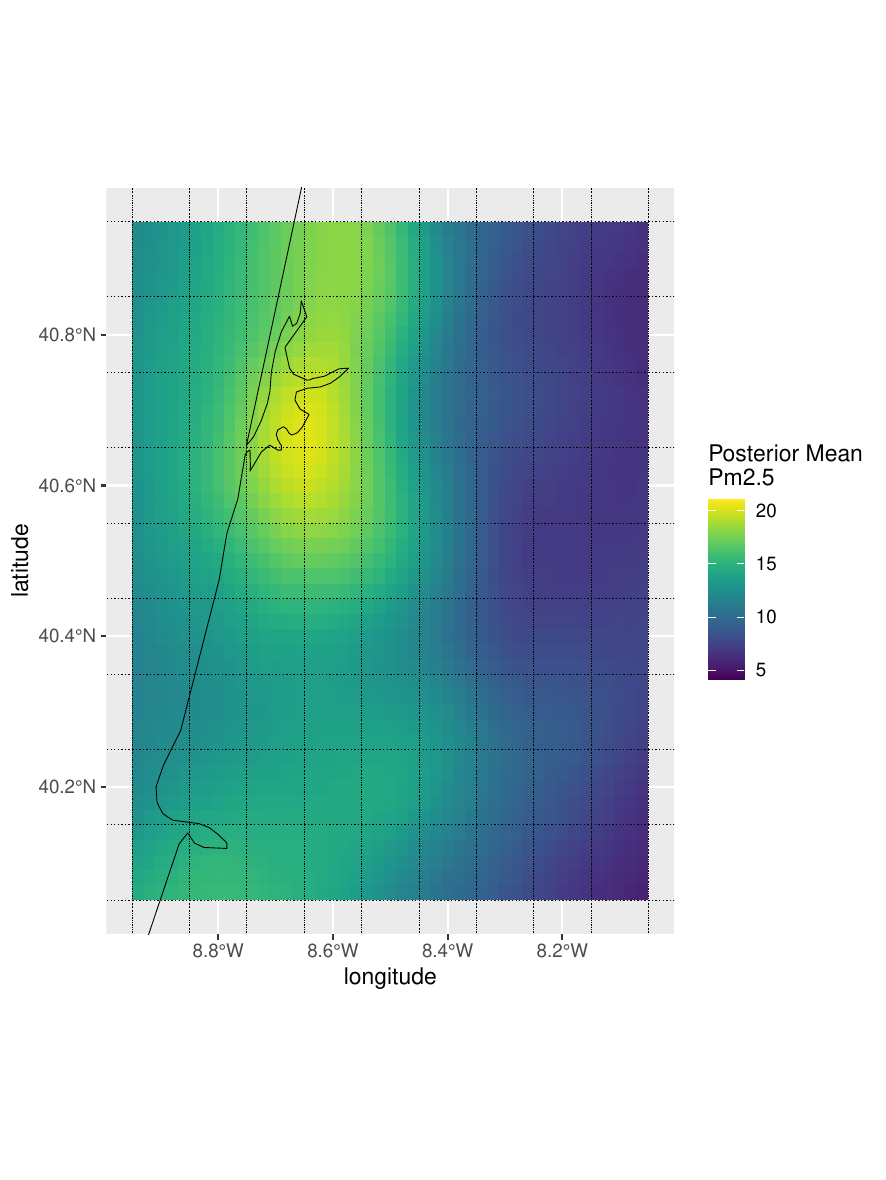}
    \end{minipage}
    \begin{minipage}{0.32\textwidth}
        \includegraphics[width=\linewidth]{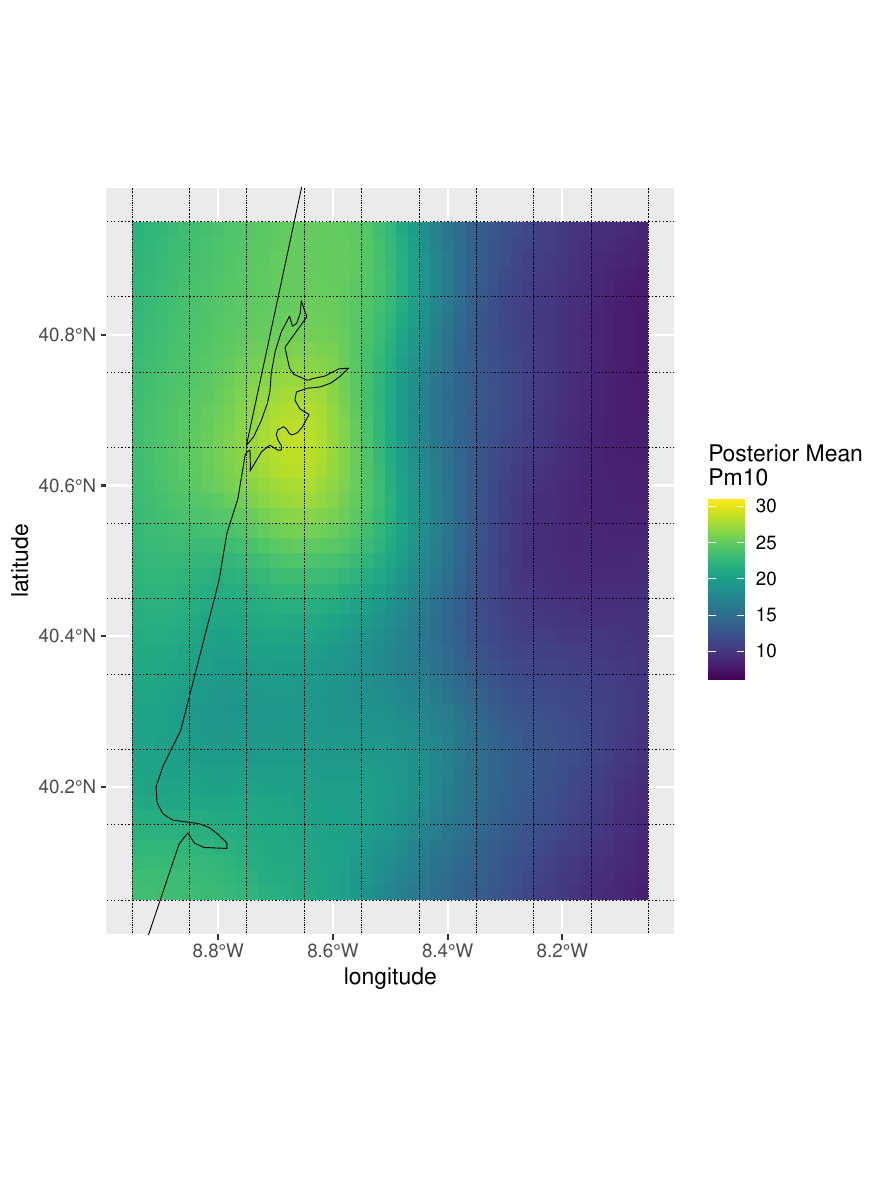}
    \end{minipage}
    \begin{minipage}{0.32\textwidth}
        \includegraphics[width=\linewidth]{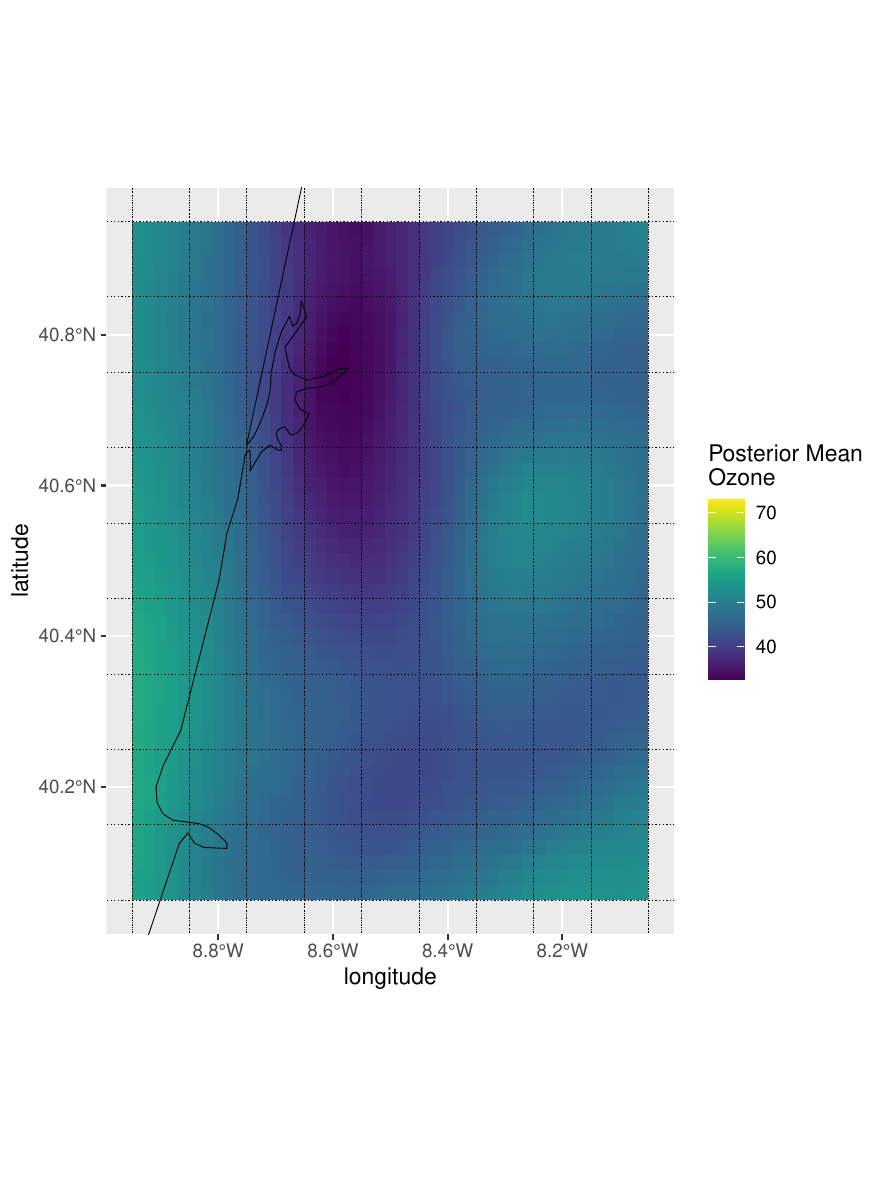}
    \end{minipage}
    \caption{Posterior mean of our model ($0.02^{\circ} \times 0.02^{\circ} \approx 2$km $\times 2$km resolution)}
\end{subfigure}
\begin{subfigure}{1\textwidth}
    \begin{minipage}{0.32\textwidth}
        \includegraphics[width=\linewidth]{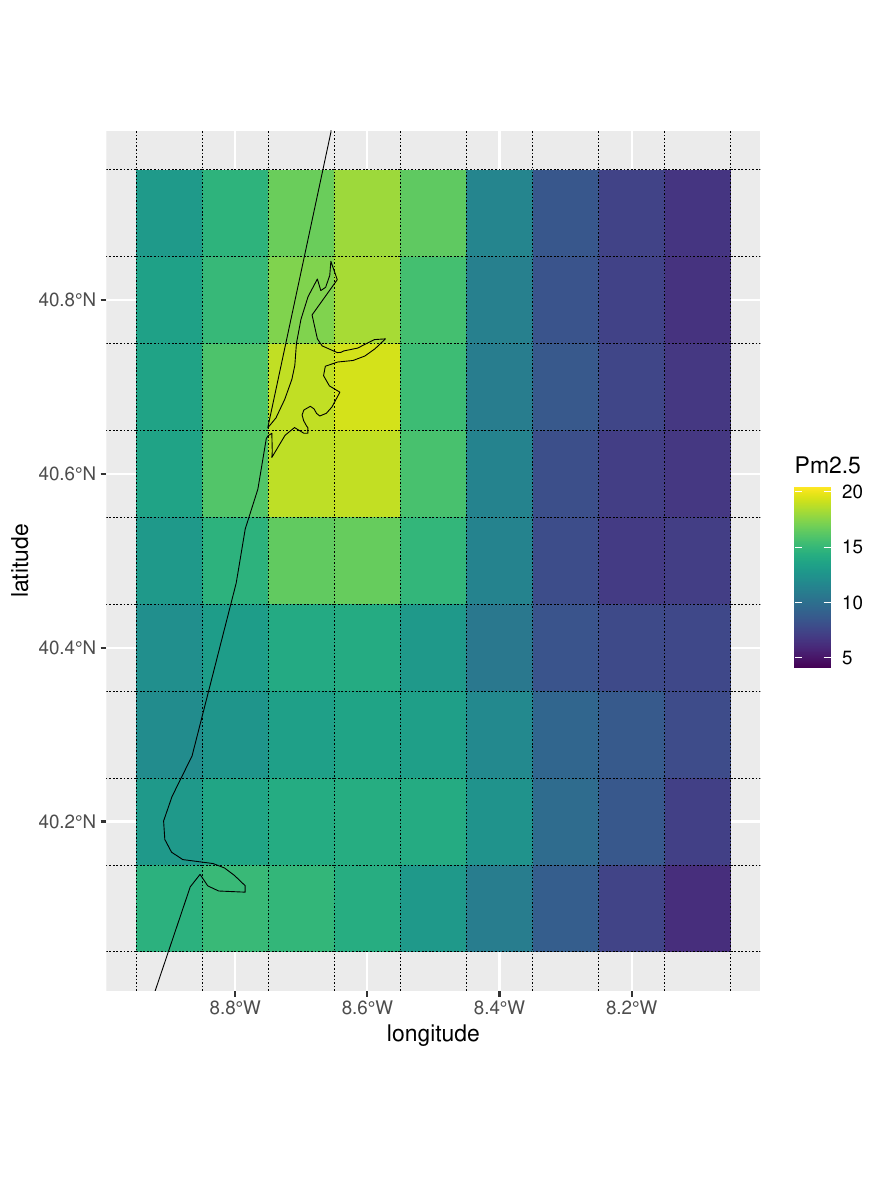}
    \end{minipage}
    \begin{minipage}{0.32\textwidth}
        \includegraphics[width=\linewidth]{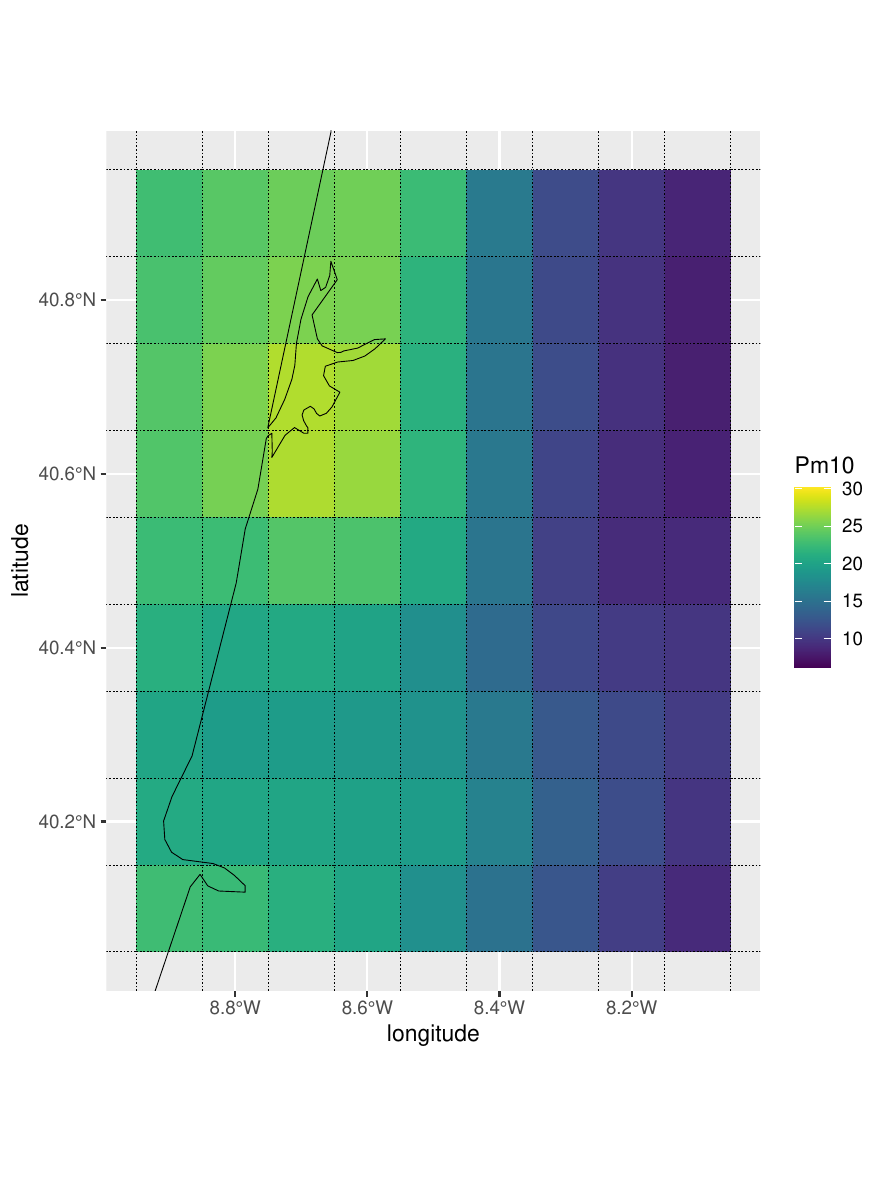}
    \end{minipage}
    \begin{minipage}{0.32\textwidth}
        \includegraphics[width=\linewidth]{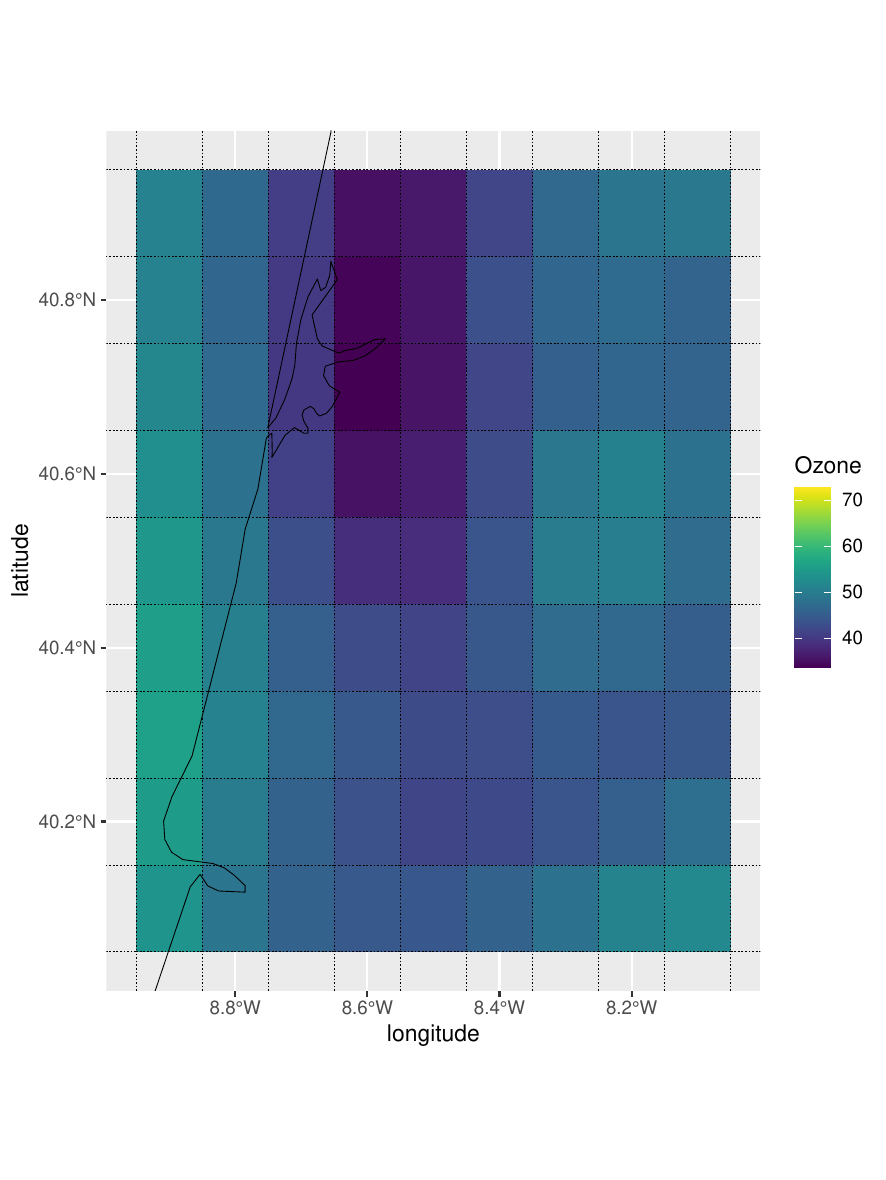}
    \end{minipage}
    \caption{True values ($0.1^{\circ} \times 0.1^{\circ} \approx 10$km $\times 10$km resolution)}
\end{subfigure}
\caption{Posterior mean of our model and true values for a zoom version of the mainland of Portugal.}
\label{fig:port-zoom}
\end{figure}

In Figure \ref{fig:port-norm}, we present the original values and the posterior mean from the disaggregation model applied to PM2.5, PM10, and ozone emissions in Portugal. This figure demonstrates that the model preserves each variable's original scale and spatial patterns while providing a smoother data representation. Importantly, the smoothing effect does not distort the distribution of minimal and maximal values, maintaining the integrity of the original dataset. For enhanced visualization, a zoomed-in version of the plot is provided in Figure \ref{fig:port-zoom}, focusing on a specific spatial range. In this zoomed view, it becomes evident that each original observation in the dataset has been subdivided into 25 smaller squares in the downscaling process, resulting in a finer and smoother representation of the data.

In Table \ref{tab:pr_fixed_parameters}, we present a summary of the fixed parameters of the model. In this table, the intercepts $\alpha_k$ represent the mean value of each of the three components. Specifically, the mean PM2.5 concentration over the spatial domain is $\alpha_1 = 8.337$ $\mu g/m^3$, the mean PM10 concentration is $\alpha_2 = 14.58$ $\mu g/m^3$, and the mean ozone concentration is $\alpha_3 = 55.716$ $\mu g/m^3$. Notably, the credible intervals for these values are consistently positive.

\begin{table}[h!]
\centering
\begin{tabular}{|c|c|c|c|c|}
\hline
\textbf{Parameter}  & \textbf{Mean} & \textbf{0.025 Quantile} & \textbf{0.5 Quantile} & \textbf{0.975 Quantile} \\ \hline
$\alpha_1$  & 8.33     & 7.83     & 8.33     & 8.84     \\ \hline
$\alpha_2$  & 14.58    & 13.95    & 14.58    & 15.22    \\ \hline
$\alpha_3$  & 55.71    & 54.54    & 55.71    & 56.89    \\ \hline
$\beta_1$   & -0.09    & -0.13    & -0.09    & -0.05    \\ \hline
$\beta_2$   & -0.14    & -0.19    & -0.14    & -0.10    \\ \hline
$\beta_3$   & 0.37     & 0.26     & 0.37     & 0.48     \\ \hline
$\lambda_1$ & 1.11     & 1.09     & 1.11     & 1.13     \\ \hline
$\lambda_2$ & -1.58    & -1.68    & -1.59    & -1.47    \\ \hline
$\lambda_3$ & 1.37     & 1.19     & 1.37     & 1.57     \\ \hline
\end{tabular}
\vspace{0.2cm}
\caption{Summary of the intercept ($\alpha_k$) parameters, the elevation effects ($\beta_k$) and the cross-dependence parameters ($\lambda_k$) for Portugal}
\label{tab:pr_fixed_parameters}
\end{table}

The elevation coefficients $\beta_k$ quantify the linear relationship between elevation and each variable. An increase in elevation by 1 km, while holding other variables constant, reduces PM2.5 emissions by $0.09$ $\mu g/m^3$, PM10 emissions by $0.14$ $\mu g/m^3$, and increases ozone emissions by $0.37$ $\mu g/m^3$. Importantly, the credible intervals for these coefficients do not cross zero, confirming the statistical significance of these effects.

\begin{table}[h!]
\renewcommand{\arraystretch}{1.3}
\centering
\begin{tabular}{|c|c|c|c|c|c|}
\hline \multicolumn{2}{|c|}{ \textbf{Correlation} } & \multirow{2}{*}{ \textbf{mean} } & \multirow{2}{*}{ \textbf{0.025 Quantile} } & \multirow{2}{*}{ \textbf{0.5 Quantile} } & \multirow{2}{*}{ \textbf{0.975 Quantile} } \\
\cline { 1 - 2 } \textbf{var1} & \textbf{var2} & & & & \\
\hline pm25 & pm10 & 0.89 & 0.71 & 0.90 & 0.98 \\
\hline pm25 & Ozone & -0.58 & -0.81 & -0.58 & -0.30 \\
\hline pm10 & Ozone & -0.40 & -0.73 & -0.41 & -0.02 \\
\hline
\end{tabular}
\vspace{0.2cm}
\caption{Summary of correlations between the response variables PM2.5, PM10, and Ozone in Portugal. The table includes the mean correlation values along with the $2.5 \%$, $50 \%$ (median), and $97.5 \%$ quantiles of their credible intervals, demonstrating the strength and significance of the relationships.}
\label{tab:correlation-portugal}
\end{table}

The coefficients $\lambda_k$ quantify the linear dependency between the variables. To derive interpretable insights from these coefficients, it is necessary to transform the variables using the methodology described in Section \ref{sec:lmc}. Table \ref{tab:correlation-portugal} summarizes the correlations between the response variables in Portugal. The results indicate a strong positive mean correlation of $0.89$ between PM2.5 and PM10, a significant negative mean correlation of $-0.58$ between PM2.5 and ozone, and a weaker negative mean correlation of $-0.40$ between PM10 and ozone. Importantly, the $95 \%$ credible intervals for all reported correlations do not include zero, providing evidence that these correlations are statistically significant.

Finally, we present the exceedance probabilities for each pollutant based on the World Health Organization (WHO) air quality guidelines released in 2021 \cite{who2021}. These guidelines provide air quality levels divided into interim targets to assess the severity of pollutant concentrations. Specifically, for PM2.5, the interim targets are 35, 25, 15, and 10 $\mu g/m^3$; for PM10, the targets are 70, 50, 30, and 20 $\mu g/m^3$; and for ozone, the targets are 100 and 70 $\mu g/m^3$.  

\begin{figure}[htp]
    \centering
    \begin{subfigure}[b]{0.48\textwidth}
        \includegraphics[width=\linewidth]{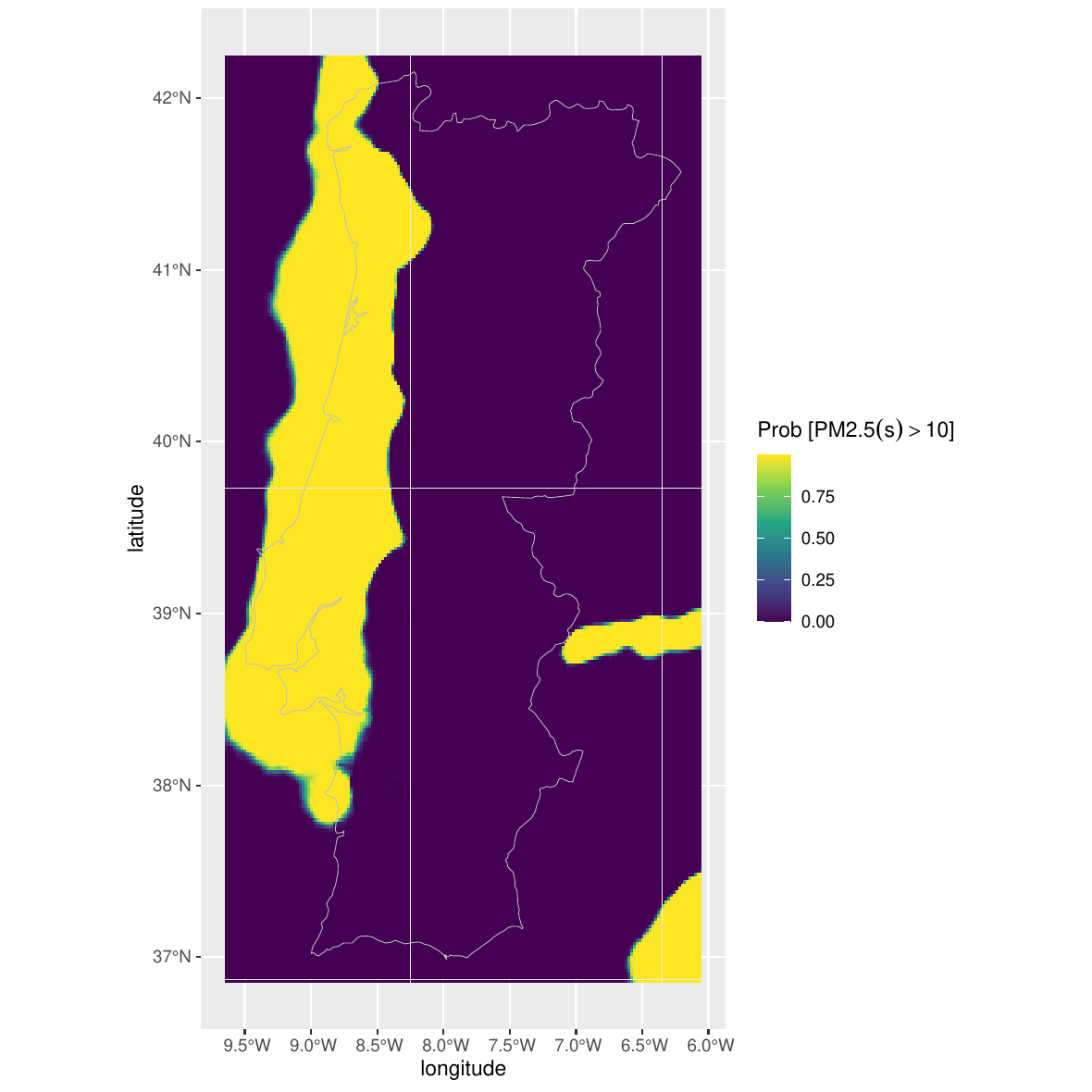}
        \caption{Threshold: 10 $\mu g/m^3$}
    \end{subfigure}
    \hfill
    \begin{subfigure}[b]{0.48\textwidth}
        \includegraphics[width=\linewidth]{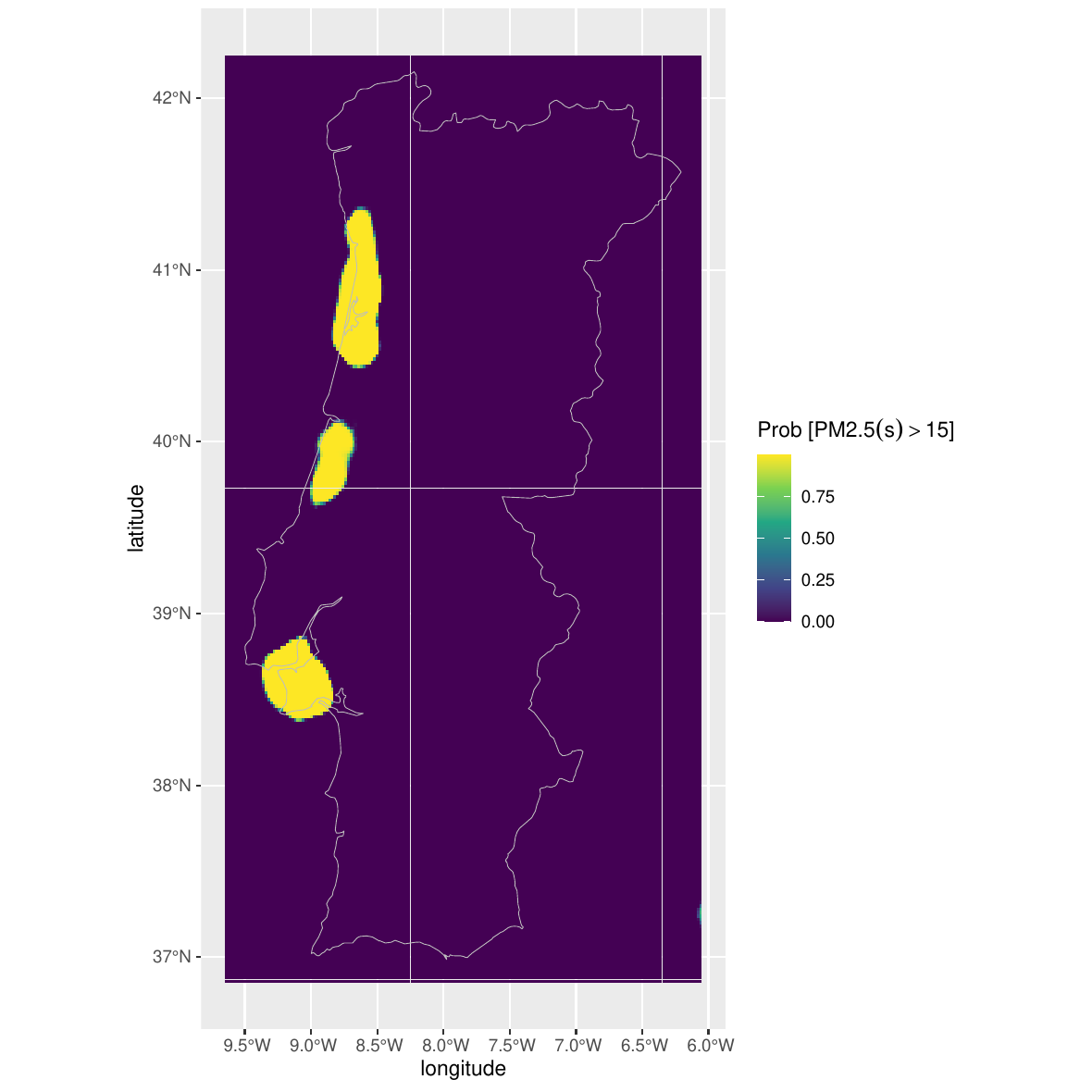}
        \caption{Threshold: 15 $\mu g/m^3$}
    \end{subfigure}
    \caption{Exceedance probabilities for PM2.5 in Portugal at two thresholds (10 and 15 $\mu g/m^3$).}
    \label{fig:exc-pm25-por}
\end{figure}

\begin{figure}[htp]
    \centering
    \begin{subfigure}[b]{0.48\textwidth}
        \includegraphics[width=\linewidth]{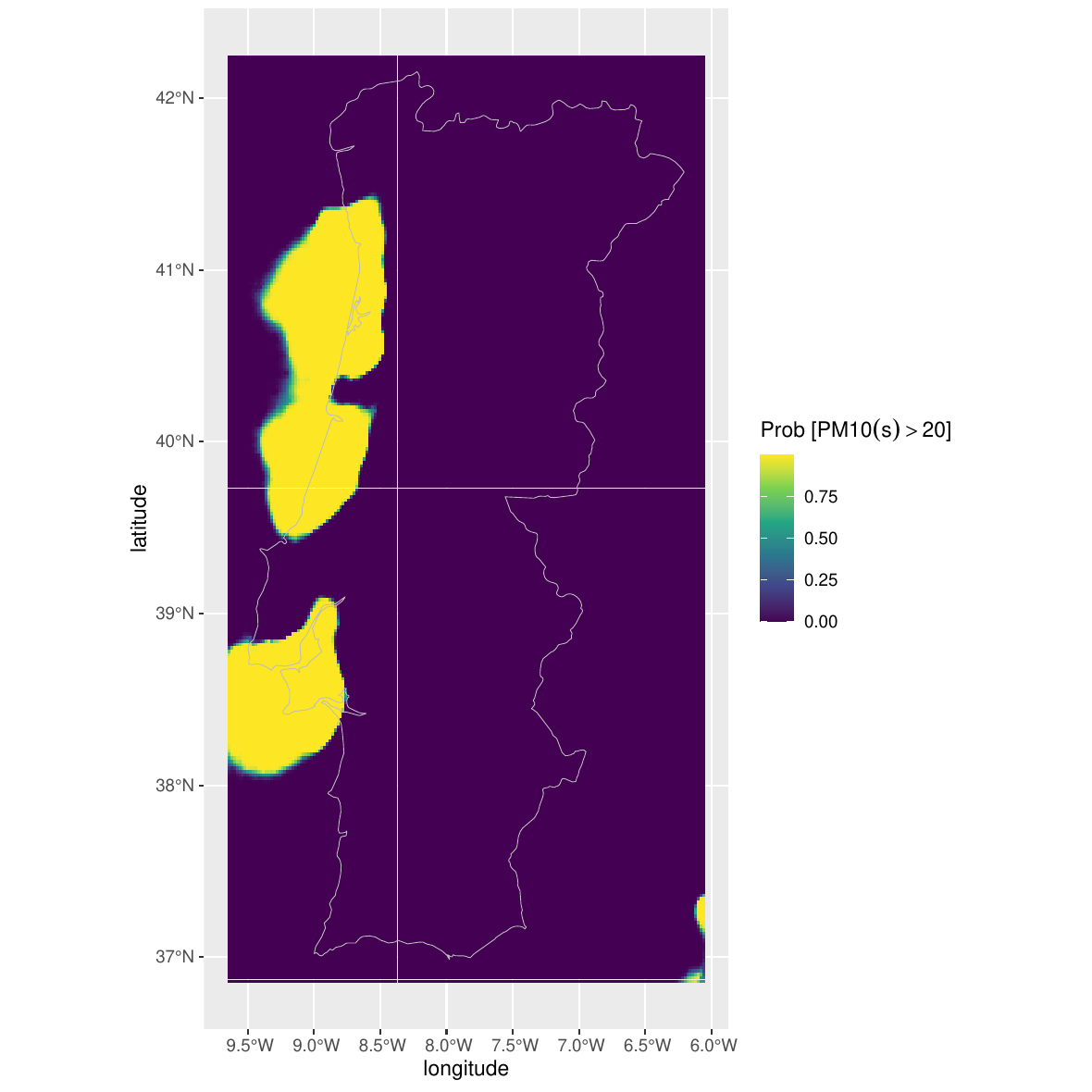}
        \caption{Threshold: 20 $\mu g/m^3$}
    \end{subfigure}
    \hfill
    \begin{subfigure}[b]{0.48\textwidth}
        \includegraphics[width=\linewidth]{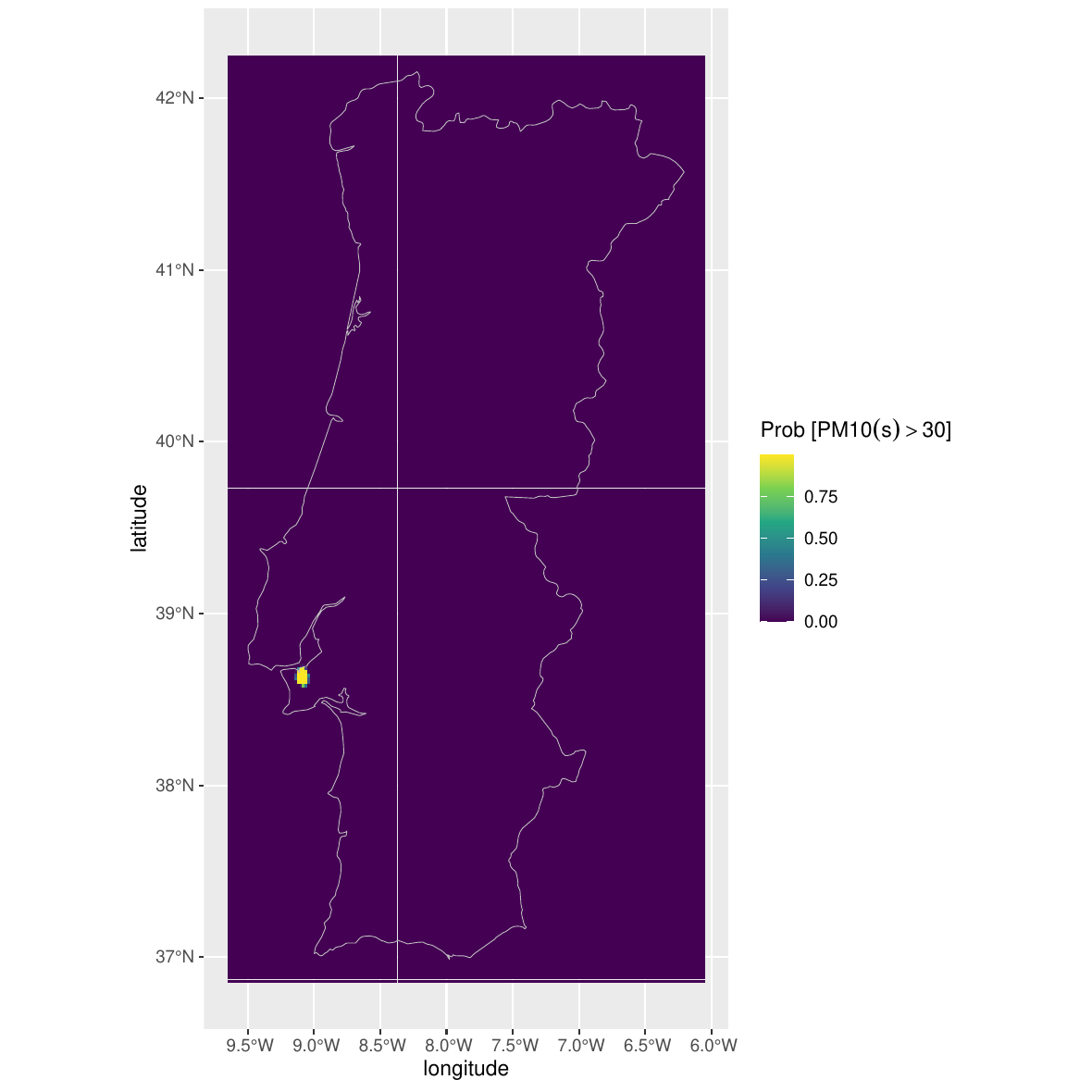}
        \caption{Threshold: 30 $\mu g/m^3$}
    \end{subfigure}
    \caption{Exceedance probabilities for PM10 in Portugal at two thresholds (20 and 30 $\mu g/m^3$).}
    \label{fig:exc-pm10-por}
\end{figure}

\begin{figure}[htp]
    \centering
    \includegraphics[width=0.6\linewidth]{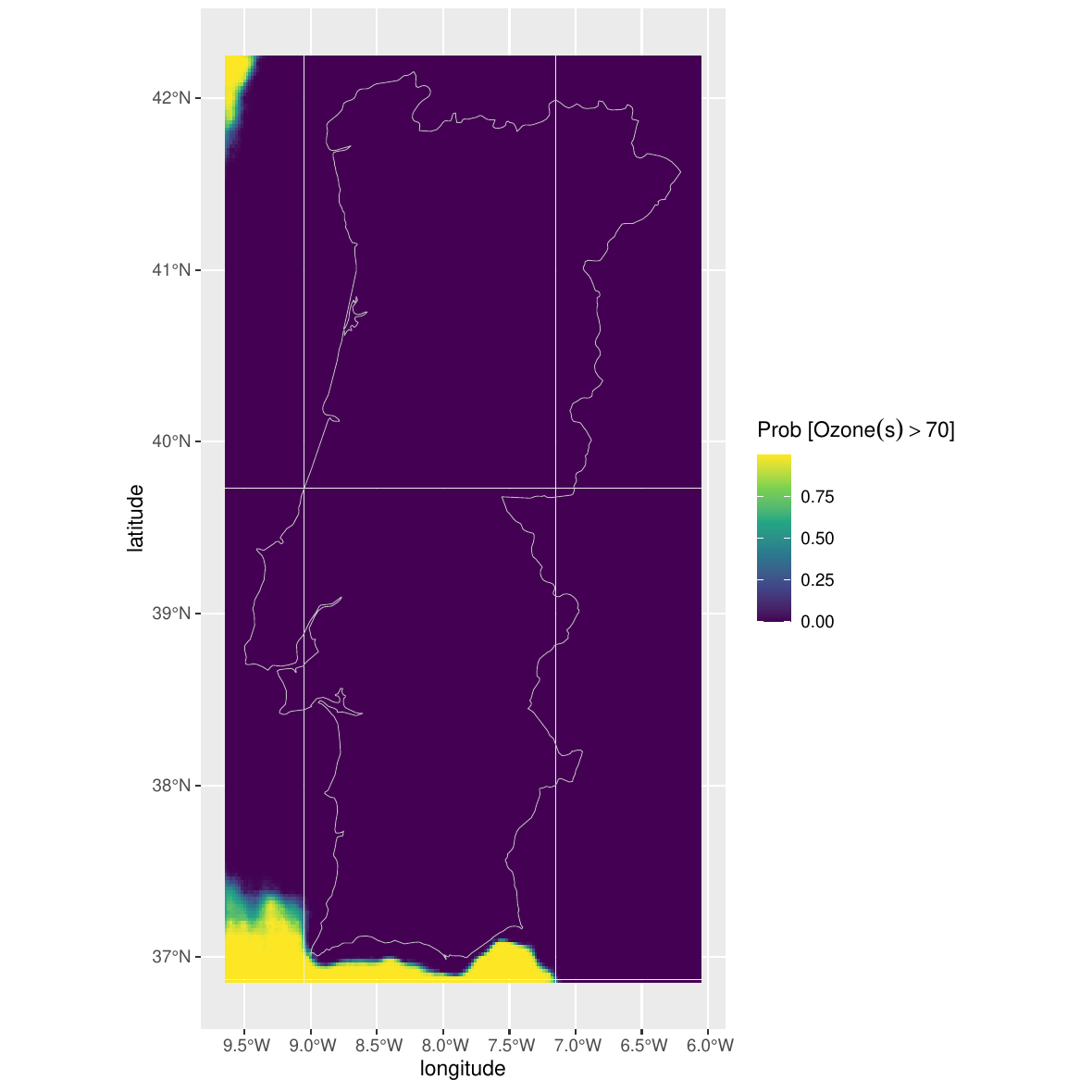}
    \caption{Exceedance probabilities for ozone in Portugal at the threshold of 70 $\mu g/m^3$.}
    \label{fig:exc-ozone-por}
\end{figure}

Figures \ref{fig:exc-pm25-por}, \ref{fig:exc-pm10-por}, and \ref{fig:exc-ozone-por} display the exceedance probabilities for these pollutants in Portugal. Each figure highlights the probabilities of pollutant levels exceeding the specified WHO interim targets that are most relevant to the observed contaminant levels.  

\subsection{Italy}

\begin{figure}[htp]
\begin{subfigure}{1\textwidth}
    \begin{minipage}{0.32\textwidth}
        \includegraphics[width=\linewidth]{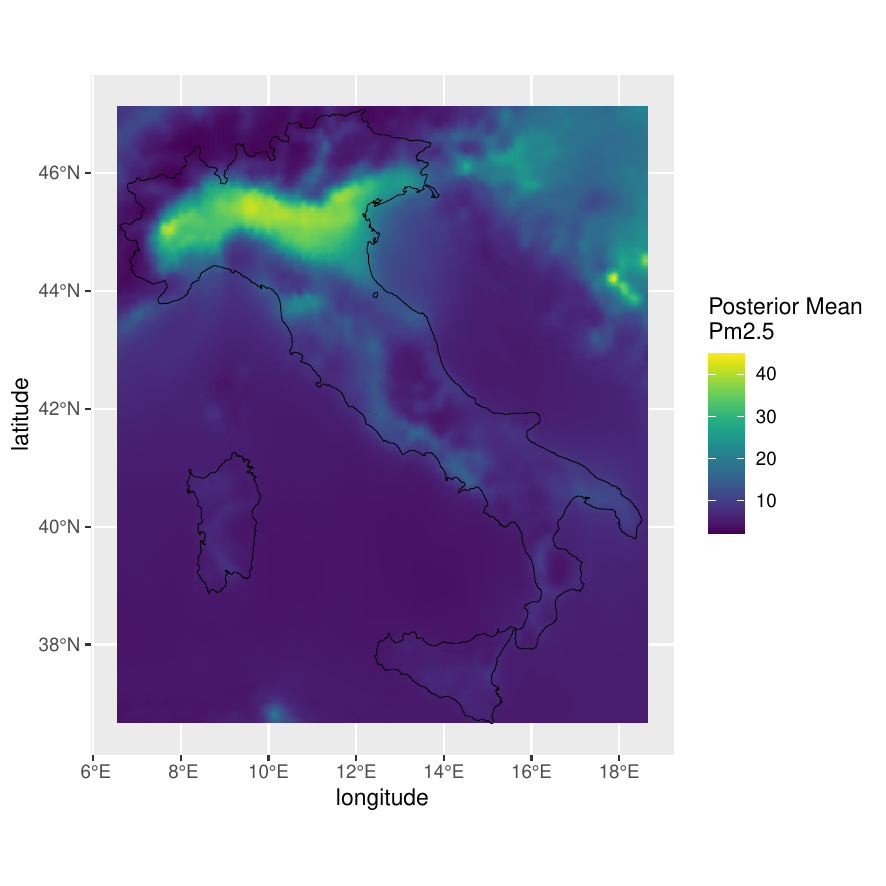}
    \end{minipage}
    \begin{minipage}{0.32\textwidth}
        \includegraphics[width=\linewidth]{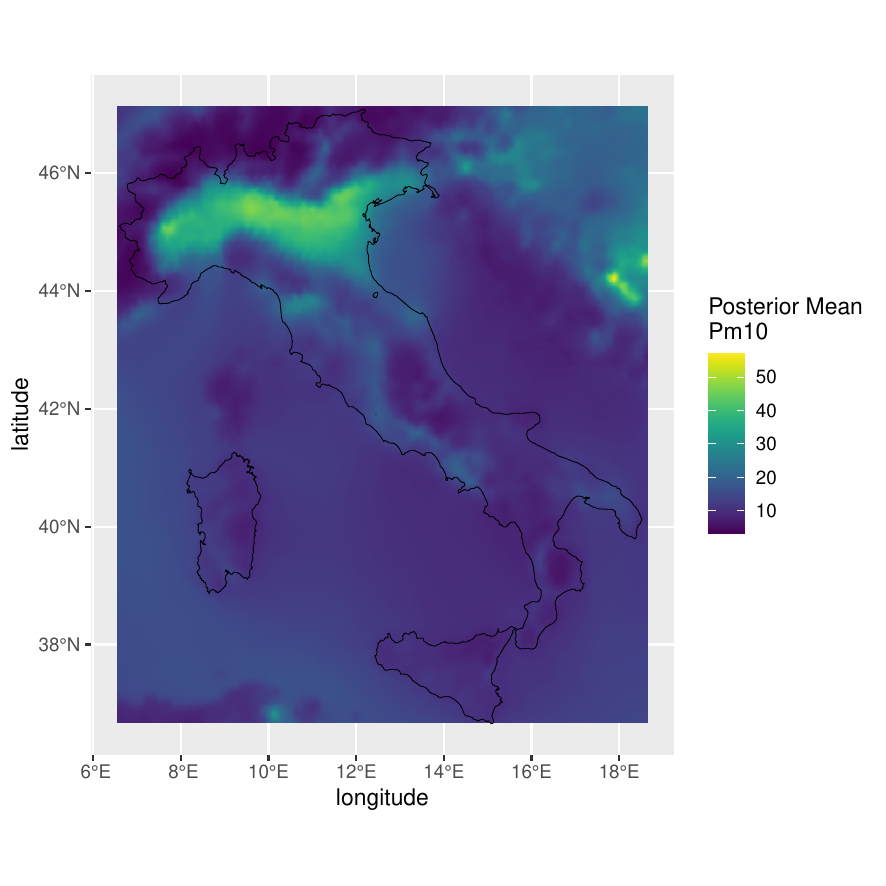}
    \end{minipage}
    \begin{minipage}{0.32\textwidth}
        \includegraphics[width=\linewidth]{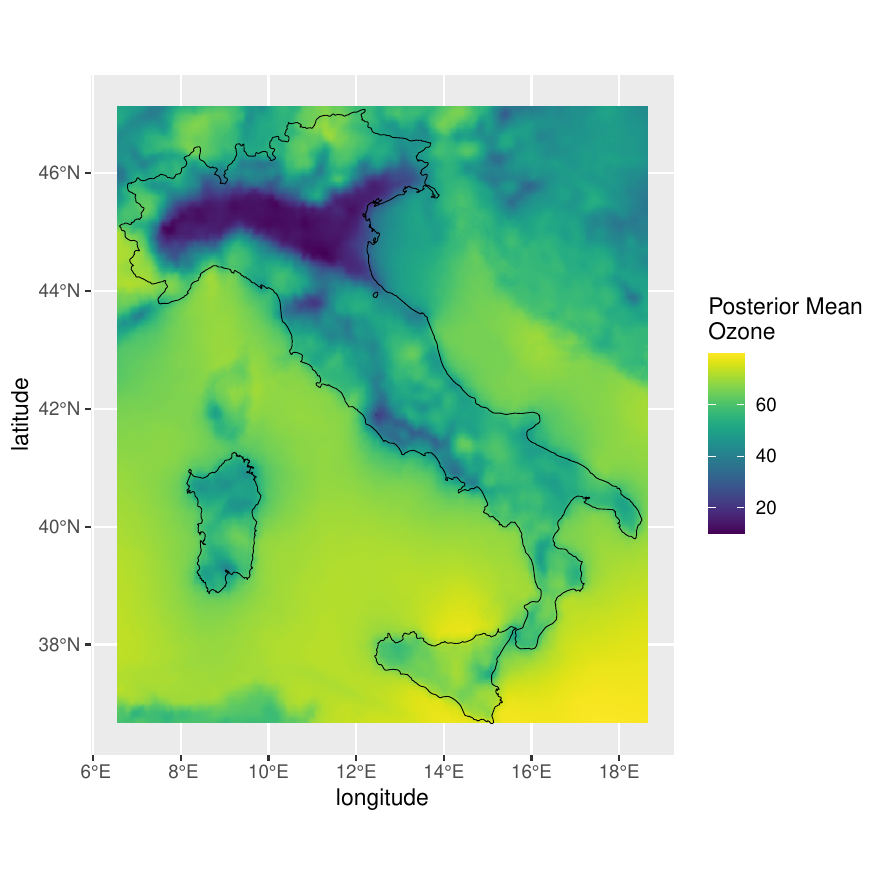}
    \end{minipage}
    \caption{Posterior mean of our model ($0.02^{\circ} \times 0.02^{\circ} \approx 2$km $\times 2$km resolution)}
\end{subfigure}
\begin{subfigure}{1\textwidth}
    \begin{minipage}{0.32\textwidth}
        \includegraphics[width=\linewidth]{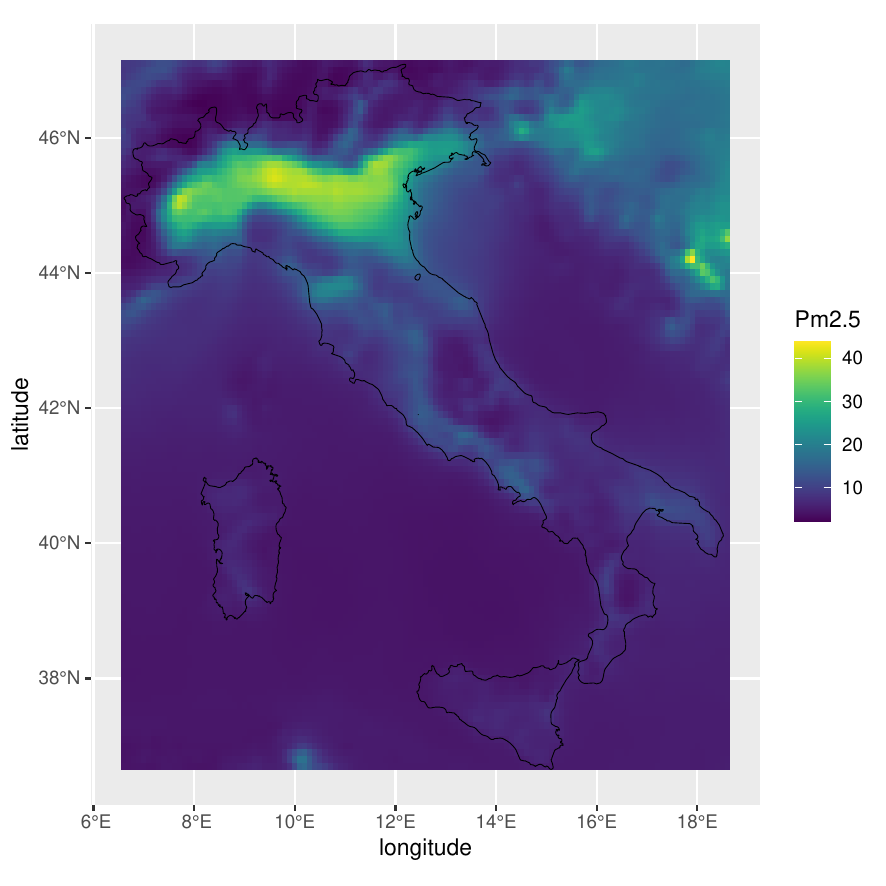}
    \end{minipage}
    \begin{minipage}{0.32\textwidth}
        \includegraphics[width=\linewidth]{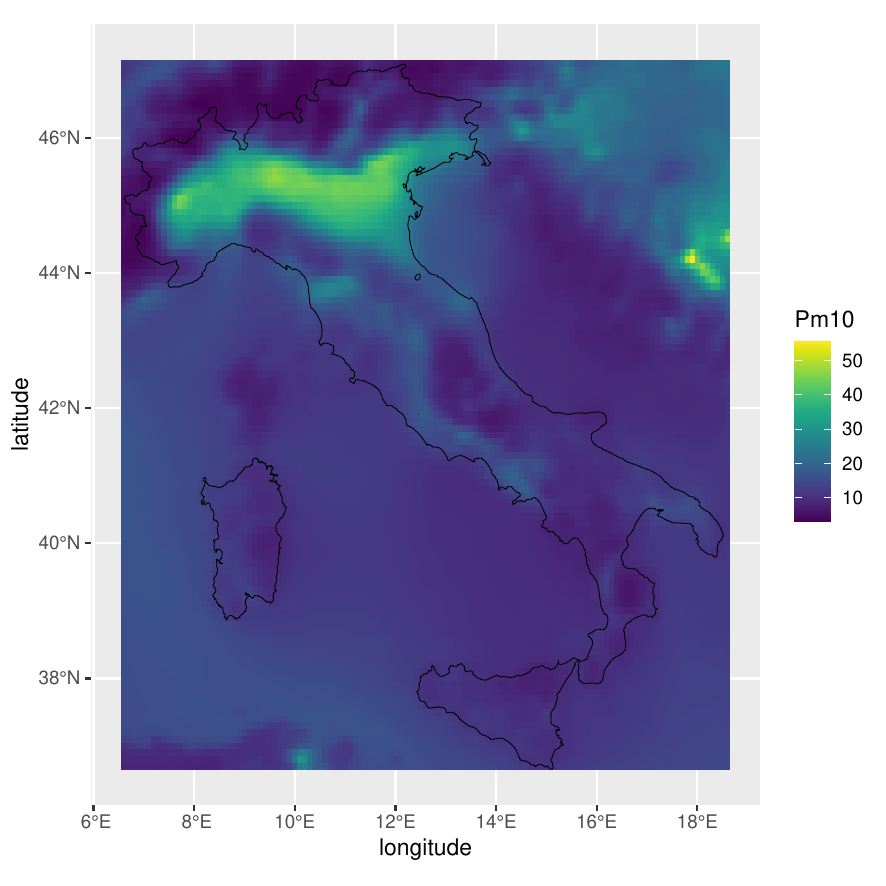}
    \end{minipage}
    \begin{minipage}{0.32\textwidth}
        \includegraphics[width=\linewidth]{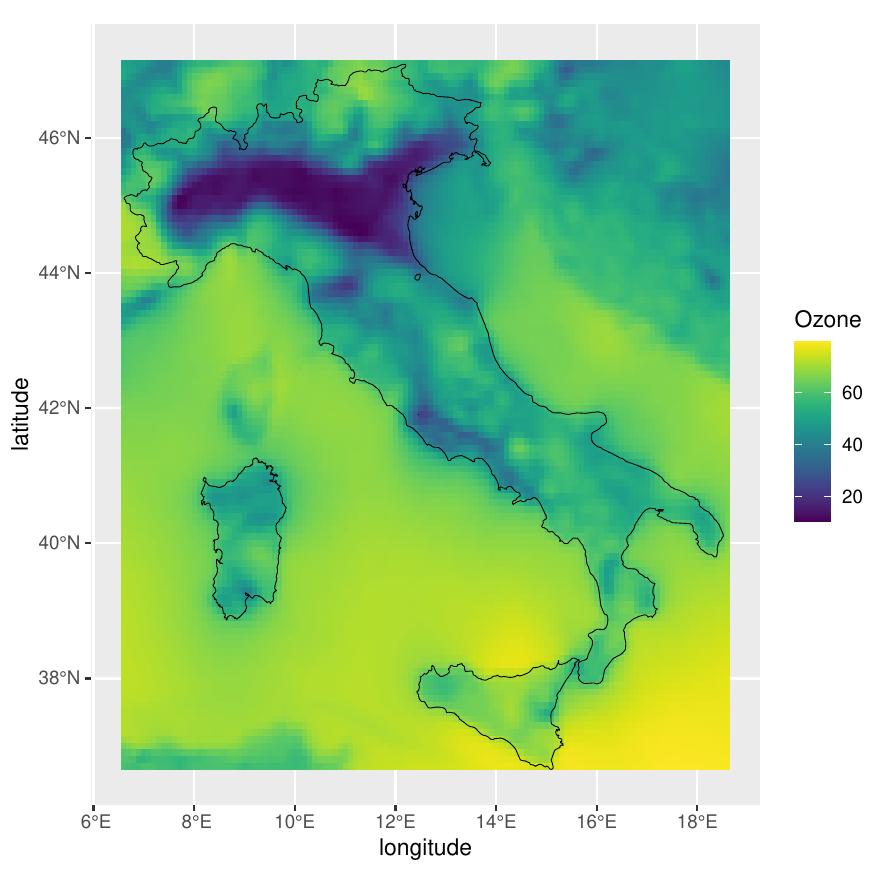}
    \end{minipage}
    \caption{True values ($0.1^{\circ} \times 0.1^{\circ} \approx 10$km $\times 10$km resolution)}
\end{subfigure}
\caption{Posterior mean of our model and true values for the mainland of Italy.}
\label{fig:ita-norm}
\end{figure}

\begin{figure}[htp]
\begin{subfigure}{1\textwidth}
    \begin{minipage}{0.32\textwidth}
        \includegraphics[width=\linewidth]{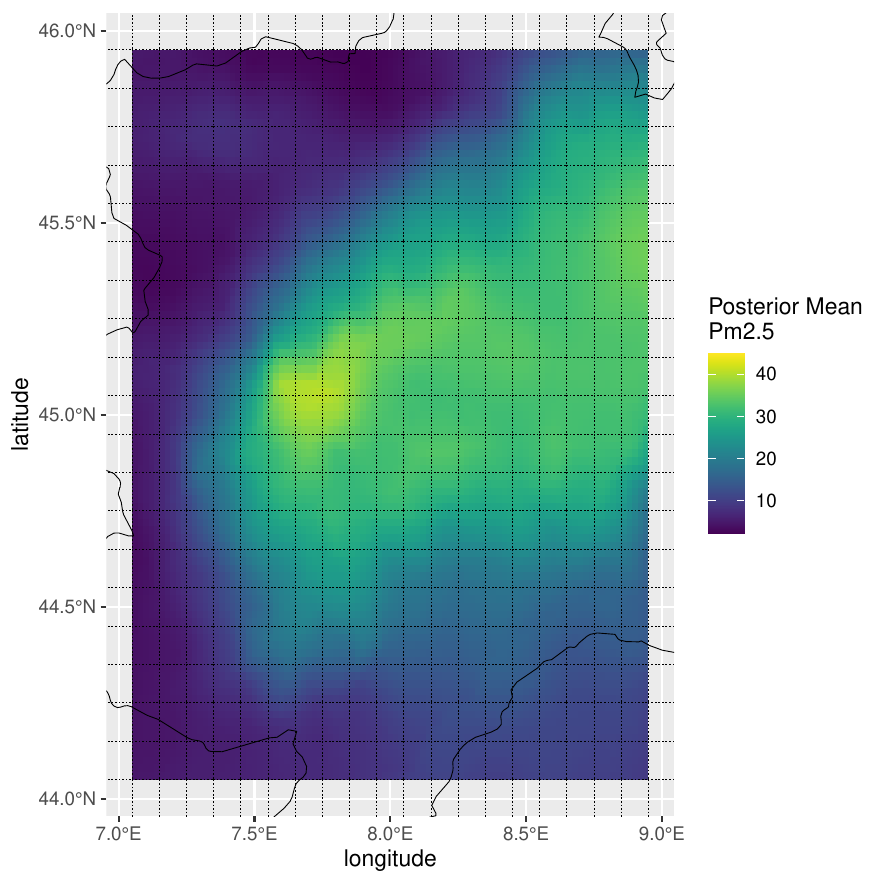}
    \end{minipage}
    \begin{minipage}{0.32\textwidth}
        \includegraphics[width=\linewidth]{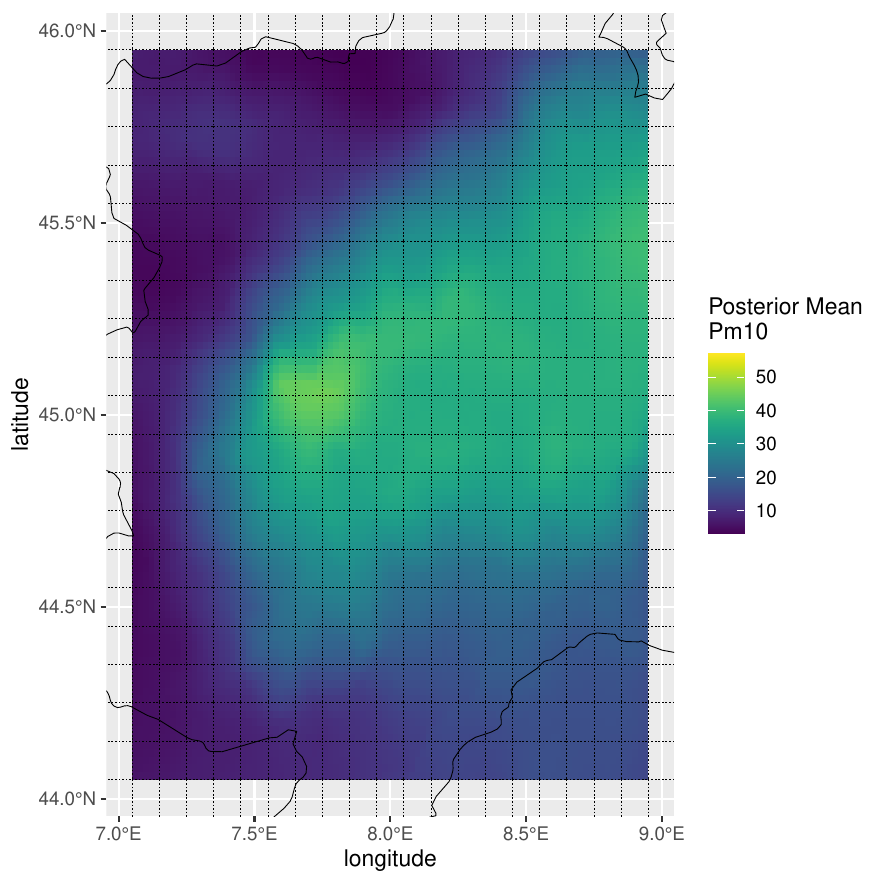}
    \end{minipage}
    \begin{minipage}{0.32\textwidth}
        \includegraphics[width=\linewidth]{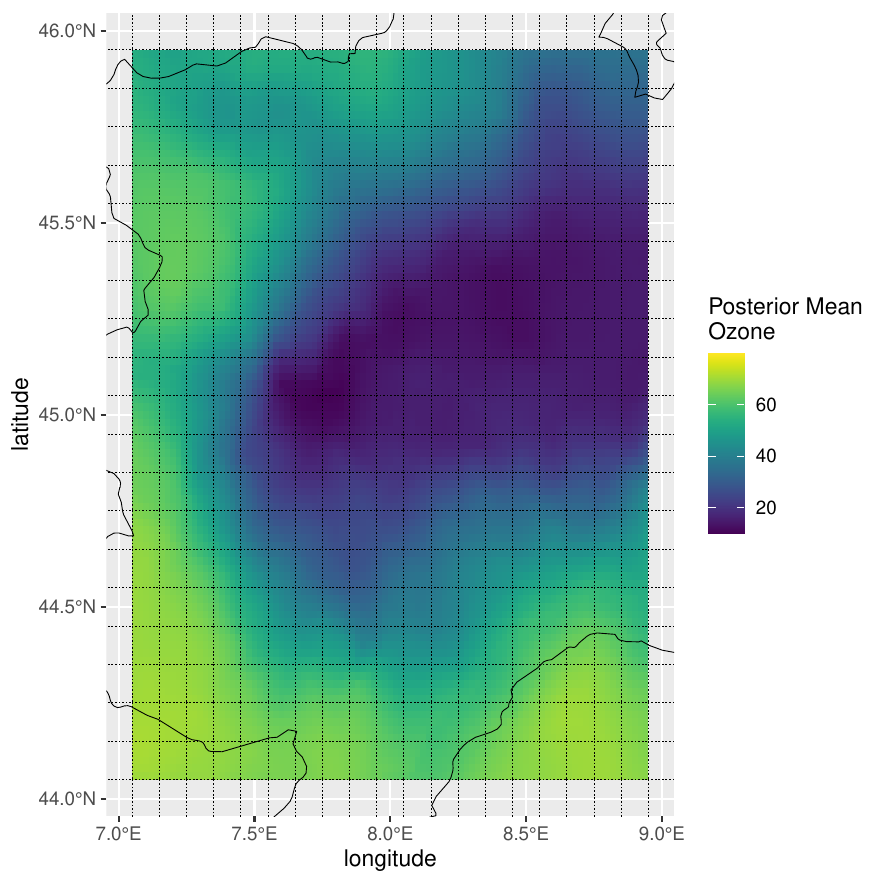}
    \end{minipage}
    \caption{Posterior mean of our model ($0.02^{\circ} \times 0.02^{\circ} \approx 2$km $\times 2$km resolution)}
\end{subfigure}
\begin{subfigure}{1\textwidth}
    \begin{minipage}{0.32\textwidth}
        \includegraphics[width=\linewidth]{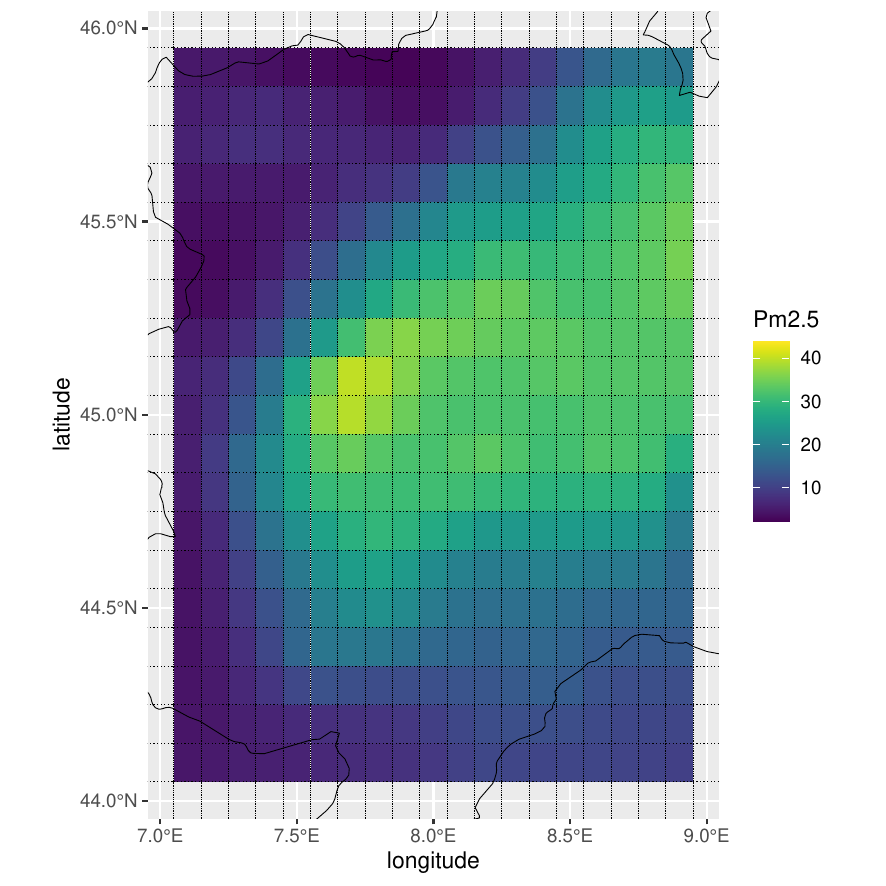}
    \end{minipage}
    \begin{minipage}{0.32\textwidth}
        \includegraphics[width=\linewidth]{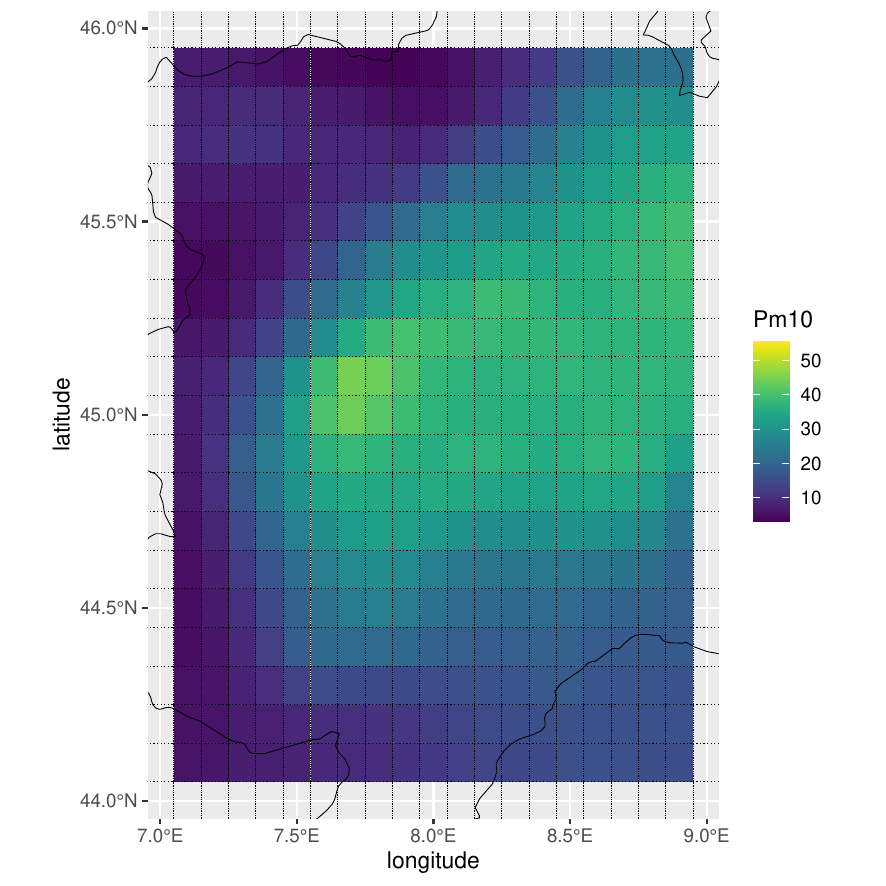}
    \end{minipage}
    \begin{minipage}{0.32\textwidth}
        \includegraphics[width=\linewidth]{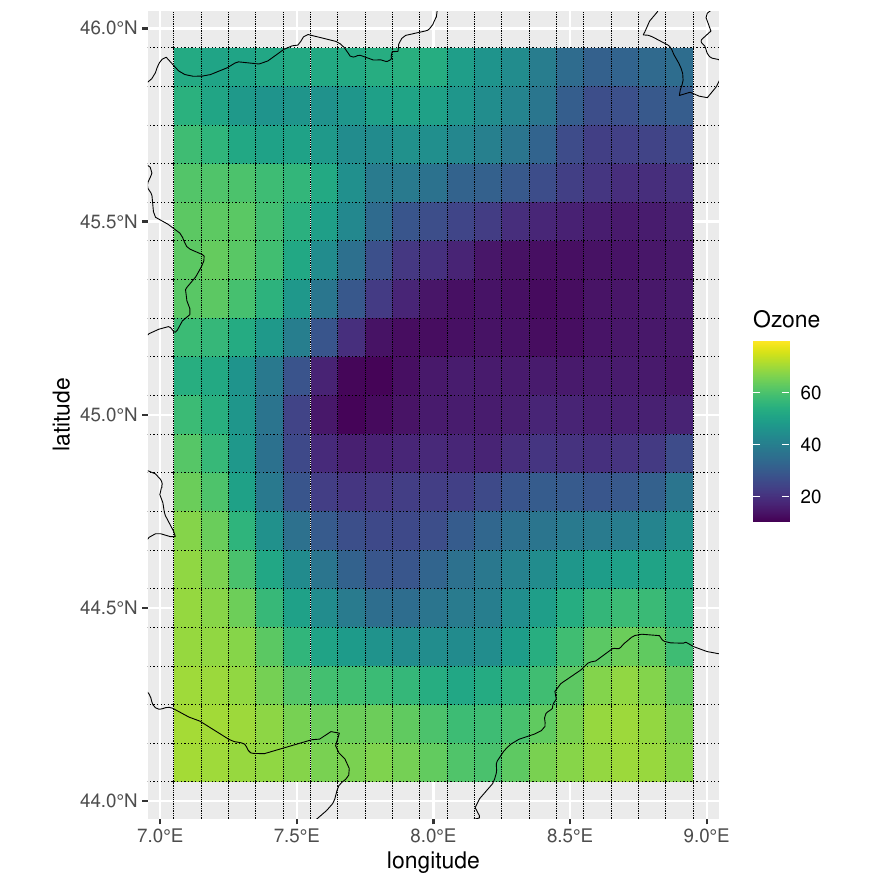}
    \end{minipage}
    \caption{True values ($0.1^{\circ} \times 0.1^{\circ} \approx 10$km $\times 10$km resolution)}
\end{subfigure}
\caption{Posterior mean of our model and true values for a zoom version of the mainland of Italy.}
\label{fig:ita-zoom}
\end{figure}

In Figure \ref{fig:ita-norm}, we present the original and posterior mean values from the disaggregation model applied to PM2.5, PM10, and ozone emissions in Italy. As well as in the example of Portugal, the model preserves each variable's original scale and spatial patterns while generating a smoother representation of the data. Figure \ref{fig:ita-zoom} displays a zoomed-in version of the plot, focusing on a $0.02^{\circ} \times 0.02^{\circ}$ in Italy to provide a clearer perspective.

In Table \ref{tab:it_fixed_parameters}, we present a summary of the fixed parameters for the model. First, the intercepts $\alpha_k$ represent the mean value of each of the three components. Specifically, the mean PM2.5 concentration over the spatial domain is $\alpha_1 = 9.07$ $\mu g/m^3$, the mean PM10 concentration is $\alpha_2 = 14.33$ $\mu g/m^3$, and the mean ozone concentration is $\alpha_3 = 58.80$ $\mu g/m^3$. 

\begin{table}[h!]
\centering
\begin{tabular}{|l|c|c|c|c|}
\hline
\textbf{Parameter}  & \textbf{Mean} & \textbf{0.025 Quantile} & \textbf{0.5 Quantile} & \textbf{0.975 Quantile} \\ \hline
$\alpha_1$  & 9.07     & 8.88     & 9.07     & 9.27     \\ \hline
$\alpha_2$  & 14.33    & 14.11    & 14.33    & 14.55    \\ \hline
$\alpha_3$  & 58.80    & 58.46    & 58.80    & 59.14    \\ \hline
$\beta_1$   & -0.21    & -0.24    & -0.21    & -0.18    \\ \hline
$\beta_2$   & -0.25    & -0.29    & -0.25    & -0.21    \\ \hline
$\beta_3$   & 0.44     & 0.38     & 0.44     & 0.50     \\ \hline
$\lambda_1$ & 1.11     & 1.10     & 1.11     & 1.11     \\ \hline
$\lambda_2$ & -1.21    & -1.23    & -1.21    & -1.19    \\ \hline
$\lambda_3$ & 0.87     & 0.78     & 0.87     & 0.95     \\ \hline
\end{tabular}
\caption{Summary of the intercept ($\alpha_k$) parameters, the elevation effects ($\beta_k$) and the cross-dependence parameters ($\lambda_k$) for Italy}
\label{tab:it_fixed_parameters}
\end{table}

The elevation coefficients $\beta_k$ indicate that an increase in elevation by 1 km, while holding other variables constant, reduces PM2.5 emissions by $0.21$ $\mu g/m^3$, PM10 emissions by $0.25$ $\mu g/m^3$, and increases ozone emissions by $0.44$ $\mu g/m^3$. Importantly, the credible intervals for these coefficients do not cross zero, confirming the statistical significance of these effects.

\begin{table}[h!]
\renewcommand{\arraystretch}{1.3}
\centering
\begin{tabular}{|c|c|c|c|c|c|}
\hline 
\multicolumn{2}{|c|}{\textbf{Correlation}} & \multirow{2}{*}{\textbf{mean}} & \multirow{2}{*}{\textbf{0.025 Quantile}} & \multirow{2}{*}{\textbf{0.5 Quantile}} & \multirow{2}{*}{\textbf{0.975 Quantile}} \\
\cline{1-2} 
\textbf{var1} & \textbf{var2} & & & & \\
\hline 
pm25 & pm10 & 0.97 & 0.91 & 0.98 & 0.99 \\
\hline 
pm25 & Ozone & -0.88 & -0.96 & -0.89 & -0.71 \\
\hline 
pm10 & Ozone & -0.83 & -0.95 & -0.86 & -0.58 \\
\hline
\end{tabular}
\vspace{0.2cm}
\caption{Summary of correlations between the response variables PM2.5, PM10, and Ozone in Italy. The table includes the mean correlation values along with the $2.5 \%$, $50 \%$ (median), and $97.5 \%$ quantiles of their credible intervals, demonstrating the strength and significance of the relationships.}
\label{tab:correlation-italy}
\end{table}

The $\lambda_k$ parameters quantify the linear dependency between the variables. However, as we did for Portugal, it is necessary to transform these variables using the methodology discussed in Section \ref{sec:lmc}. Table \ref{tab:correlation-italy} summarizes the correlations between the response variables in Italy. The results indicate a strong positive mean correlation of $0.97$ between PM2.5 and PM10, a strong negative mean correlation of $-0.88$ between PM2.5 and ozone, and also a strong negative mean correlation of $-0.83$ between PM10 and ozone. Importantly, the $95 \%$ credible intervals for all reported correlations do not include zero, providing evidence that these correlations are statistically significant. These results are consistent with the correlation values observed in Portugal, providing further evidence that the relationships between these pollutants are stable across regions and can be captured effectively using the proposed model.

\begin{figure}[htp]
    \centering
    \begin{subfigure}[b]{0.31\textwidth}
        \includegraphics[width=\linewidth]{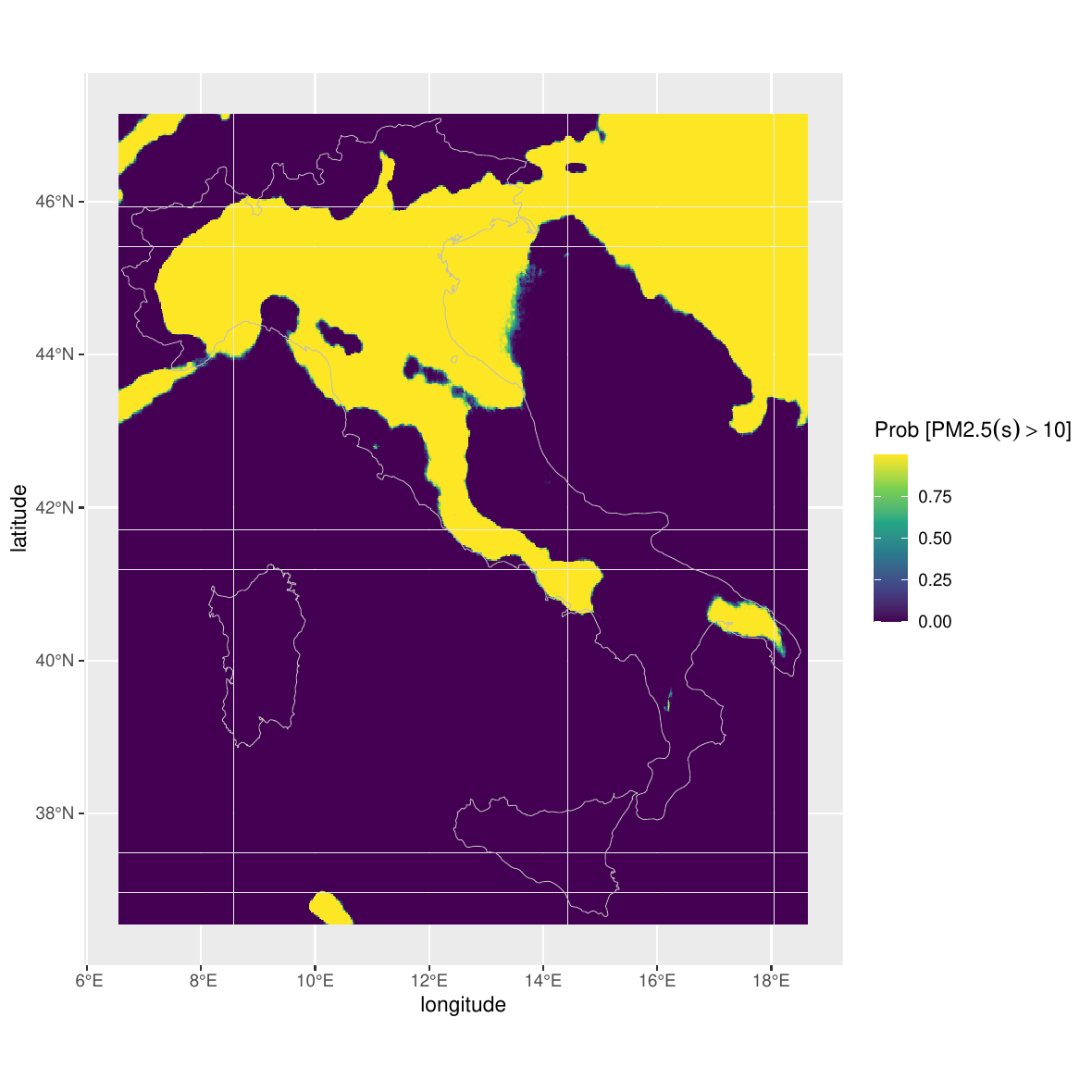}
        \caption{Threshold: 10 $\mu g/m^3$}
    \end{subfigure}
    \hfill
    \begin{subfigure}[b]{0.31\textwidth}
        \includegraphics[width=\linewidth]{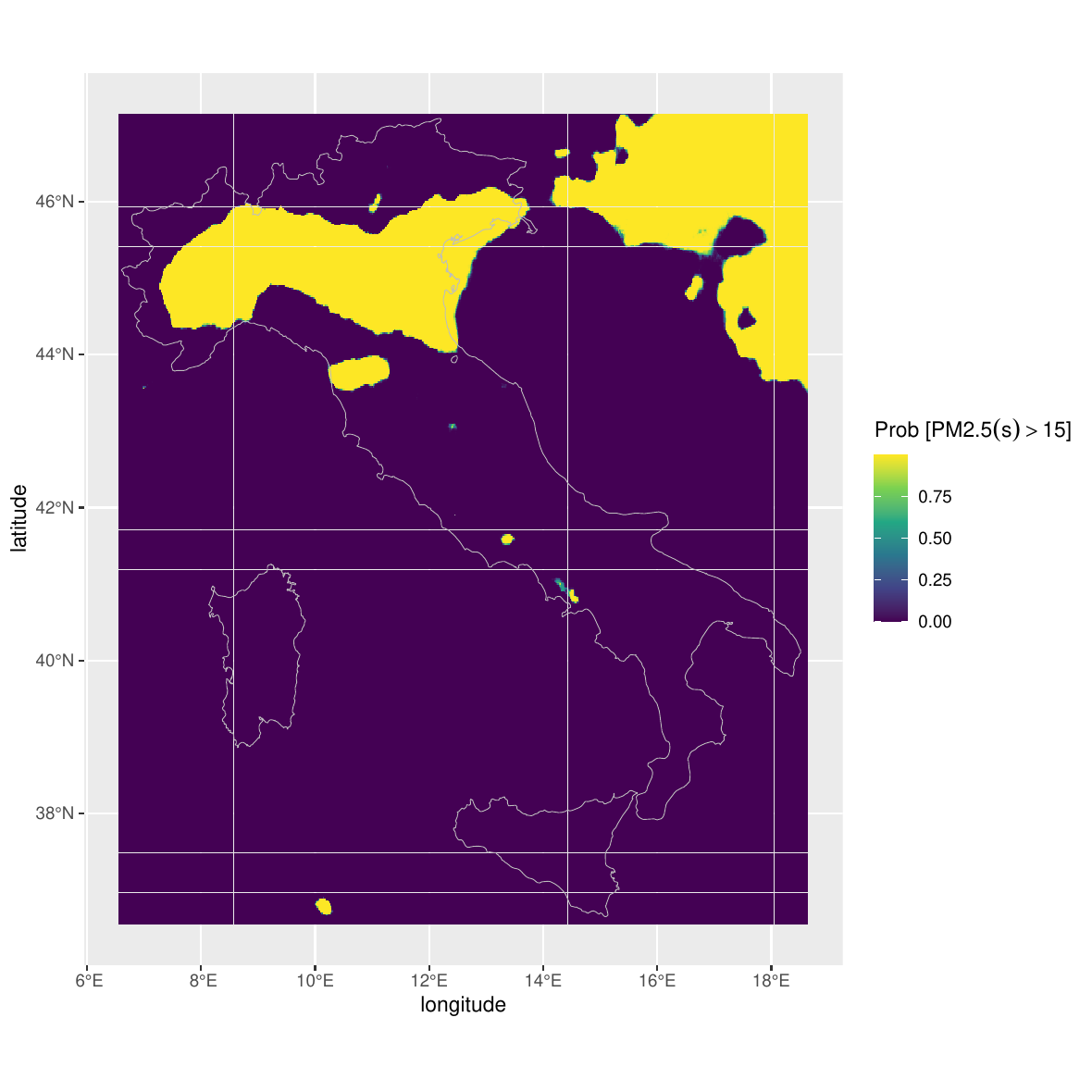}
        \caption{Threshold: 15 $\mu g/m^3$}
    \end{subfigure}
    \hfill
    \begin{subfigure}[b]{0.31\textwidth}
        \includegraphics[width=\linewidth]{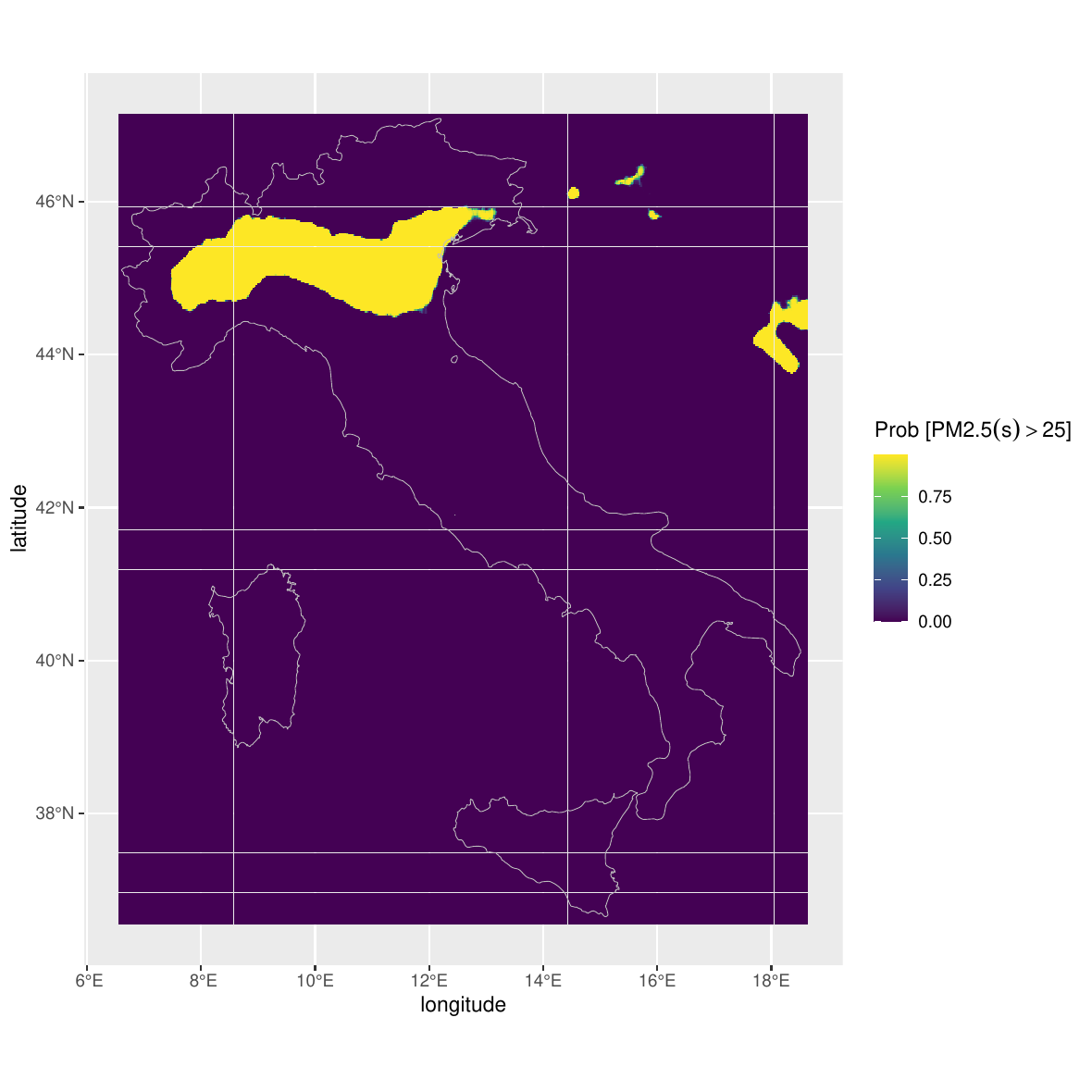}
        \caption{Threshold: 25 $\mu g/m^3$}
    \end{subfigure}
    \caption{Exceedance probabilities for PM2.5 in Italy at three thresholds (10, 15, and 25 $\mu g/m^3$).}
    \label{fig:exc-pm25-ita}
\end{figure}

\begin{figure}[htp]
    \centering
    \begin{subfigure}[b]{0.31\textwidth}
        \includegraphics[width=\linewidth]{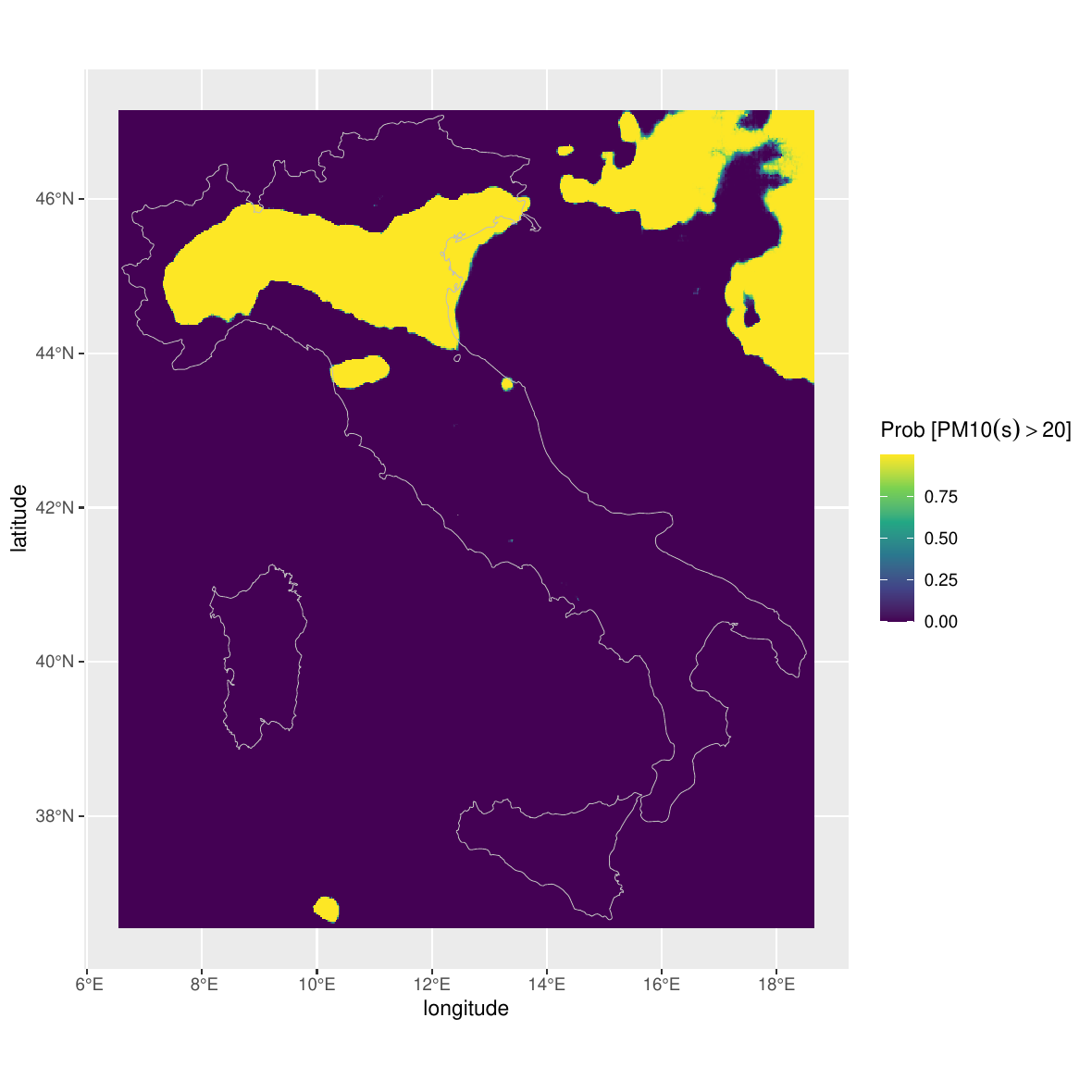}
        \caption{Threshold: 20 $\mu g/m^3$}
    \end{subfigure}
    \hfill
    \begin{subfigure}[b]{0.31\textwidth}
        \includegraphics[width=\linewidth]{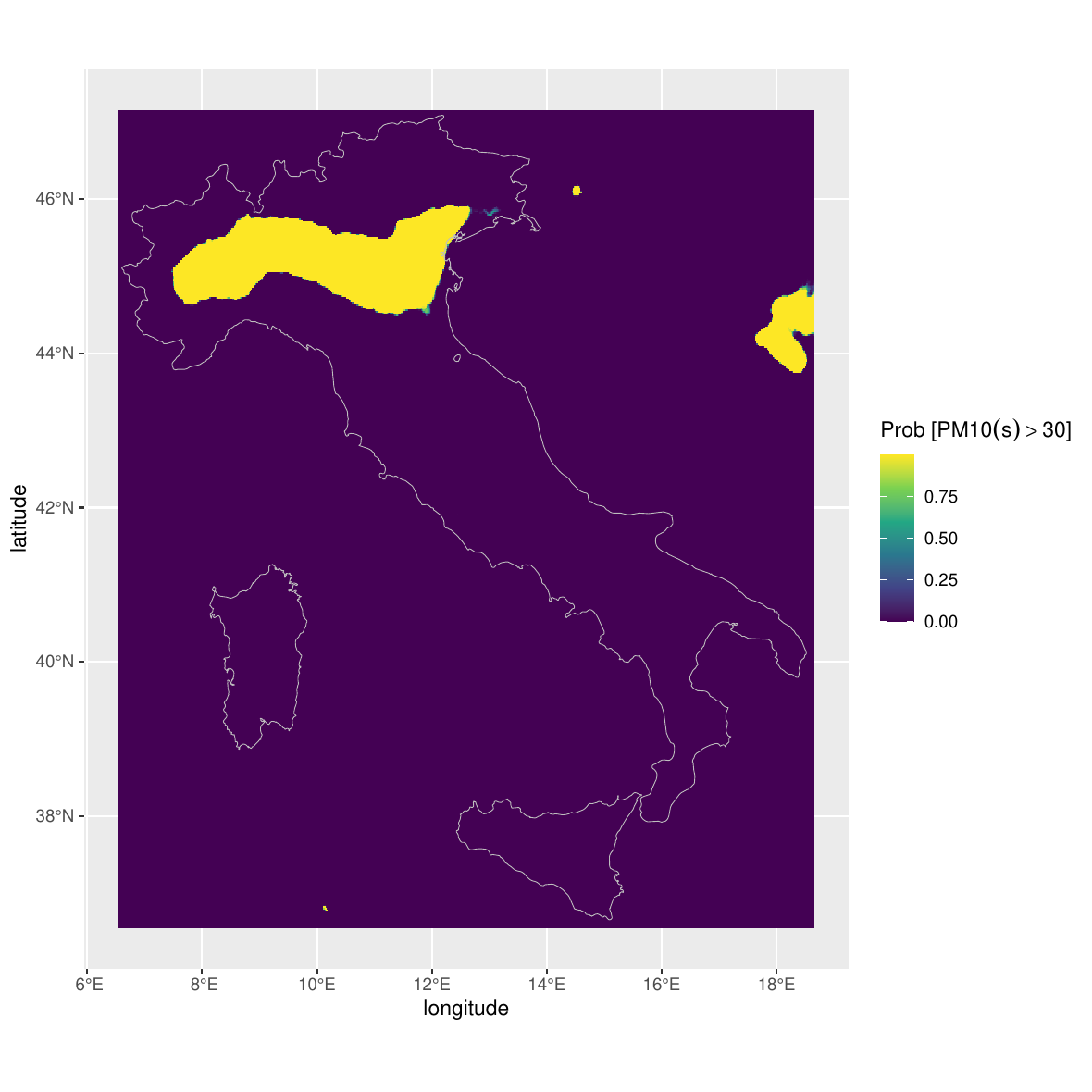}
        \caption{Threshold: 30 $\mu g/m^3$}
    \end{subfigure}
    \hfill
    \begin{subfigure}[b]{0.31\textwidth}
        \includegraphics[width=\linewidth]{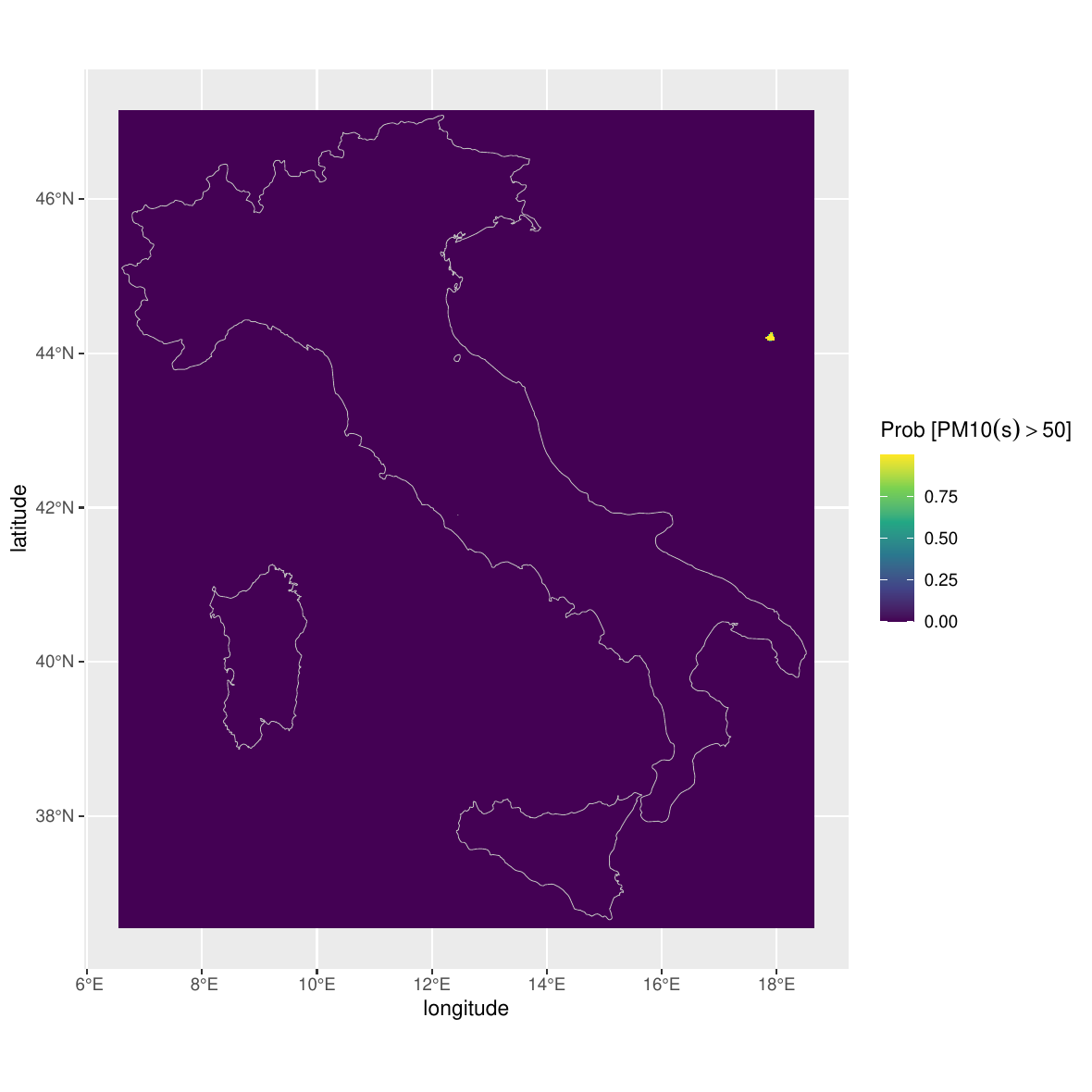}
        \caption{Threshold: 50 $\mu g/m^3$}
    \end{subfigure}
    \caption{Exceedance probabilities for PM10 in Italy at three thresholds (20, 30, and 50 $\mu g/m^3$).}
    \label{fig:exc-pm10-ita}
\end{figure}

\begin{figure}[h]
    \centering
    \includegraphics[width=0.6\linewidth]{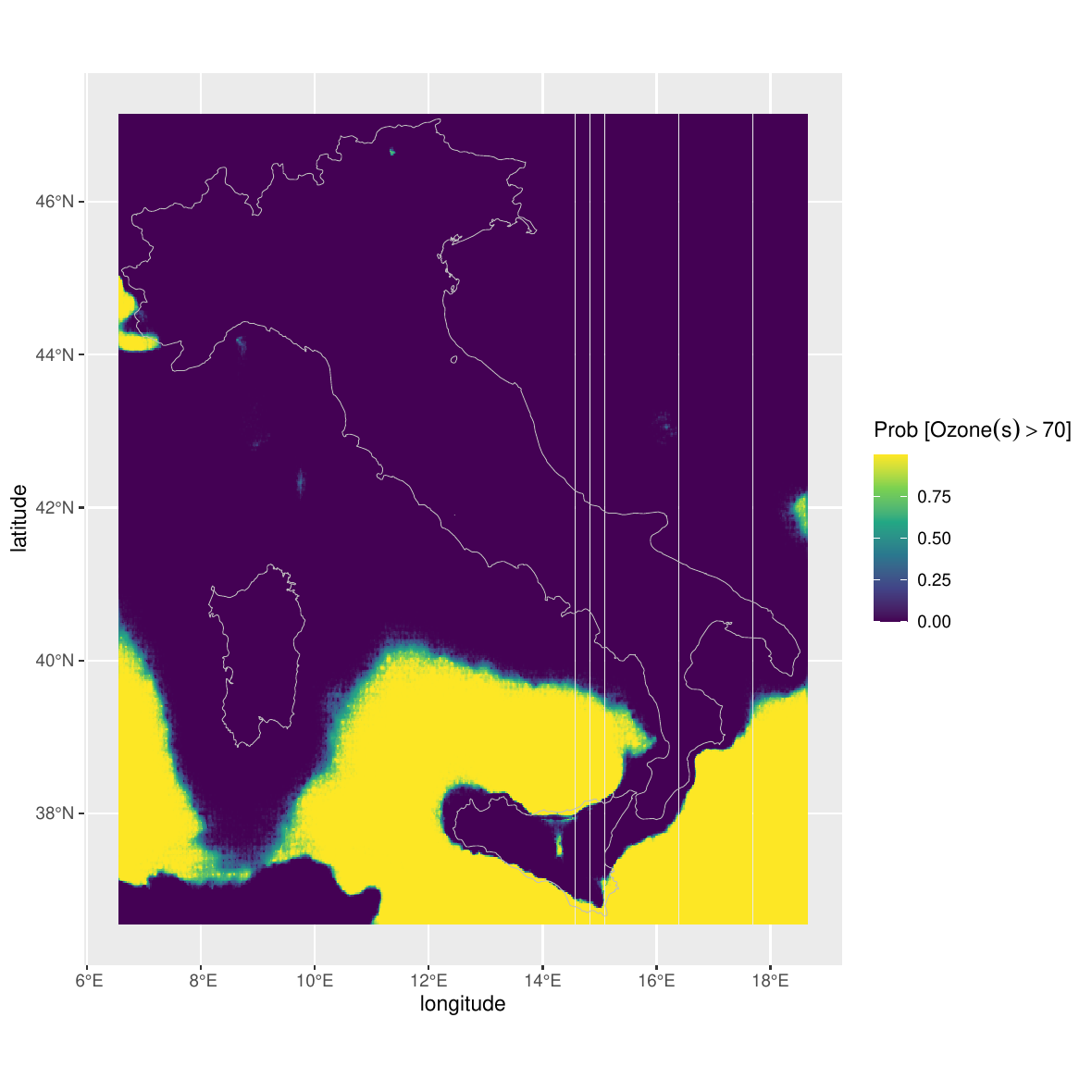}
    \caption{Exceedance probabilities for ozone in Italy at the threshold of 70 $\mu g/m^3$.}
    \label{fig:exc-ozone-ita}
\end{figure}

Finally, as with Portugal, we present the exceedance probabilities for each pollutant in Italy based on the World Health Organization (WHO) air quality guidelines released in 2021 \cite{who2021}. These guidelines define interim targets to assess air quality levels and potential health impacts. Figures \ref{fig:exc-pm25-ita}, \ref{fig:exc-pm10-ita}, and \ref{fig:exc-ozone-ita} illustrate the exceedance probabilities for PM2.5, PM10, and ozone, respectively. Each figure highlights the probabilities of pollutant levels exceeding the relevant WHO interim thresholds, reflecting the spatial variability of air quality in the region.

\section{Conclusion and discussion}
\label{sec:con}

In this paper, we propose a computationally efficient multivariate model for spatial data disaggregation, which leverages shared information among multiple variables. The approach assumes a continuous latent field, with observations representing averaged observation in regions, and incorporates covariates to enhance flexibility and interpretability. By modeling relationships both between covariates and response variables and among multiple response variables, the framework provides valuable insights not only to the relationships between covariates and the response variables but also to the associations among multiple response variables themselves. The use of fast Bayesian inference techniques such as \texttt{R-INLA} ensures scalability, making the model suitable for analyzing large spatial datasets in both regular and irregular regions.

Using a simulation study, we demonstrated that our model can effectively approximate the latent space from aggregated data, regardless of whether the region is divided into regular or irregular areas. The results showed that increasing the number of areas consistently reduces the RMSE between the model estimates and the true latent space values for both regular and irregular configurations. However, we observed that regular regions generally produce more accurate approximations than irregular regions with the same number of areas. We also evidenced that the best case scenarios had regular areas with the bigger number of regions possible. 

We applied our model to air pollution data, specifically focusing on PM2.5, PM10, and Ozone concentrations in Portugal and Italy. Through this application, we demonstrated the model’s capability to produce finer estimates, improving the spatial resolution from an initial $0.1^{\circ}$ (approximately 10 km) to $0.02^{\circ}$ (approximately 2 km). Additionally, we showcased how the model incorporates covariates to explain the response variables better, enhancing its interpretative power. Importantly, the fixed parameter estimates provided valuable insights into the relationships between the response variables and their covariates, as well as the interactions among the response variables themselves. These results highlight the model’s potential for improving spatial resolution and providing interpretable conclusions in multivariate settings.

Our analysis of air pollutants in Portugal and Italy  
identified a strong positive linear relationship between PM2.5 and PM10 and a negative linear relationship between PM2.5 and ozone in both countries. Additionally, covariate analysis showed a positive linear association between elevation and PM2.5 and PM10, and a negative association between elevation and ozone. These consistent trends across Portugal and Italy suggest that the relationships among air pollutants and their dependence on elevation are driven by similar underlying processes, despite regional variations.

This study assumed a normal distribution for the response variable with an identity link function. Future work will focus on extending the framework to handle more general cases, such as Poisson-distributed data with non-linear link functions, which are particularly relevant for modeling count data in fields like epidemiology and ecology \cite{breslow1993applied, moraga2019geospatial}. Additionally, we aim to explore applications to other distributions within the exponential family, such as the binomial or gamma distributions, to address a wider range of data types, including proportions and skewed continuous data \cite{mccullagh1989generalized}.
We also plan to extend the framework to handle variables measured at different spatial resolutions, such as point-level and area-level data, while accounting for sampling mechanisms \cite{zhongetal24}.
These extensions will further enhance the versatility of the model and enable its application to more complex multivariate and spatiotemporal data structures, building on advancements in generalized linear and additive modeling frameworks \cite{wood2017generalized, rue_et_al_r_inla_2009}.
 
In conclusion, this study introduces a robust multivariate model for the disaggregation of spatial data, demonstrating its versatility and effectiveness through both simulation studies and real-world applications. By leveraging shared information among multiple variables, the model achieves significant improvements in spatial resolution while maintaining interpretability and computational efficiency. The ability to incorporate covariates enhances its explanatory power, offering insights into both the relationships between covariates and response variables and the interdependencies among the response variables themselves.
This approach has been used here to model air pollution, but is adaptable to other settings, offering a flexible approach to understand the spatial distribution and interactions of variables across diverse fields.

\clearpage

\bibliographystyle{apalike}
\bibliography{biblio.bib,biblio-pollutants}       

\clearpage

\section*{Appendix}
\label{sec:appendix}

In this Appendix, we present the results of the simulation study from Section \ref{sec:sim}. Each scenario includes 9 images arranged in a 3 x 3 grid. The columns show the predictions from our disaggregation model, the predictions from the areal model, and a realization of the simulated continuous latent field, while the rows correspond to the three variables used in the simulation.

    
\begin{figure}[htp]
\centering
\large{\textbf{Scenario 1}\par\medskip}
\begin{subfigure}{1\textwidth}
    \begin{minipage}{0.32\textwidth}
        \includegraphics[width=\linewidth]{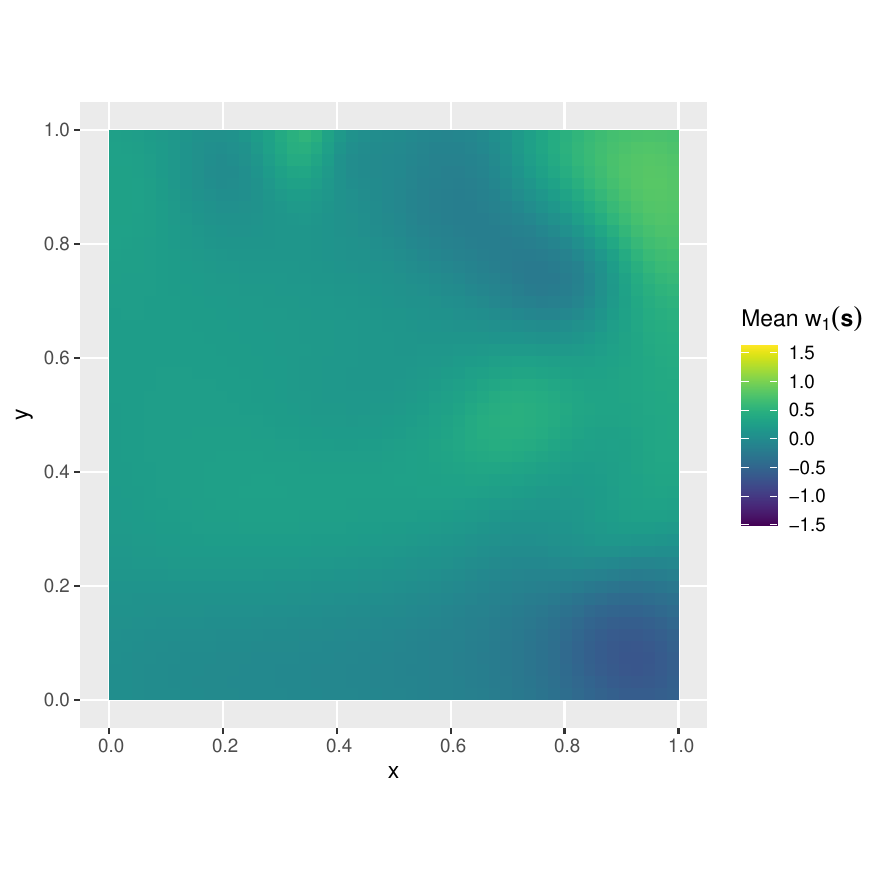}
    \end{minipage}
    \begin{minipage}{0.32\textwidth}
        \includegraphics[width=\linewidth]{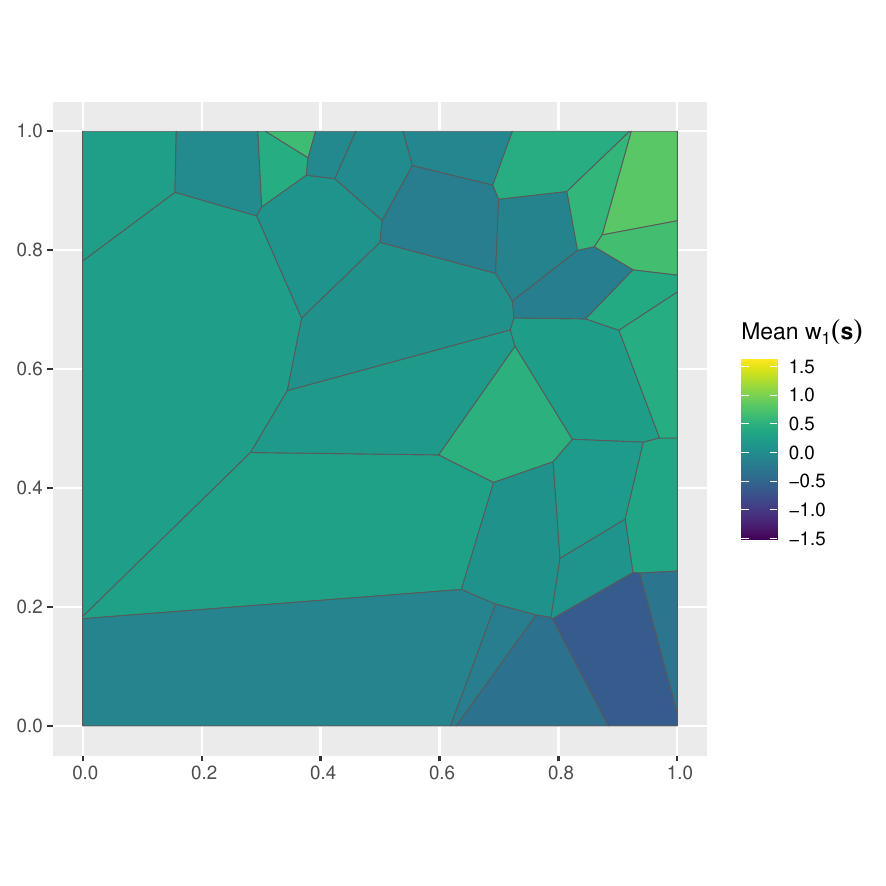}
    \end{minipage}
    \begin{minipage}{0.32\textwidth}
        \includegraphics[width=\linewidth]{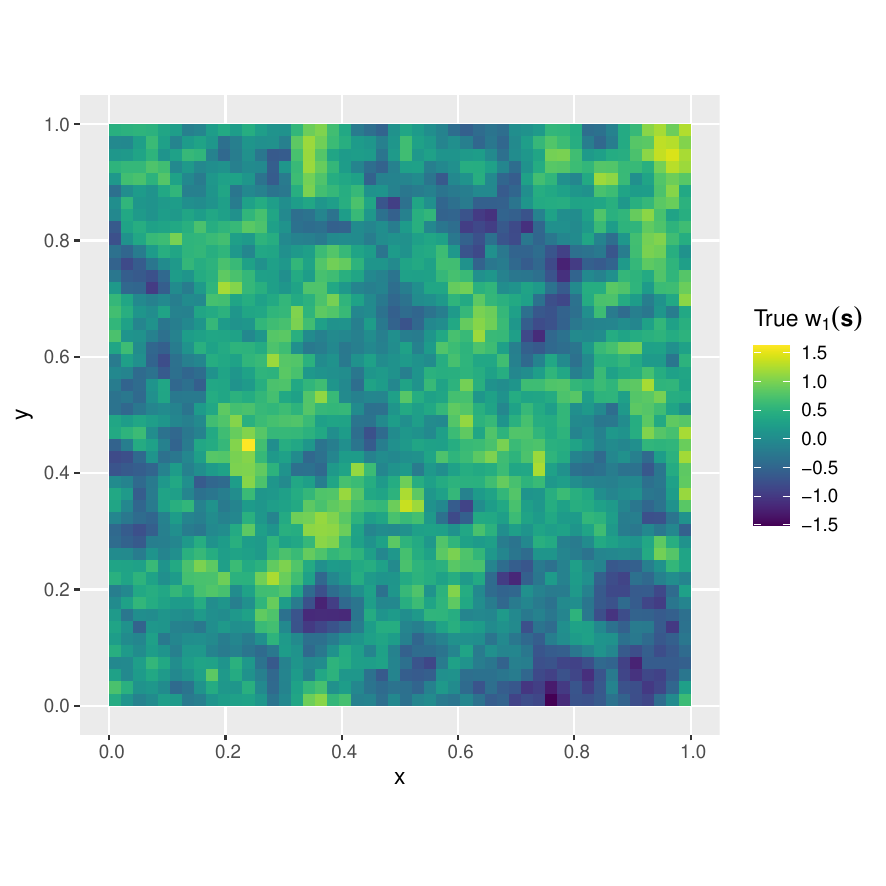}
    \end{minipage}
    \caption{$W_1(\textbf{s})$}
\end{subfigure}
\hfill
\begin{subfigure}{1\textwidth}
    \begin{minipage}{0.32\textwidth}
        \includegraphics[width=\linewidth]{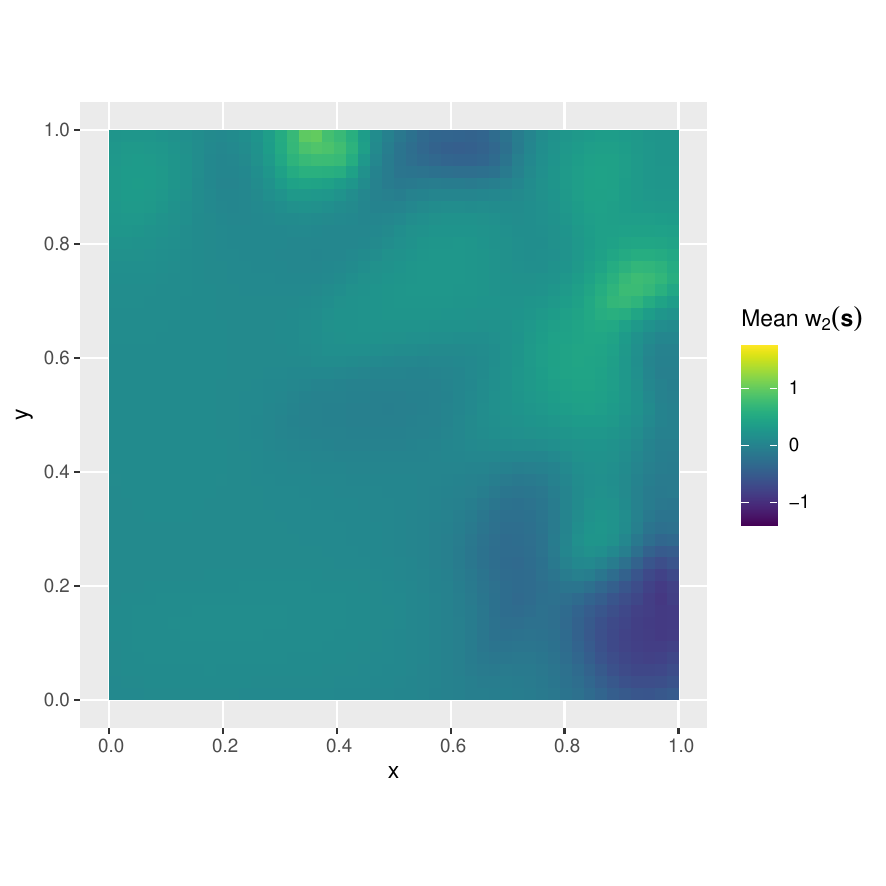}
    \end{minipage}
    \begin{minipage}{0.32\textwidth}
        \includegraphics[width=\linewidth]{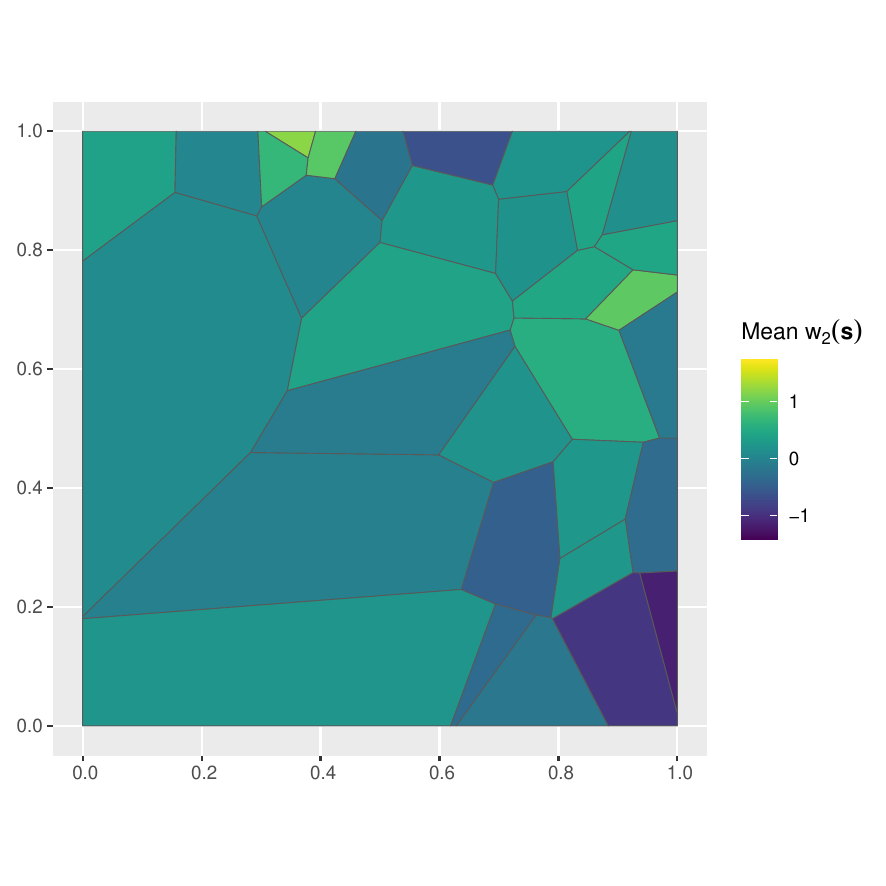}
    \end{minipage}
    \begin{minipage}{0.32\textwidth}
        \includegraphics[width=\linewidth]{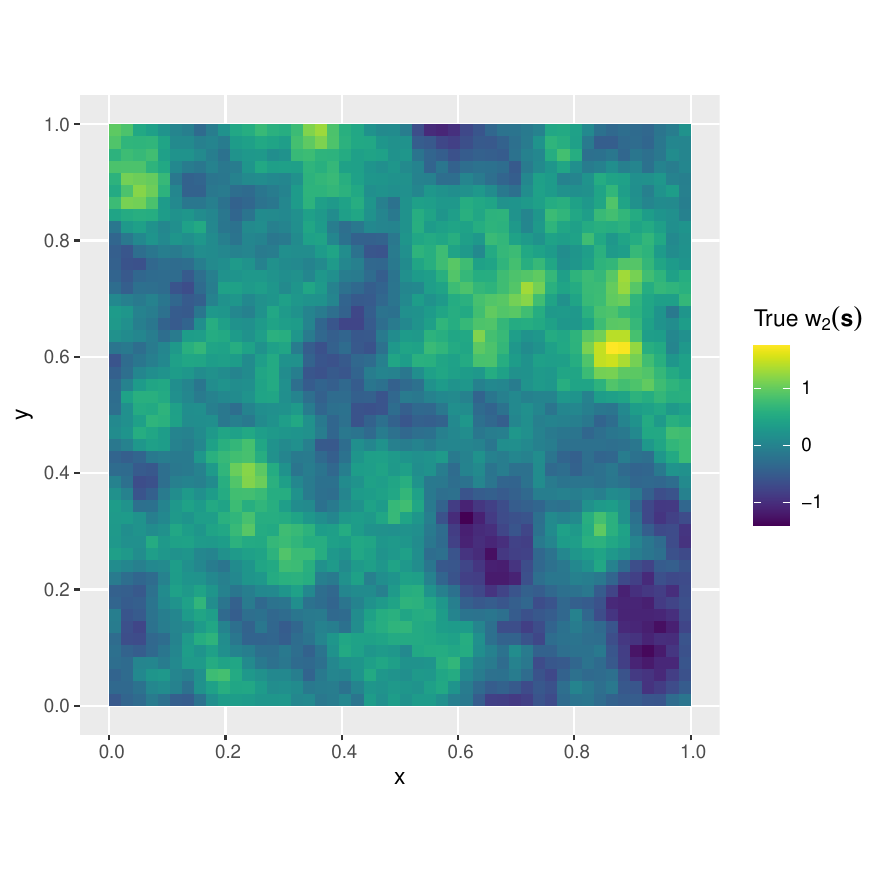}
    \end{minipage}
    \caption{$W_2(\textbf{s})$}
\end{subfigure}
\hfill
\begin{subfigure}{1\textwidth}
    \begin{minipage}{0.32\textwidth}
        \includegraphics[width=\linewidth]{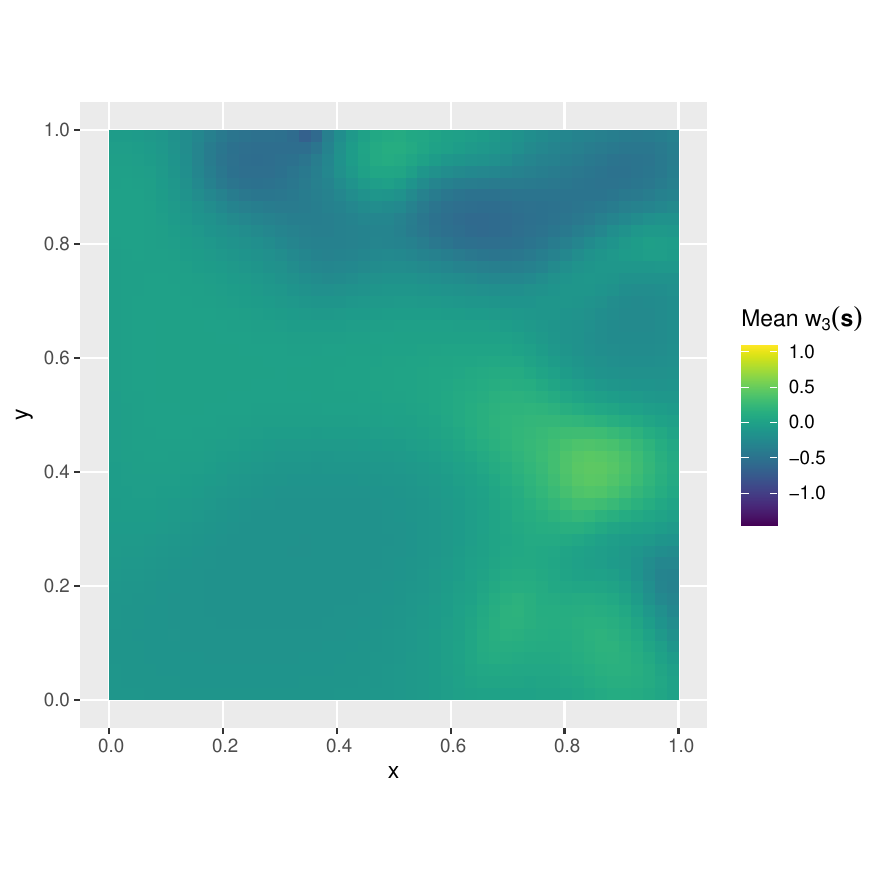}
    \end{minipage}
    \begin{minipage}{0.32\textwidth}
        \includegraphics[width=\linewidth]{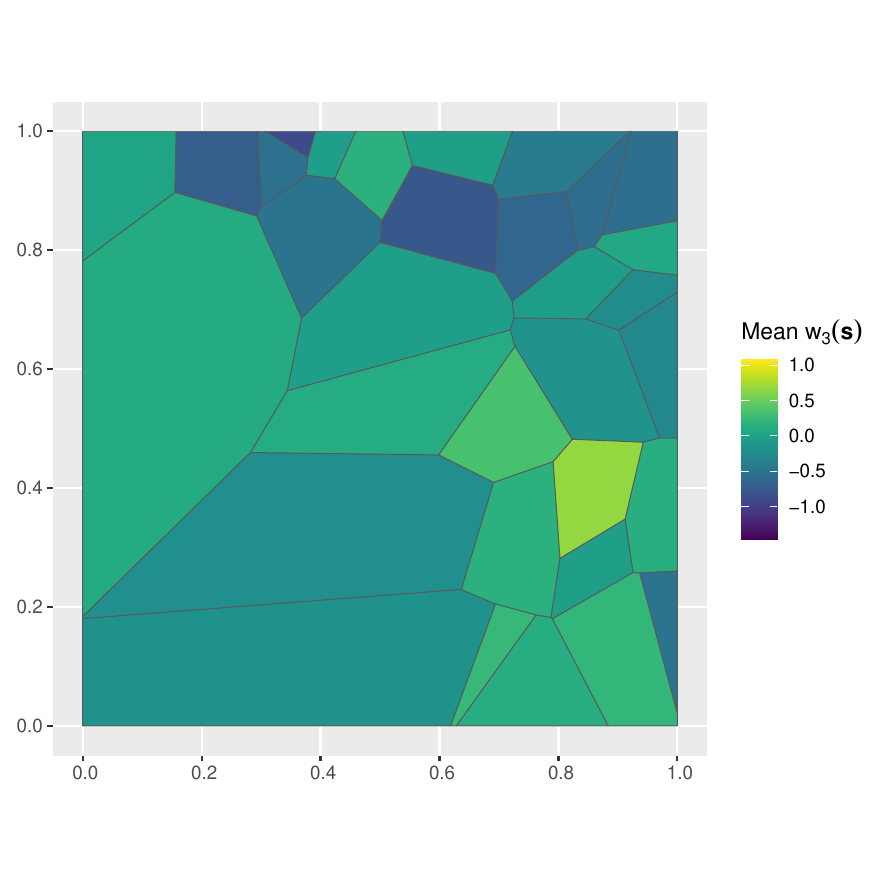}
    \end{minipage}
    \begin{minipage}{0.32\textwidth}
        \includegraphics[width=\linewidth]{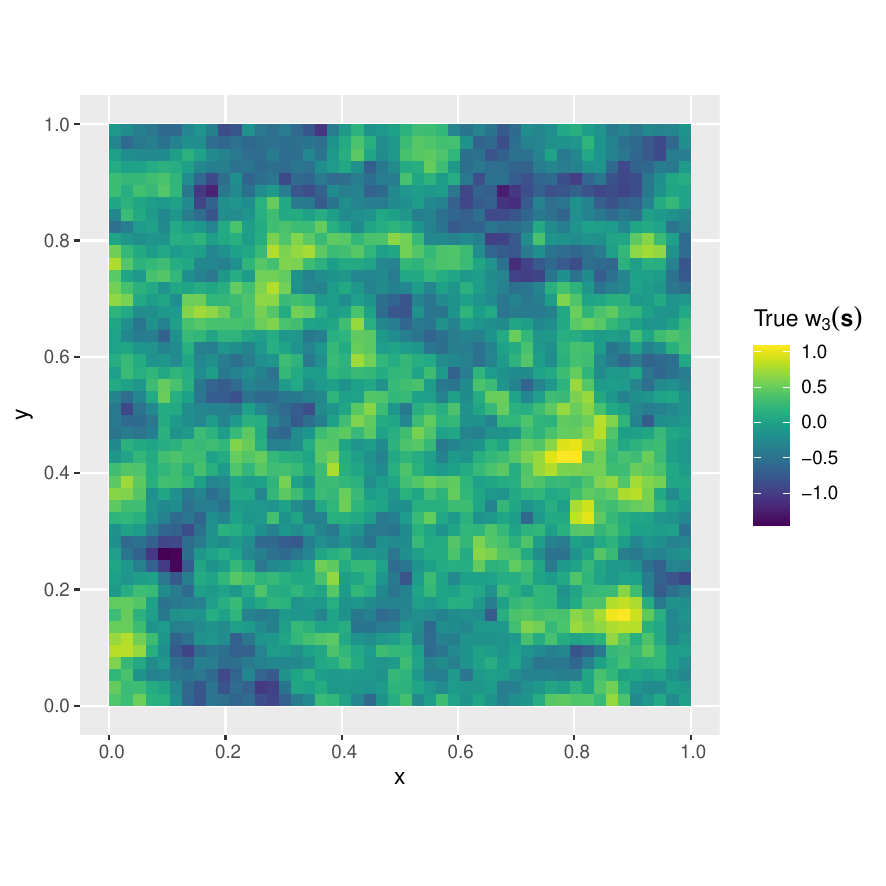}
    \end{minipage}
    \caption{$W_3(\textbf{s})$}
\end{subfigure}
\caption{The left column displays the predictions from the disaggregation model, the central column shows the predictions from the areal model, and the right column presents the realization of the latent space. Each row corresponds to one of the three components of the latent space.}
\label{fig:ex1_s_1}
\end{figure}


\begin{figure}[htp]
\centering
\large{\textbf{Scenario 2}\par\medskip}
\begin{subfigure}{1\textwidth}
    \begin{minipage}{0.32\textwidth}
        \includegraphics[width=\linewidth]{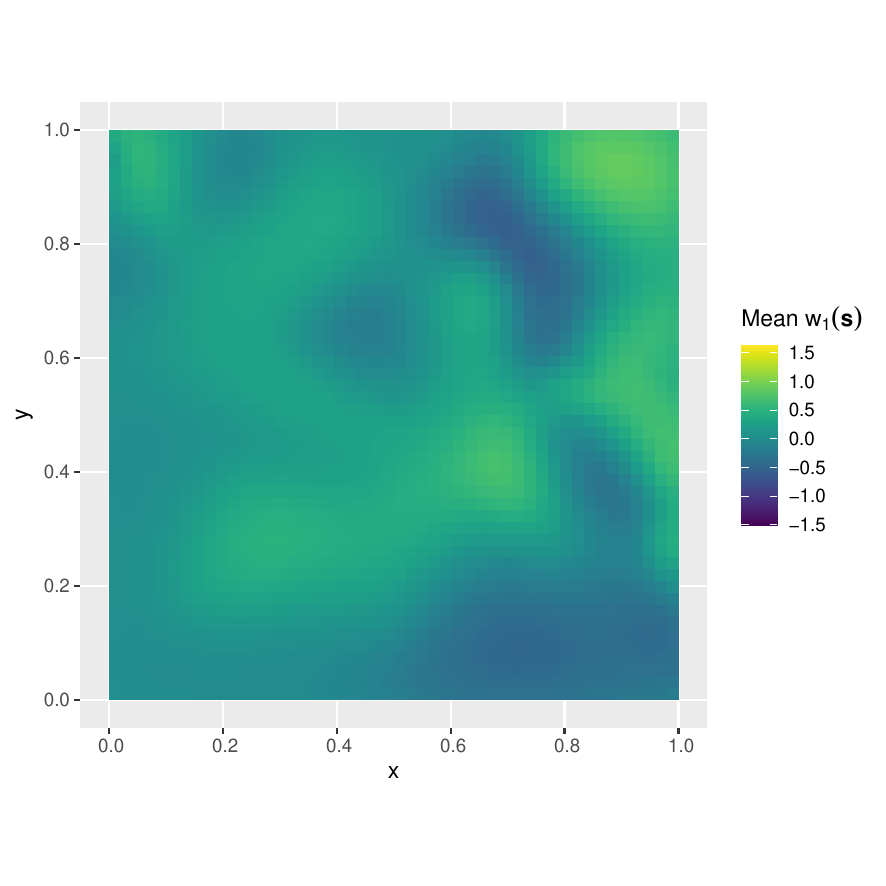}
    \end{minipage}
    \begin{minipage}{0.32\textwidth}
        \includegraphics[width=\linewidth]{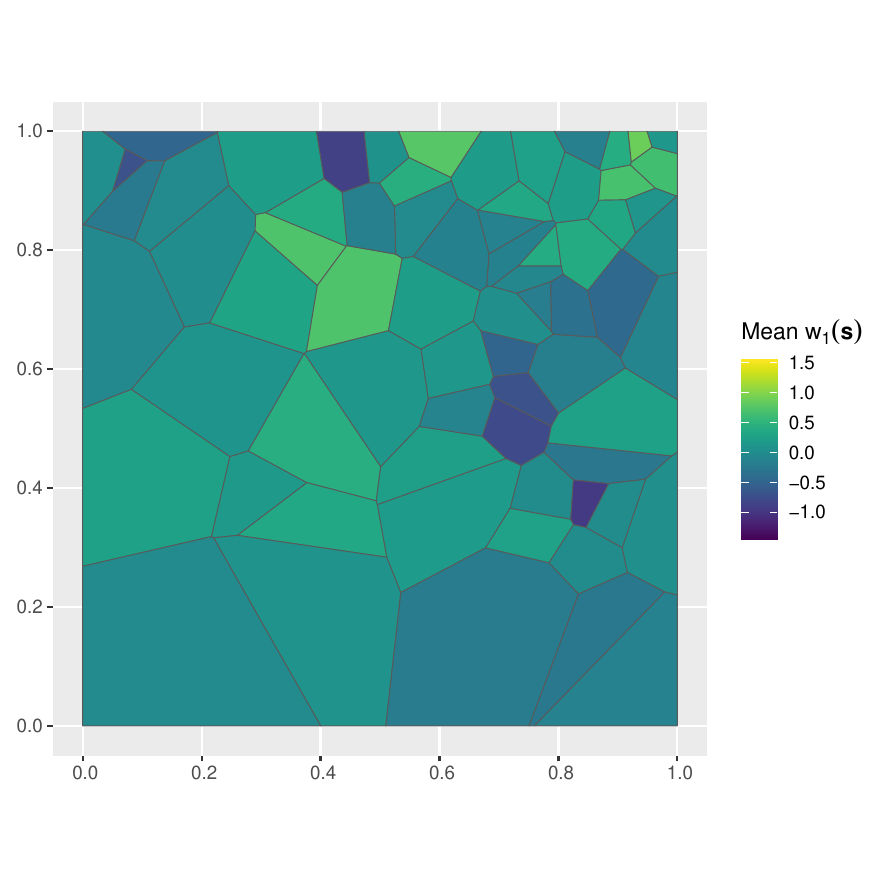}
    \end{minipage}
    \begin{minipage}{0.32\textwidth}
        \includegraphics[width=\linewidth]{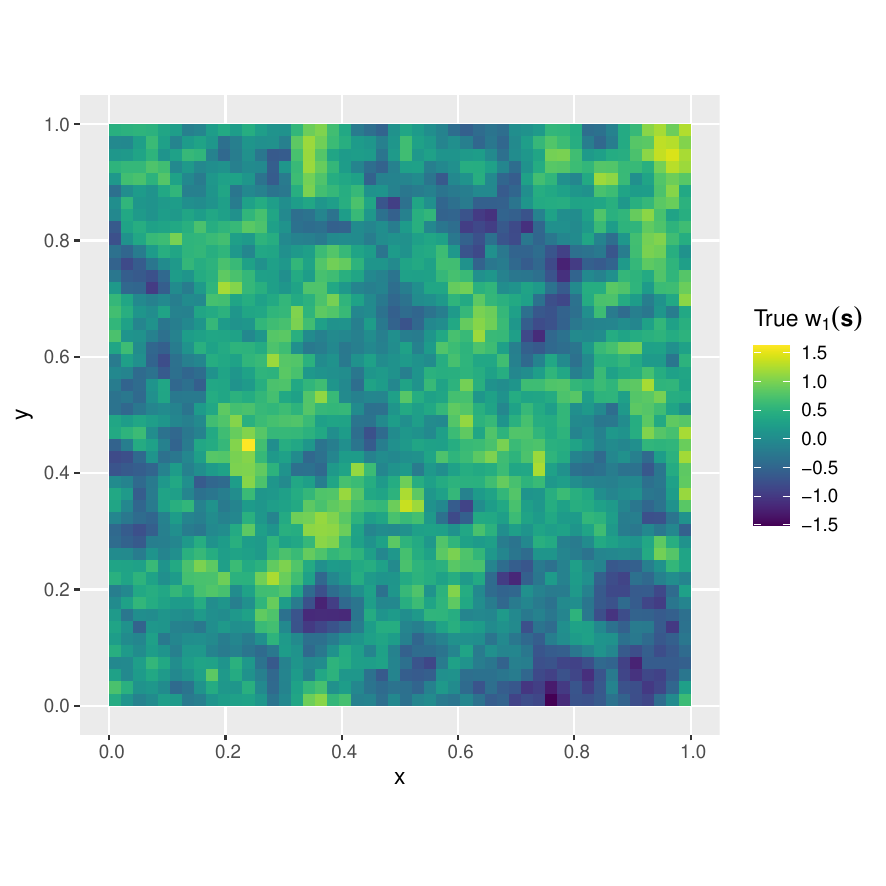}
    \end{minipage}
    \caption{$W_1(\textbf{s})$}
\end{subfigure}
\hfill
\begin{subfigure}{1\textwidth}
    \begin{minipage}{0.32\textwidth}
        \includegraphics[width=\linewidth]{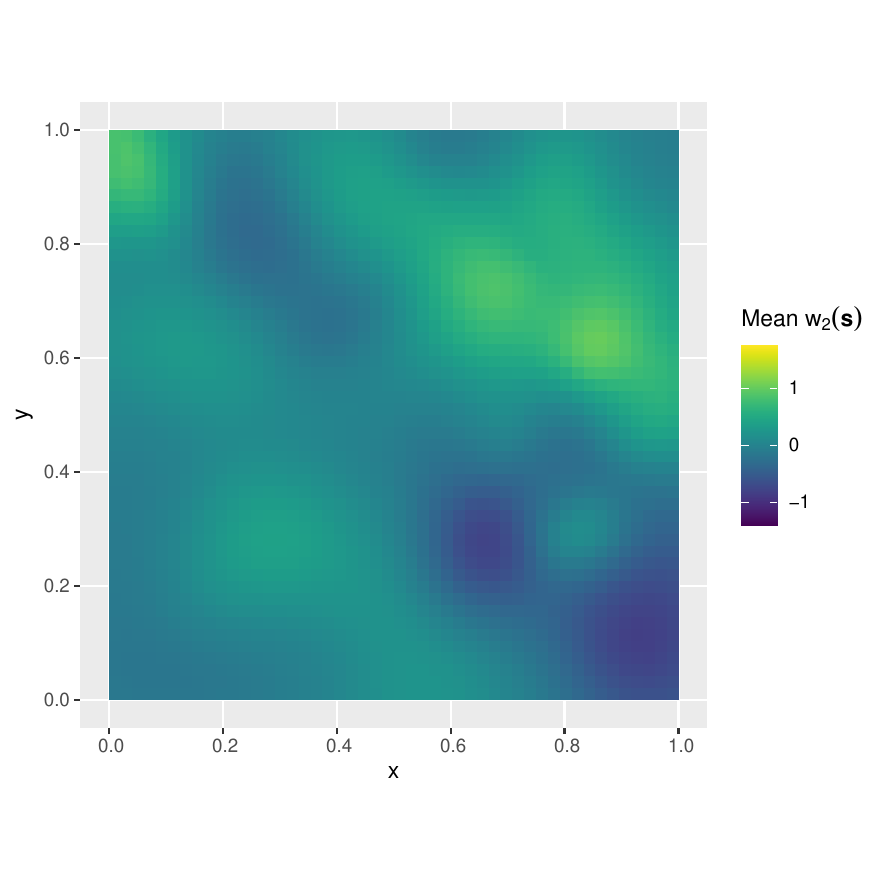}
    \end{minipage}
    \begin{minipage}{0.32\textwidth}
        \includegraphics[width=\linewidth]{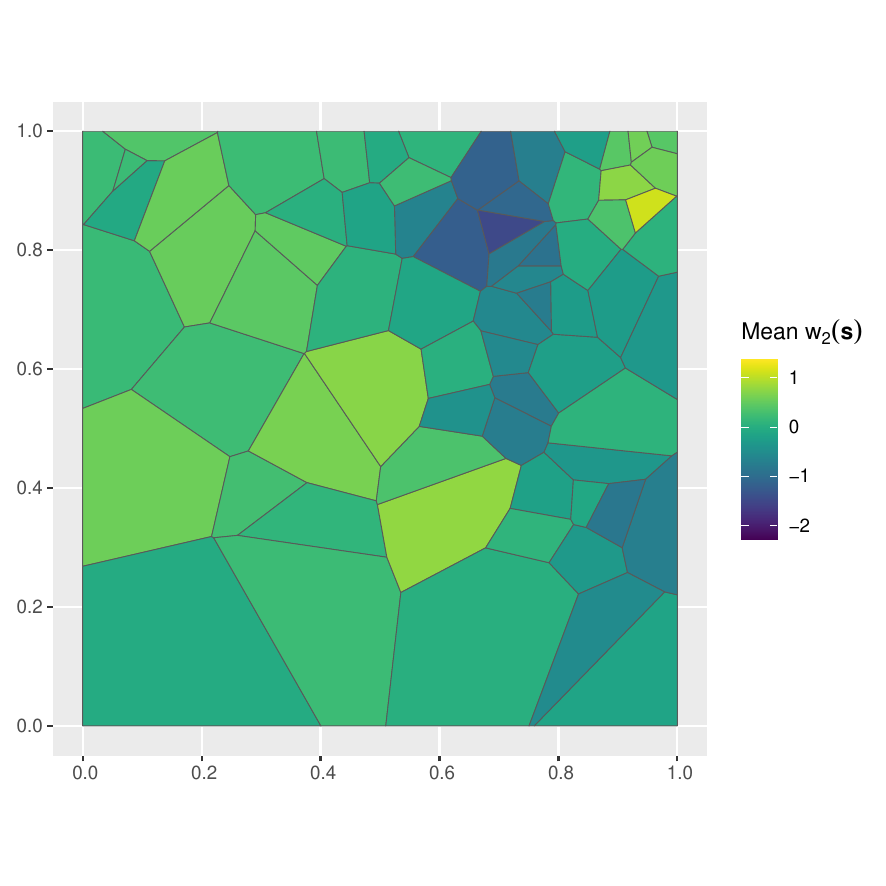}
    \end{minipage}
    \begin{minipage}{0.32\textwidth}
        \includegraphics[width=\linewidth]{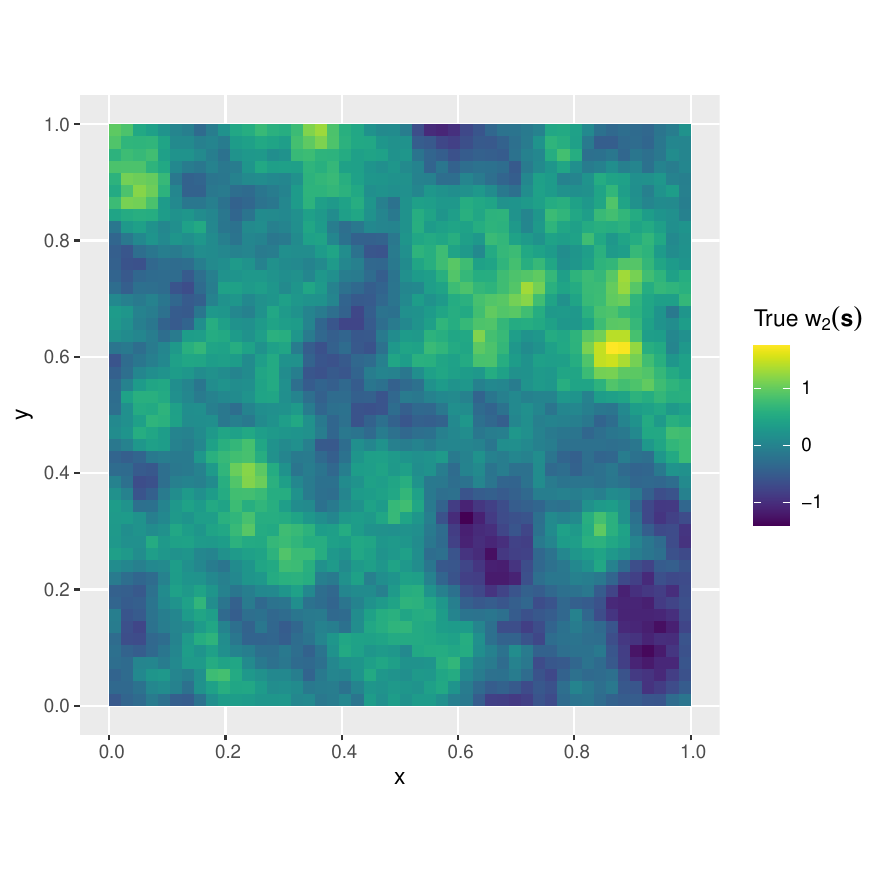}
    \end{minipage}
    \caption{$W_2(\textbf{s})$}
\end{subfigure}
\hfill
\begin{subfigure}{1\textwidth}
    \begin{minipage}{0.32\textwidth}
        \includegraphics[width=\linewidth]{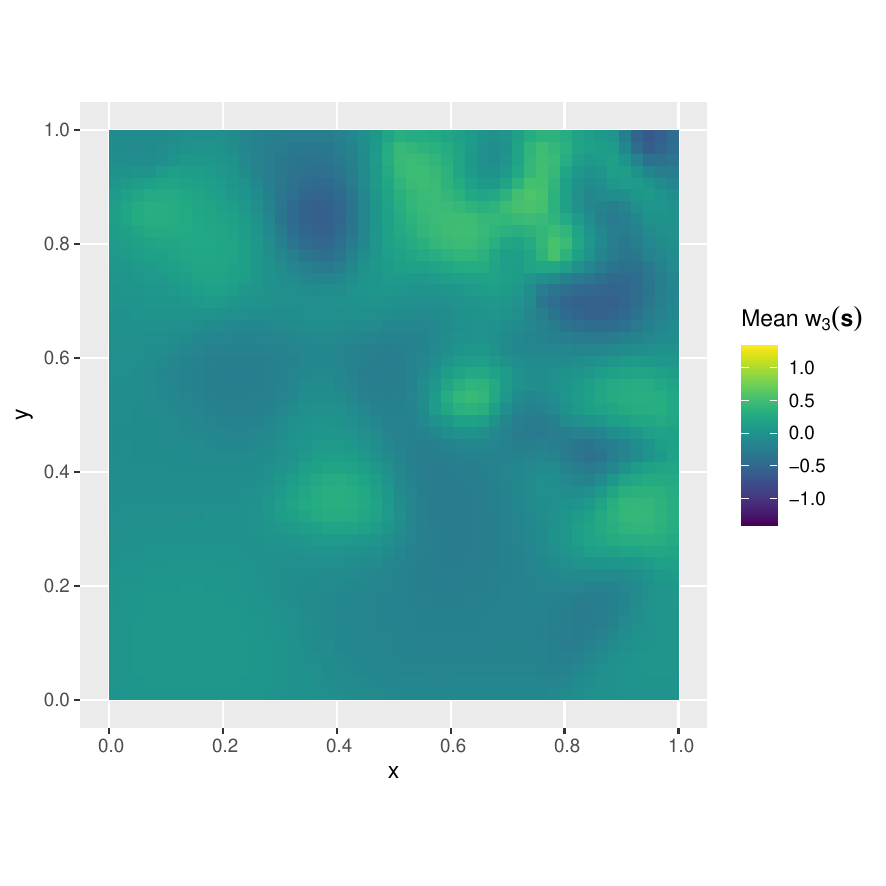}
    \end{minipage}
    \begin{minipage}{0.32\textwidth}
        \includegraphics[width=\linewidth]{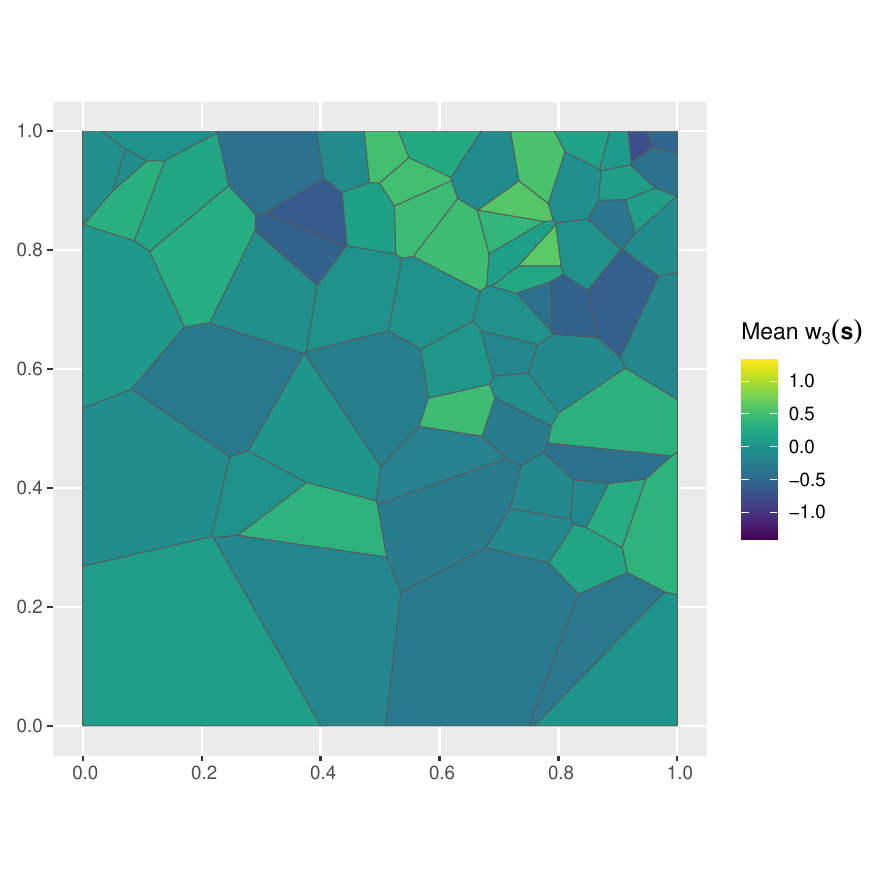}
    \end{minipage}
    \begin{minipage}{0.32\textwidth}
        \includegraphics[width=\linewidth]{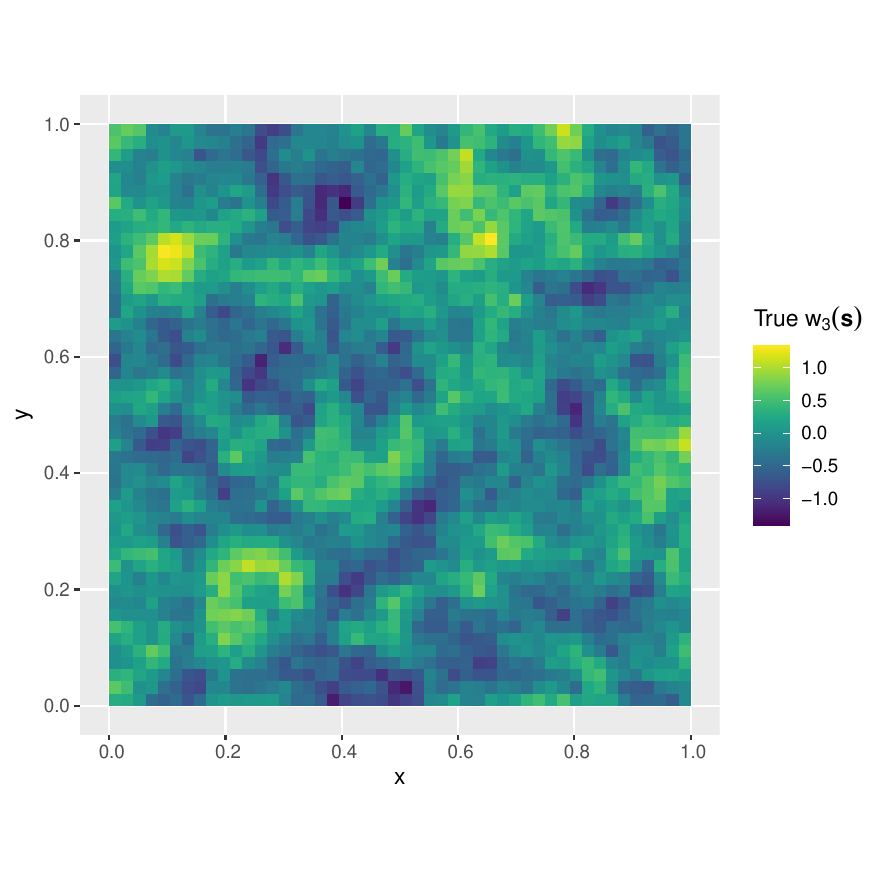}
    \end{minipage}
    \caption{$W_3(\textbf{s})$}
\end{subfigure}
\caption{The left column displays the predictions from the disaggregation model, the central column shows the predictions from the areal model, and the right column presents the realization of the latent space. Each row corresponds to one of the three components of the latent space.}
\label{fig:ex1_s_2}
\end{figure}


\begin{figure}[htp]
\centering
\large{\textbf{Scenario 3}\par\medskip}
\begin{subfigure}{1\textwidth}
    \begin{minipage}{0.32\textwidth}
        \includegraphics[width=\linewidth]{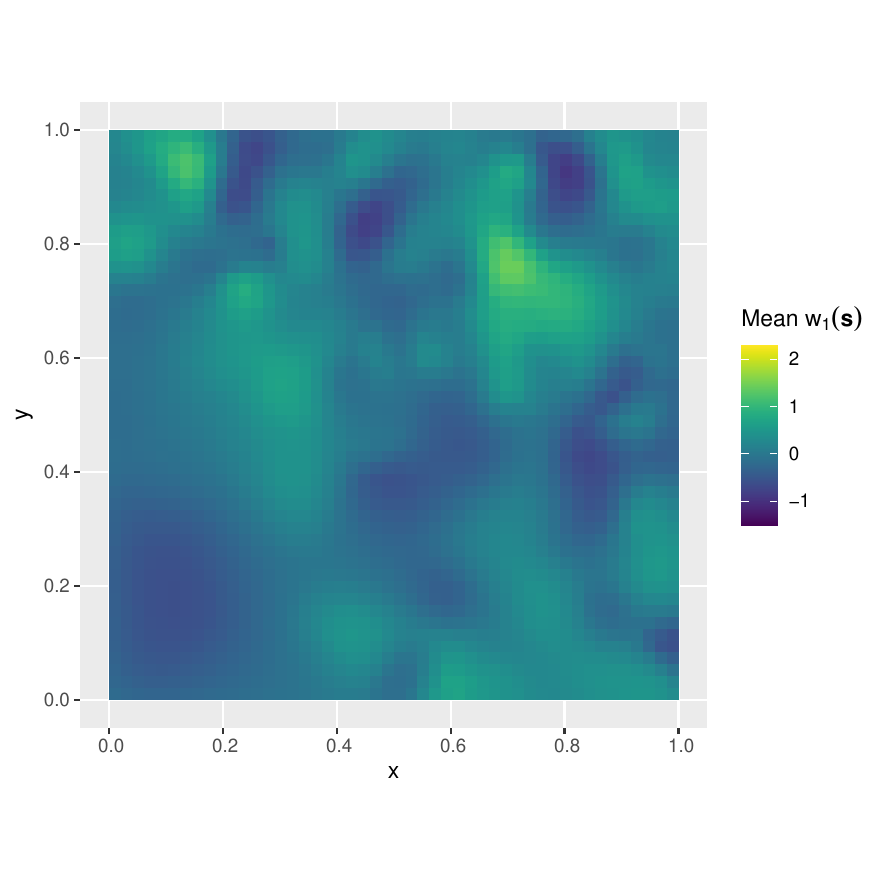}
    \end{minipage}
    \begin{minipage}{0.32\textwidth}
        \includegraphics[width=\linewidth]{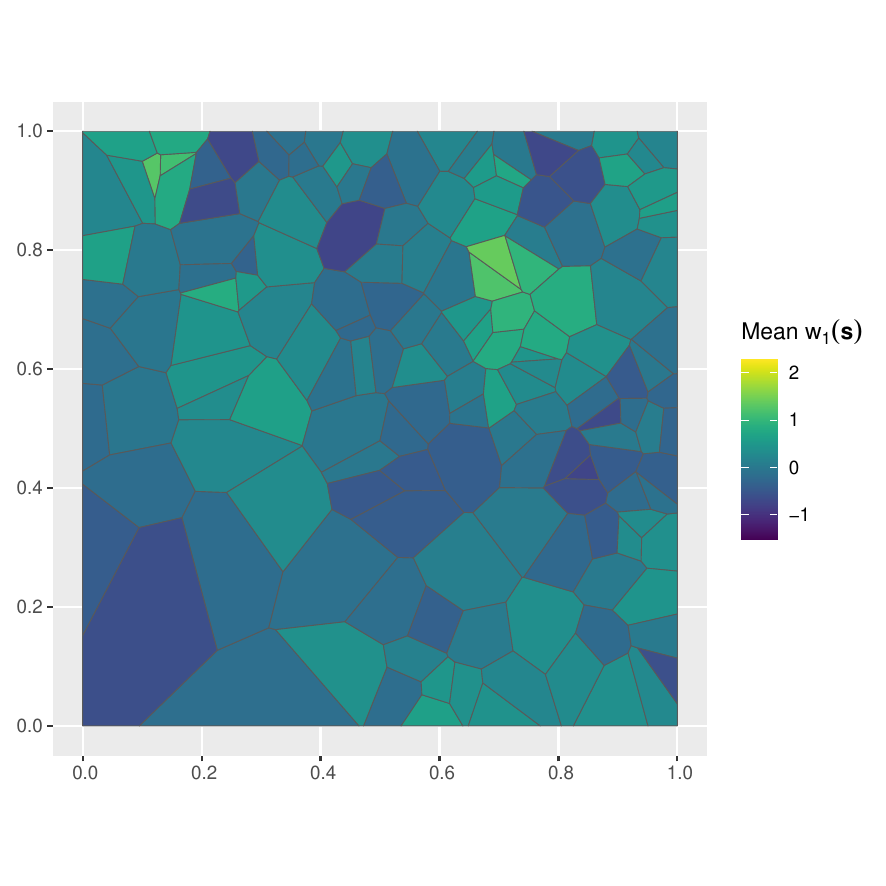}
    \end{minipage}
    \begin{minipage}{0.32\textwidth}
        \includegraphics[width=\linewidth]{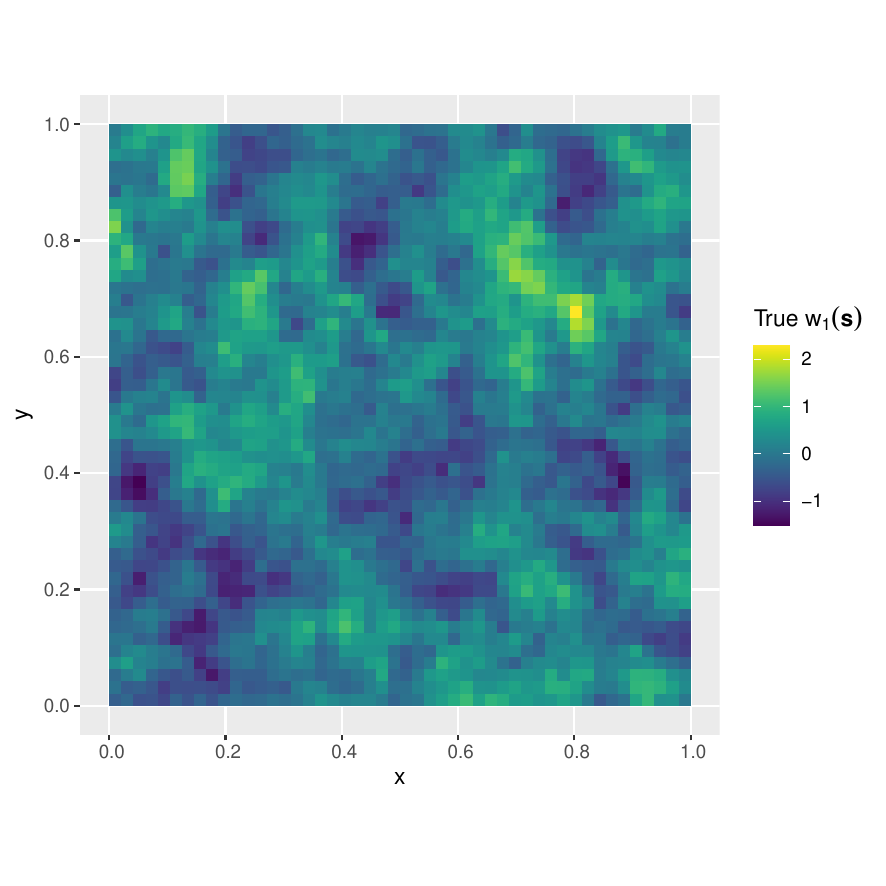}
    \end{minipage}
    \caption{$W_1(\textbf{s})$}
\end{subfigure}
\hfill
\begin{subfigure}{1\textwidth}
    \begin{minipage}{0.32\textwidth}
        \includegraphics[width=\linewidth]{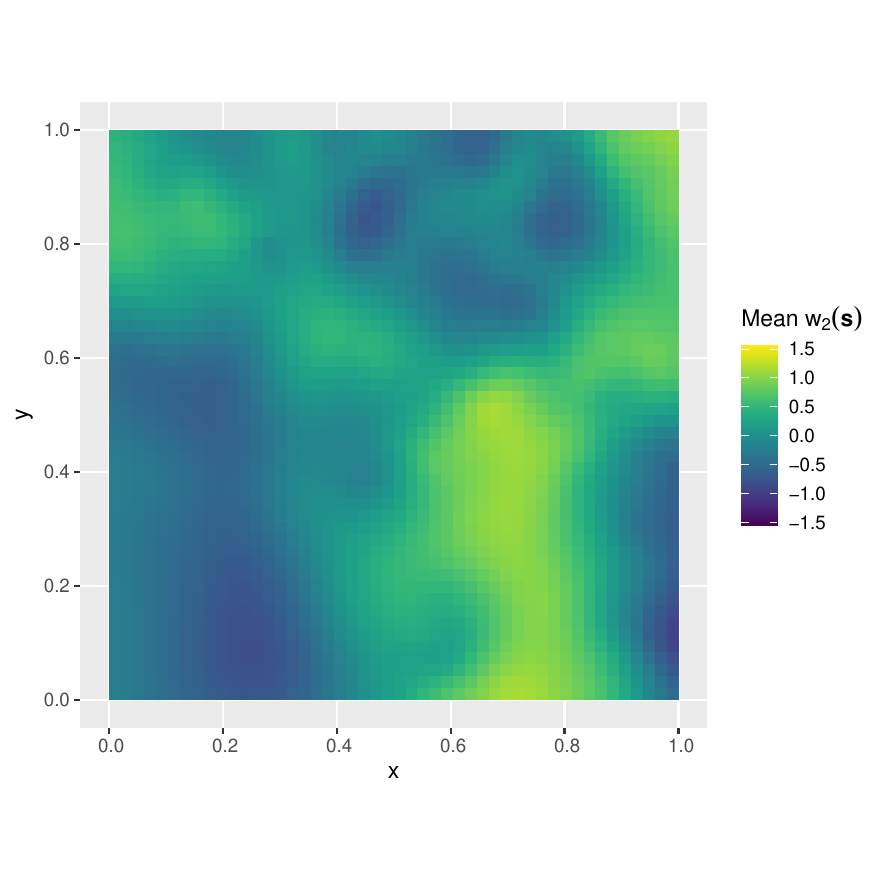}
    \end{minipage}
    \begin{minipage}{0.32\textwidth}
        \includegraphics[width=\linewidth]{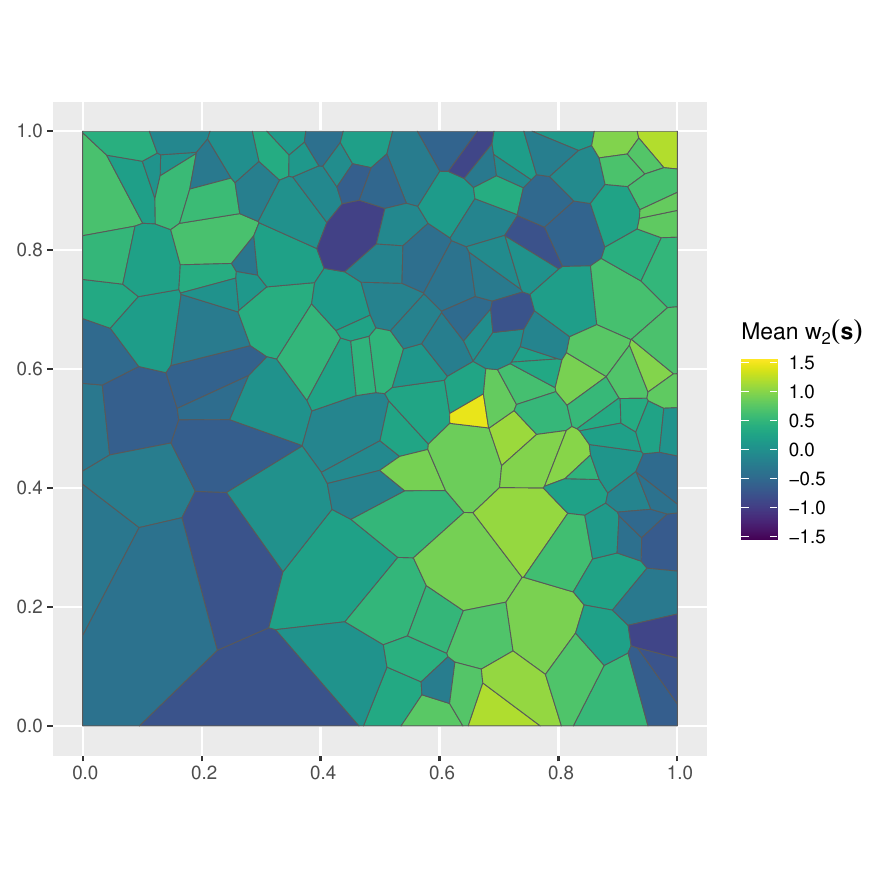}
    \end{minipage}
    \begin{minipage}{0.32\textwidth}
        \includegraphics[width=\linewidth]{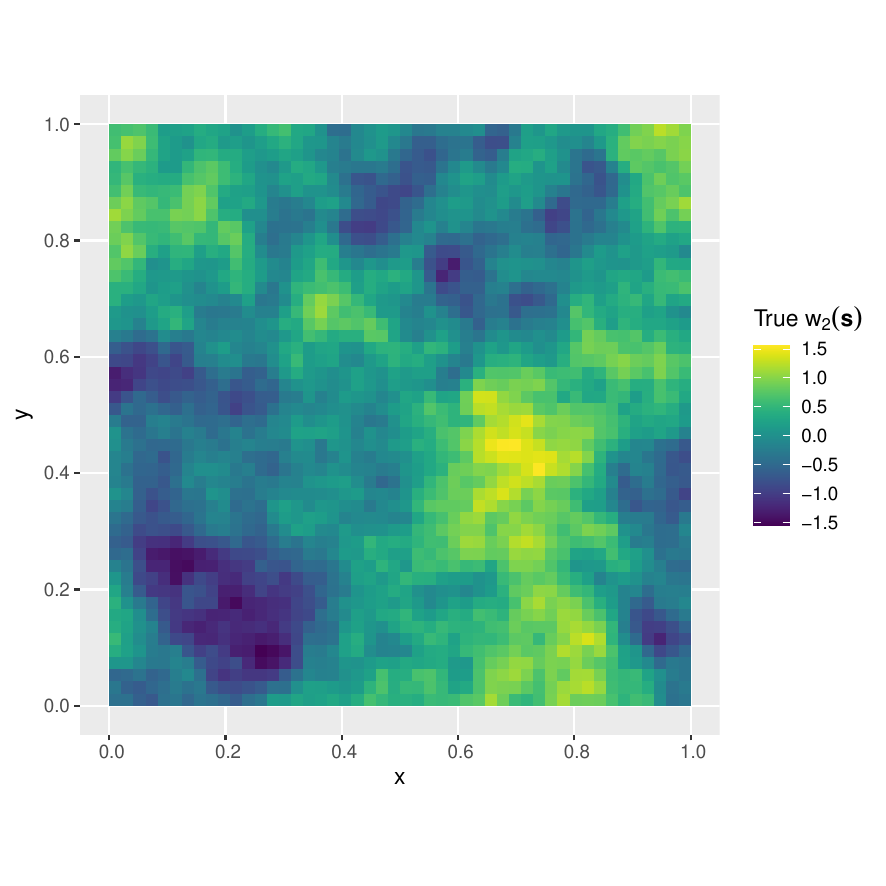}
    \end{minipage}
    \caption{$W_2(\textbf{s})$}
\end{subfigure}
\hfill
\begin{subfigure}{1\textwidth}
    \begin{minipage}{0.32\textwidth}
        \includegraphics[width=\linewidth]{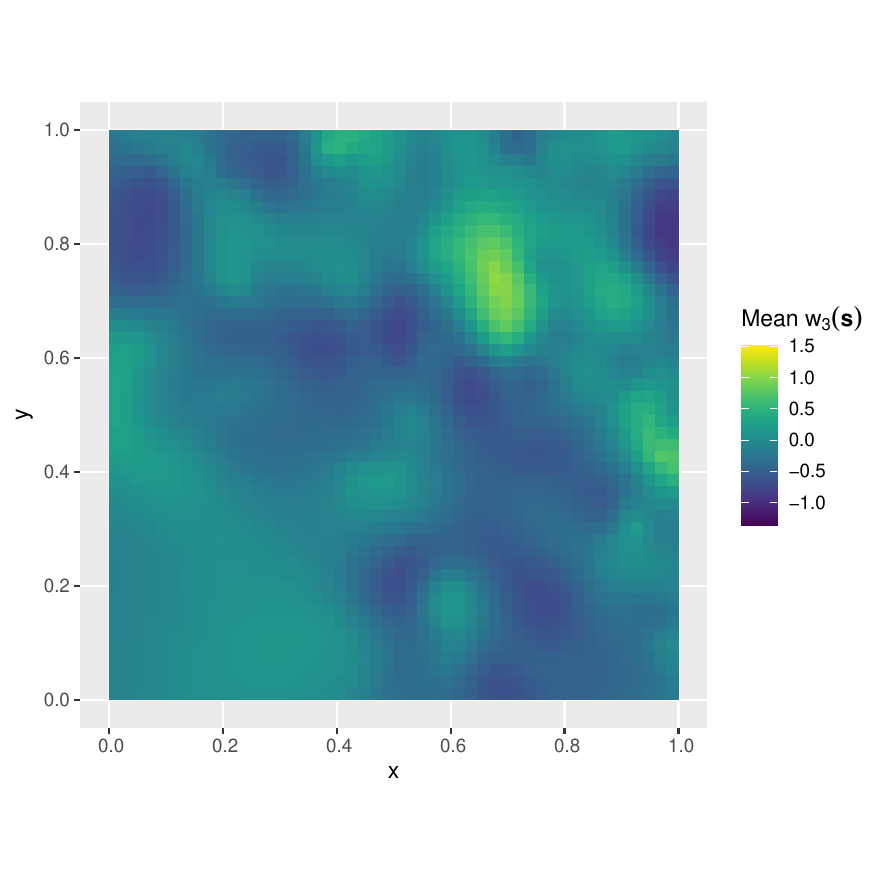}
    \end{minipage}
    \begin{minipage}{0.32\textwidth}
        \includegraphics[width=\linewidth]{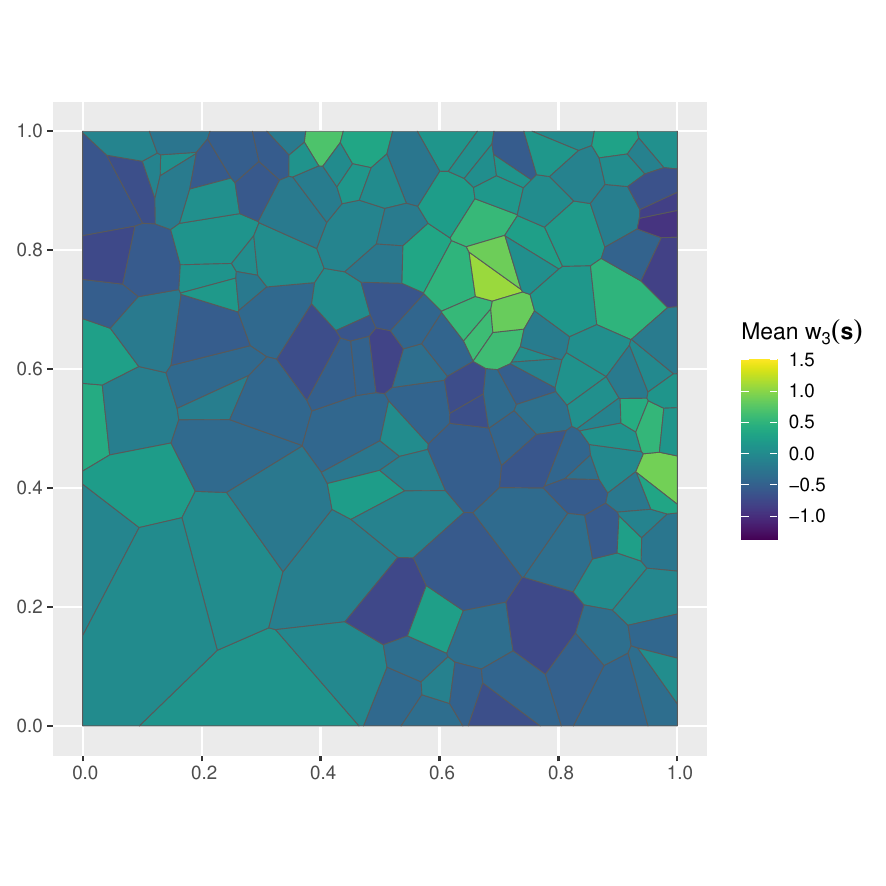}
    \end{minipage}
    \begin{minipage}{0.32\textwidth}
        \includegraphics[width=\linewidth]{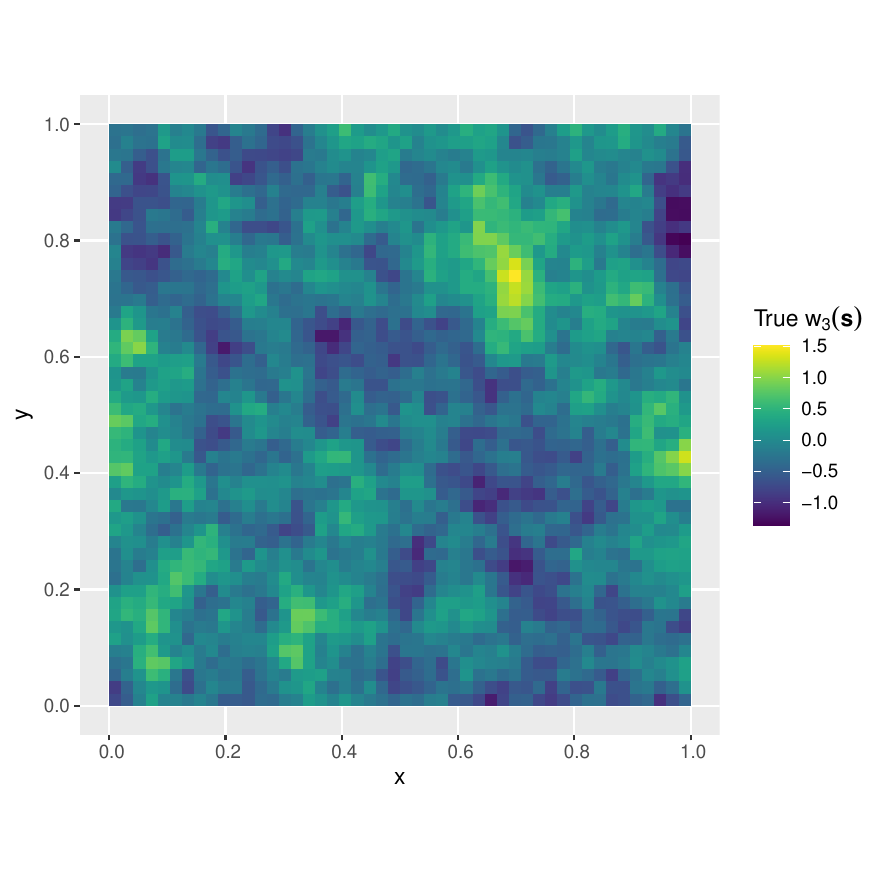}
    \end{minipage}
    \caption{$W_3(\textbf{s})$}
\end{subfigure}
\caption{The left column displays the predictions from the disaggregation model, the central column shows the predictions from the areal model, and the right column presents the realization of the latent space. Each row corresponds to one of the three components of the latent space.}
\label{fig:ex1_s_3}
\end{figure}


\begin{figure}[htp]
\centering
\large{\textbf{Scenario 4}\par\medskip}
\begin{subfigure}{1\textwidth}
    \begin{minipage}{0.32\textwidth}
        \includegraphics[width=\linewidth]{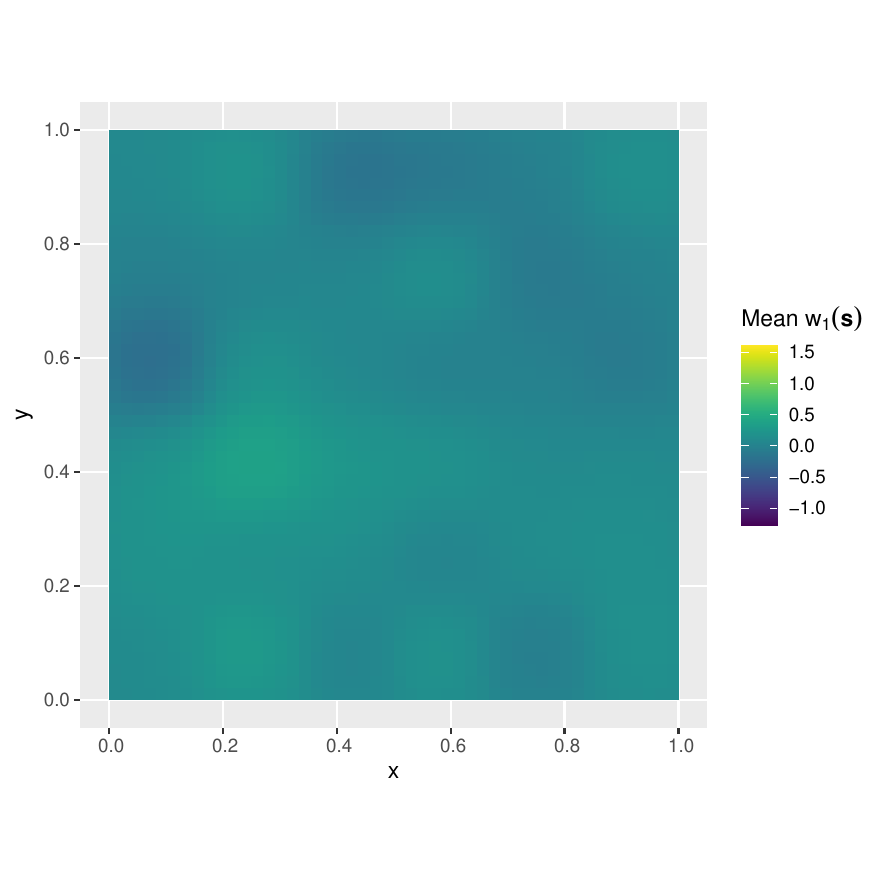}
    \end{minipage}
    \begin{minipage}{0.32\textwidth}
        \includegraphics[width=\linewidth]{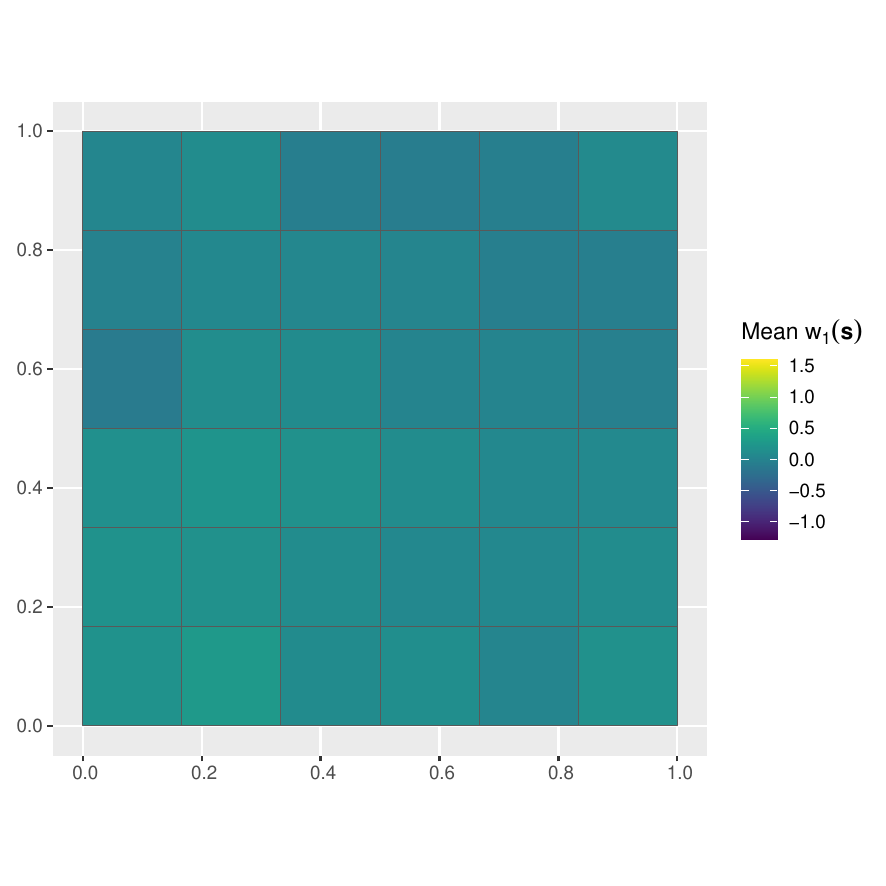}
    \end{minipage}
    \begin{minipage}{0.32\textwidth}
        \includegraphics[width=\linewidth]{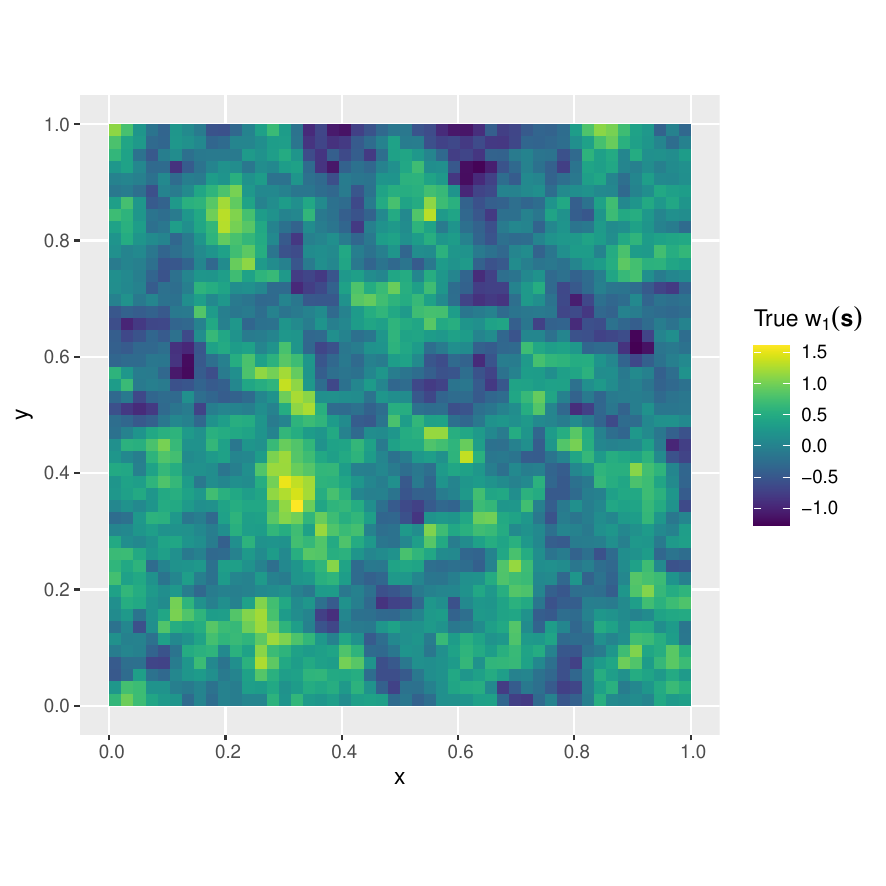}
    \end{minipage}
    \caption{$W_1(\textbf{s})$}
\end{subfigure}
\hfill
\begin{subfigure}{1\textwidth}
    \begin{minipage}{0.32\textwidth}
        \includegraphics[width=\linewidth]{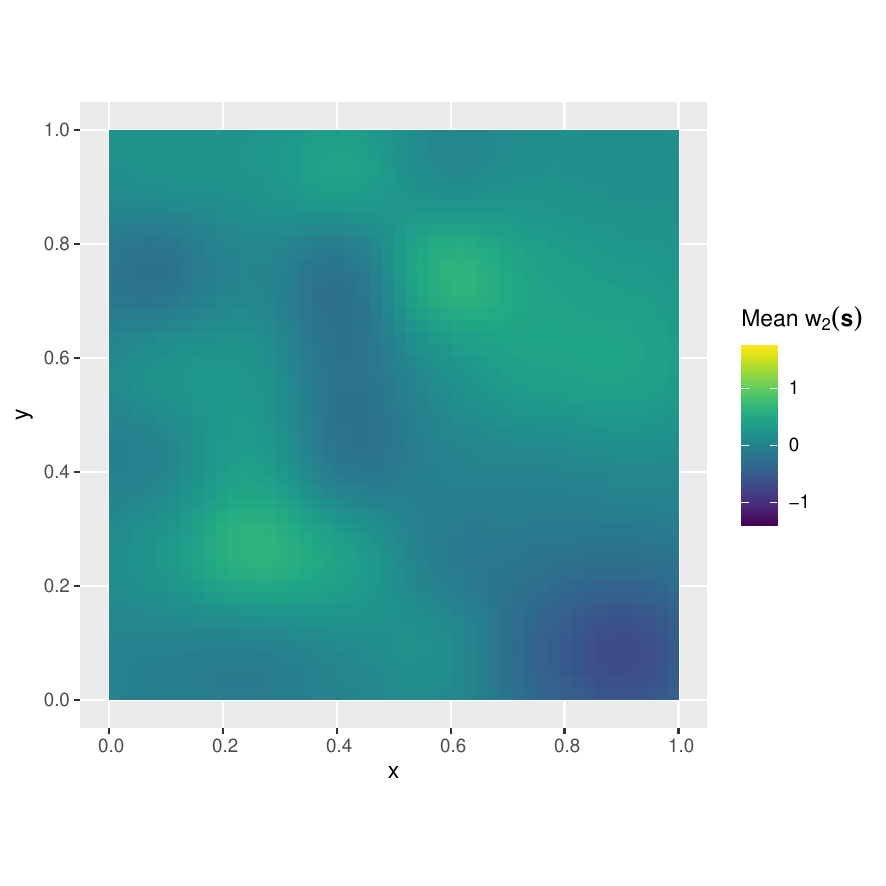}
    \end{minipage}
    \begin{minipage}{0.32\textwidth}
        \includegraphics[width=\linewidth]{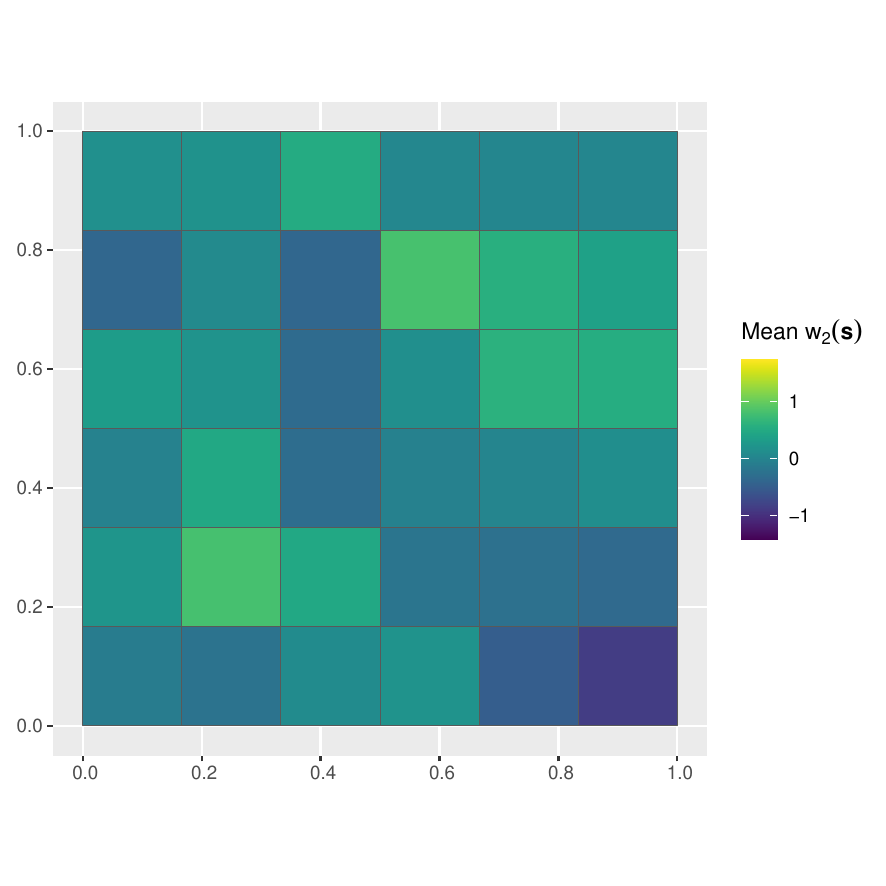}
    \end{minipage}
    \begin{minipage}{0.32\textwidth}
        \includegraphics[width=\linewidth]{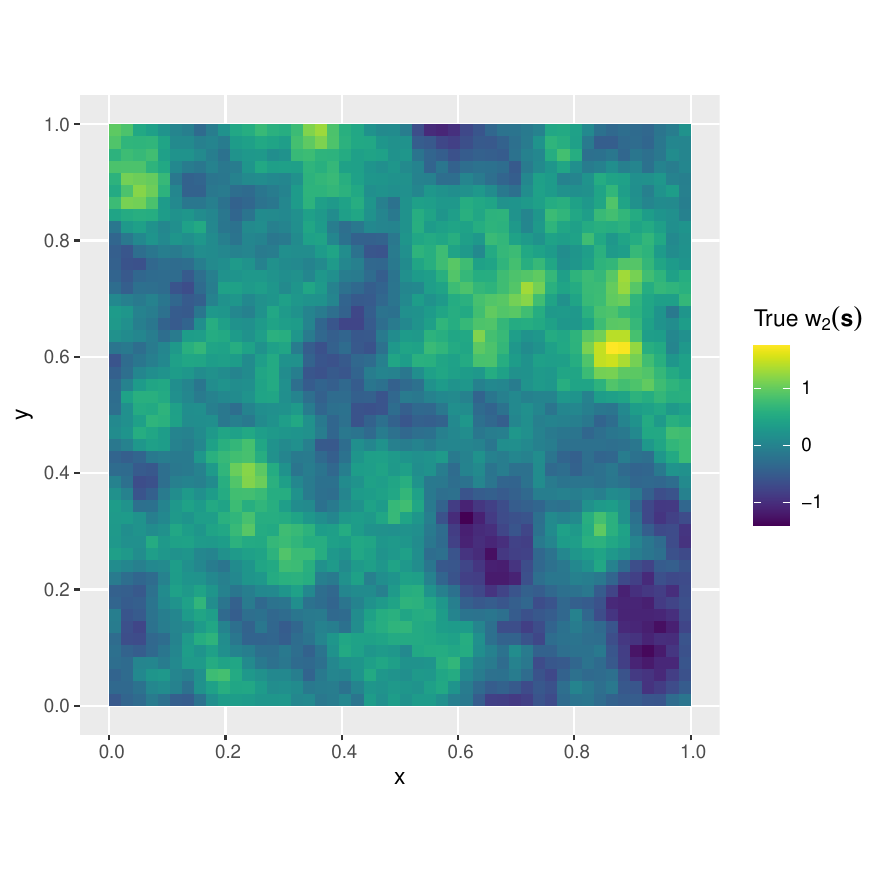}
    \end{minipage}
    \caption{$W_2(\textbf{s})$}
\end{subfigure}
\hfill
\begin{subfigure}{1\textwidth}
    \begin{minipage}{0.32\textwidth}
        \includegraphics[width=\linewidth]{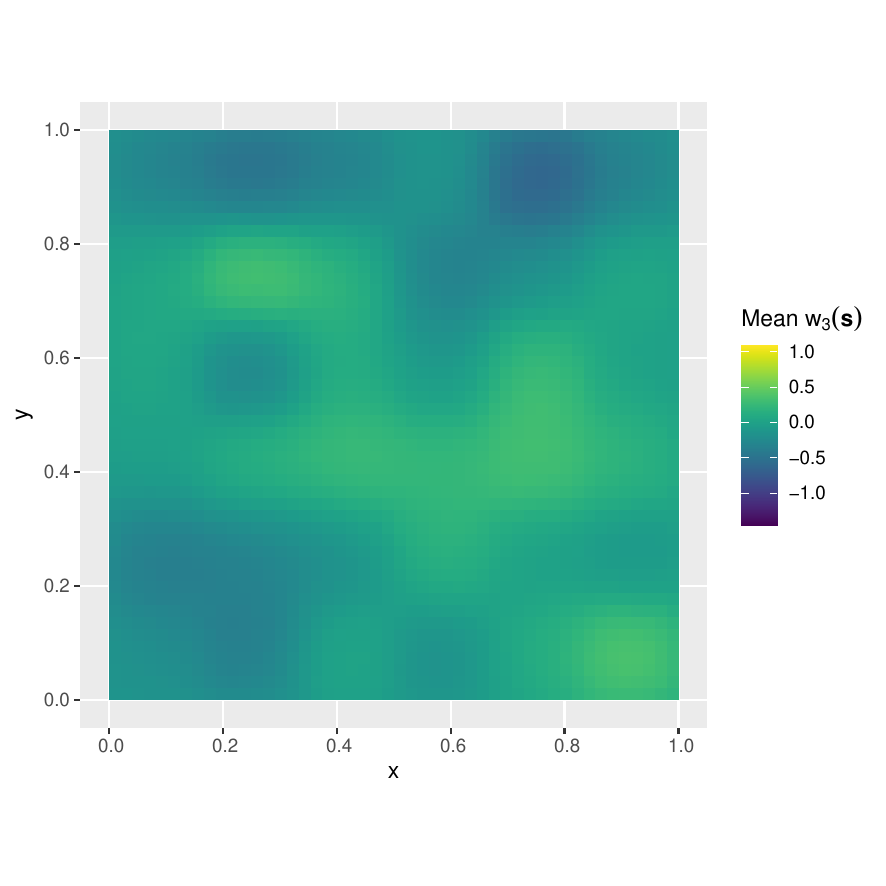}
    \end{minipage}
    \begin{minipage}{0.32\textwidth}
        \includegraphics[width=\linewidth]{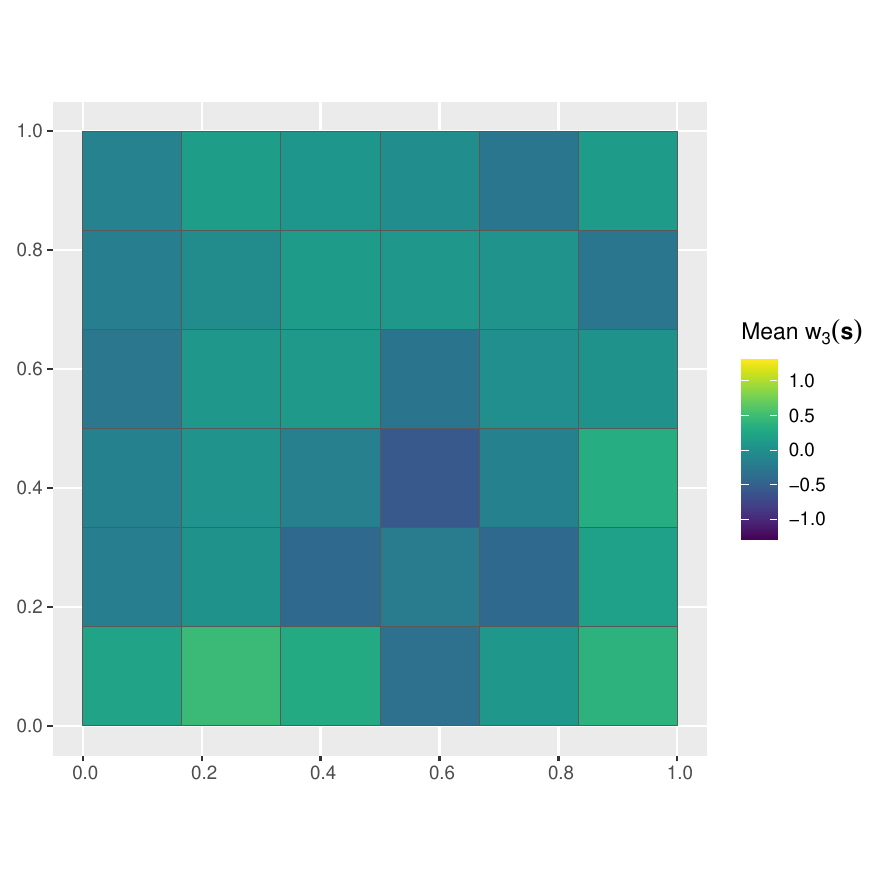}
    \end{minipage}
    \begin{minipage}{0.32\textwidth}
        \includegraphics[width=\linewidth]{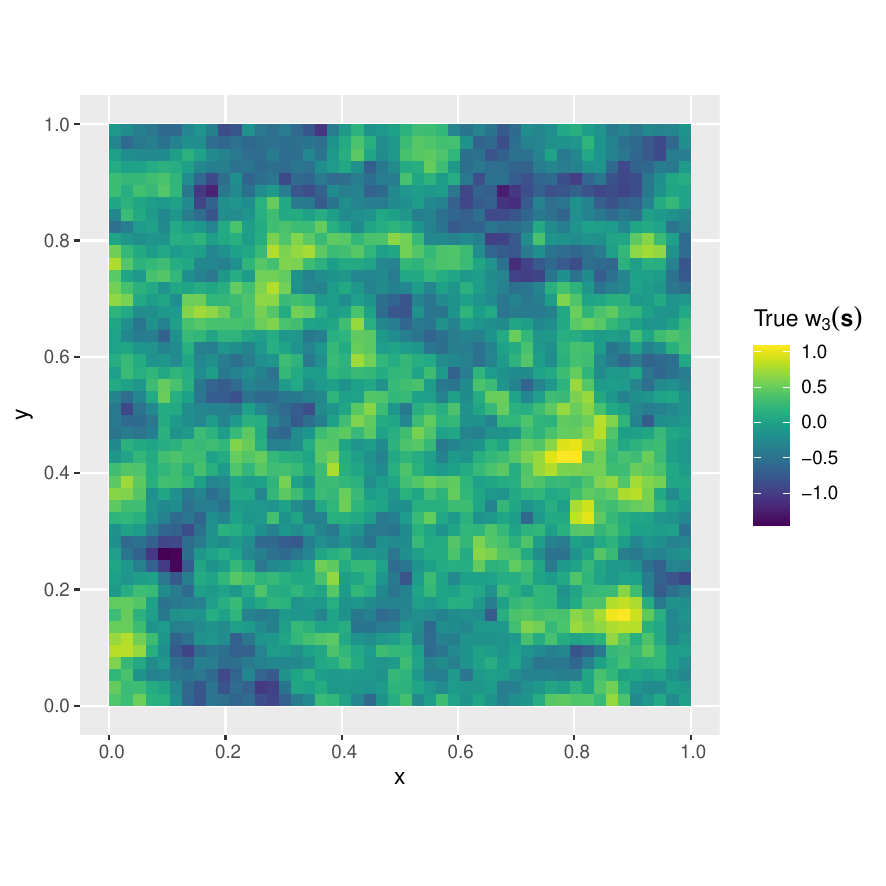}
    \end{minipage}
    \caption{$W_3(\textbf{s})$}
\end{subfigure}
\caption{The left column displays the predictions from the disaggregation model, the central column shows the predictions from the areal model, and the right column presents the realization of the latent space. Each row corresponds to one of the three components of the latent space.}
\label{fig:ex1_s_4}
\end{figure}


\begin{figure}[htp]
\centering
\large{\textbf{Scenario 5}\par\medskip}
\begin{subfigure}{1\textwidth}
    \begin{minipage}{0.32\textwidth}
        \includegraphics[width=\linewidth]{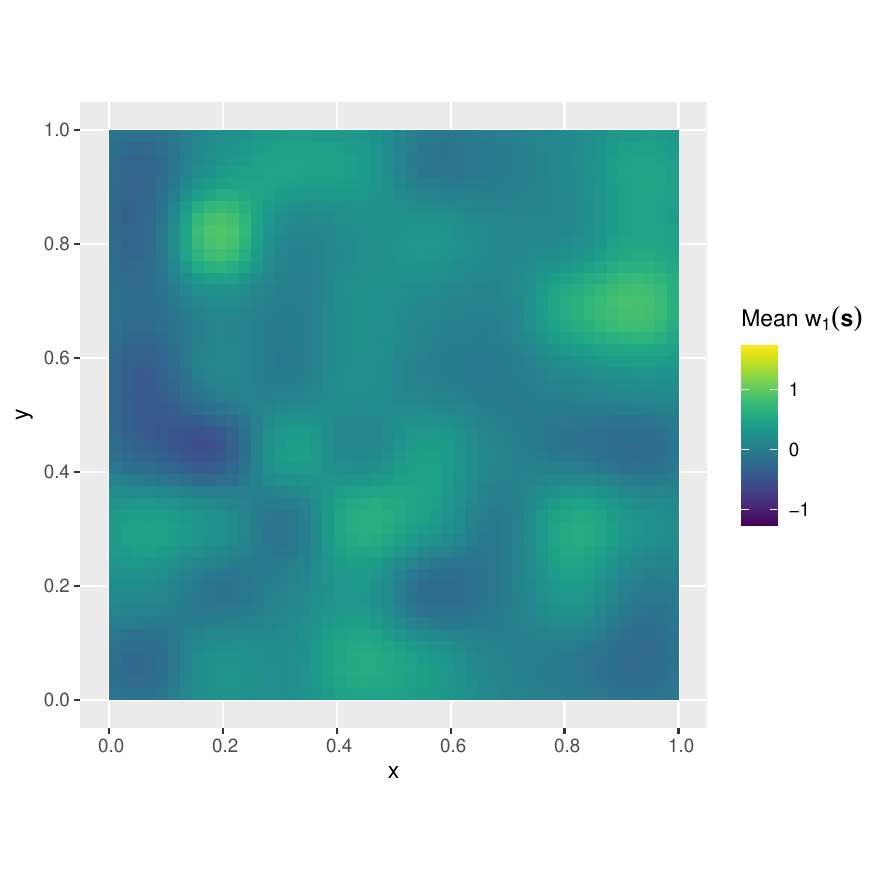}
    \end{minipage}
    \begin{minipage}{0.32\textwidth}
        \includegraphics[width=\linewidth]{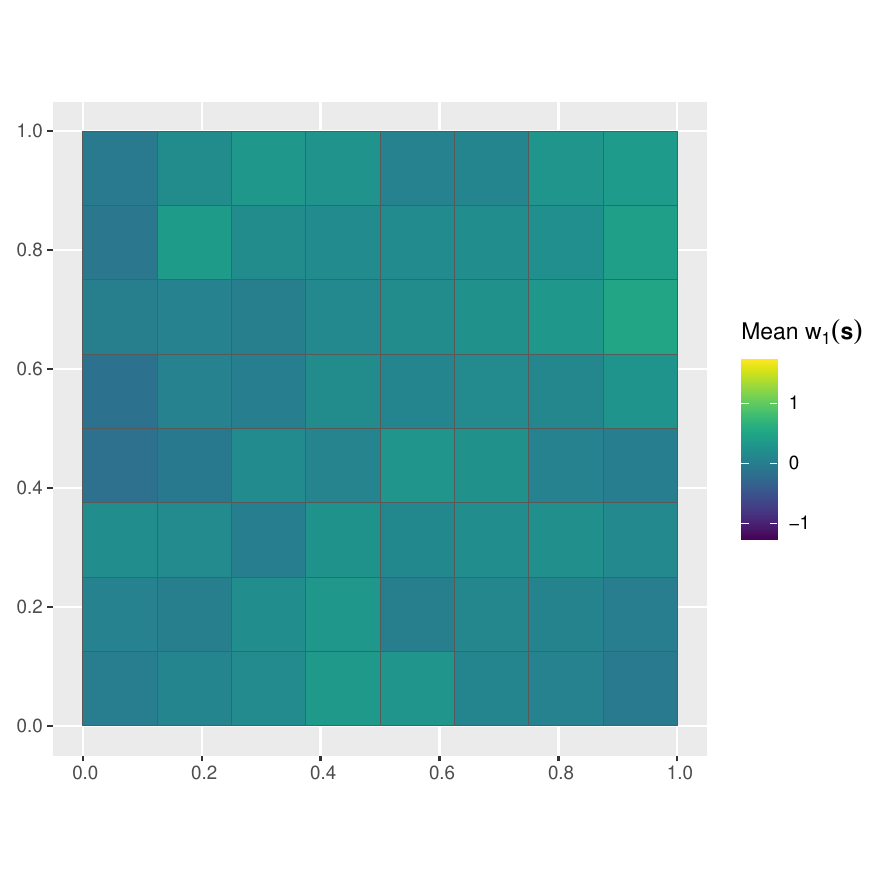}
    \end{minipage}
    \begin{minipage}{0.32\textwidth}
        \includegraphics[width=\linewidth]{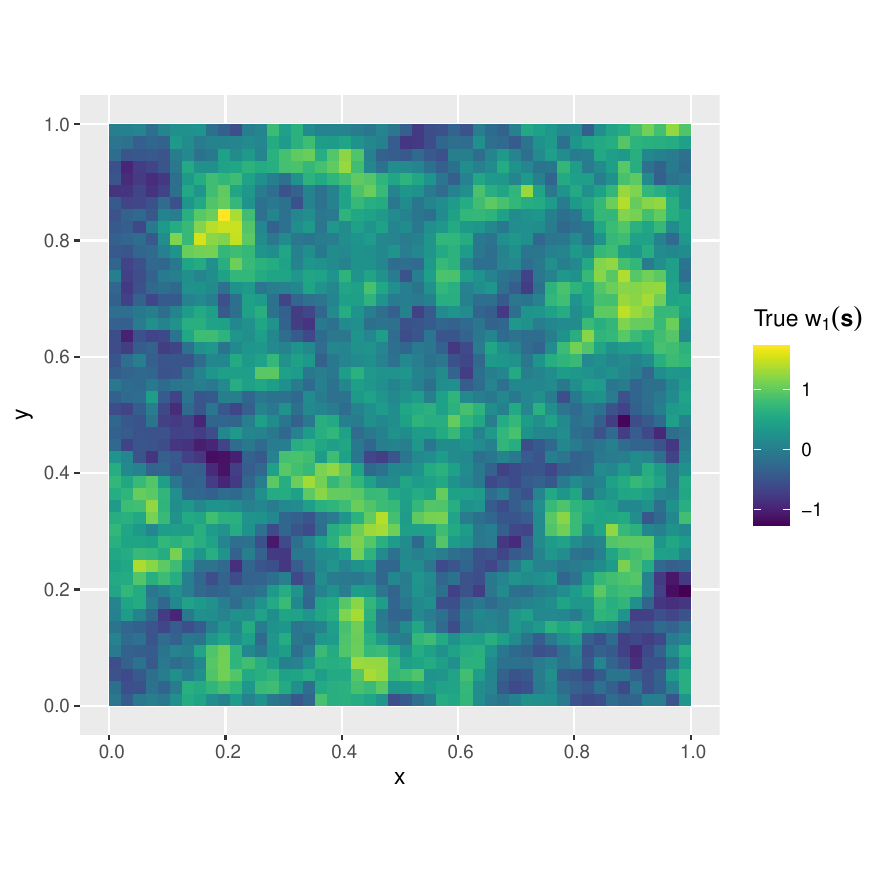}
    \end{minipage}
    \caption{$W_1(\textbf{s})$}
\end{subfigure}
\hfill
\begin{subfigure}{1\textwidth}
    \begin{minipage}{0.32\textwidth}
        \includegraphics[width=\linewidth]{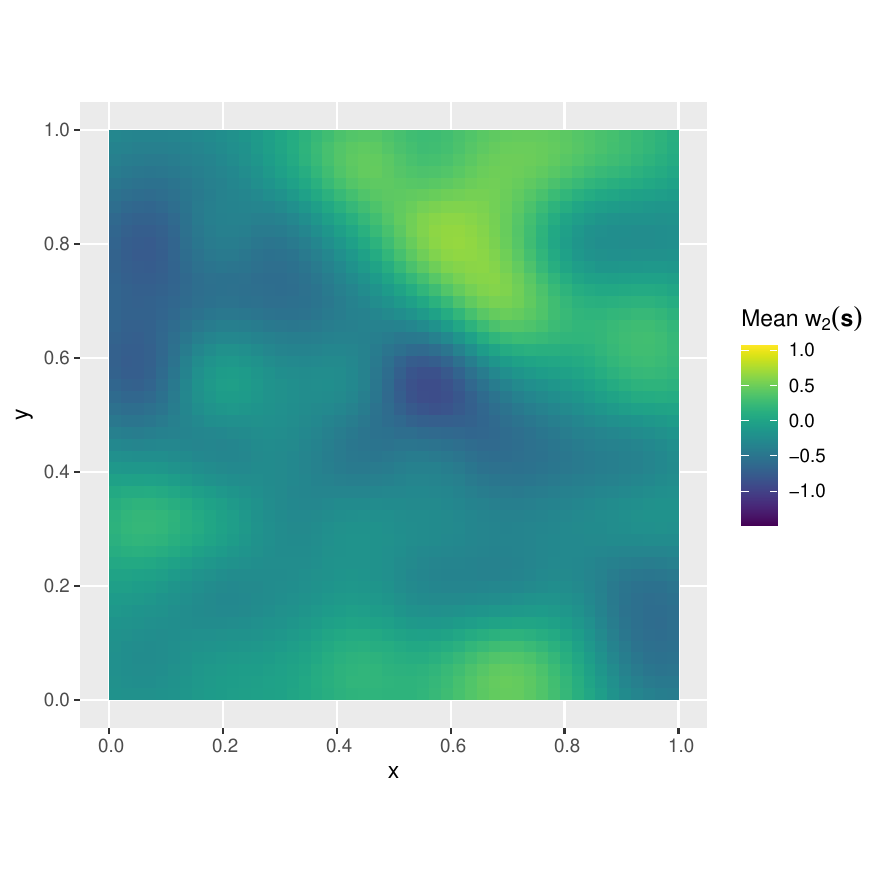}
    \end{minipage}
    \begin{minipage}{0.32\textwidth}
        \includegraphics[width=\linewidth]{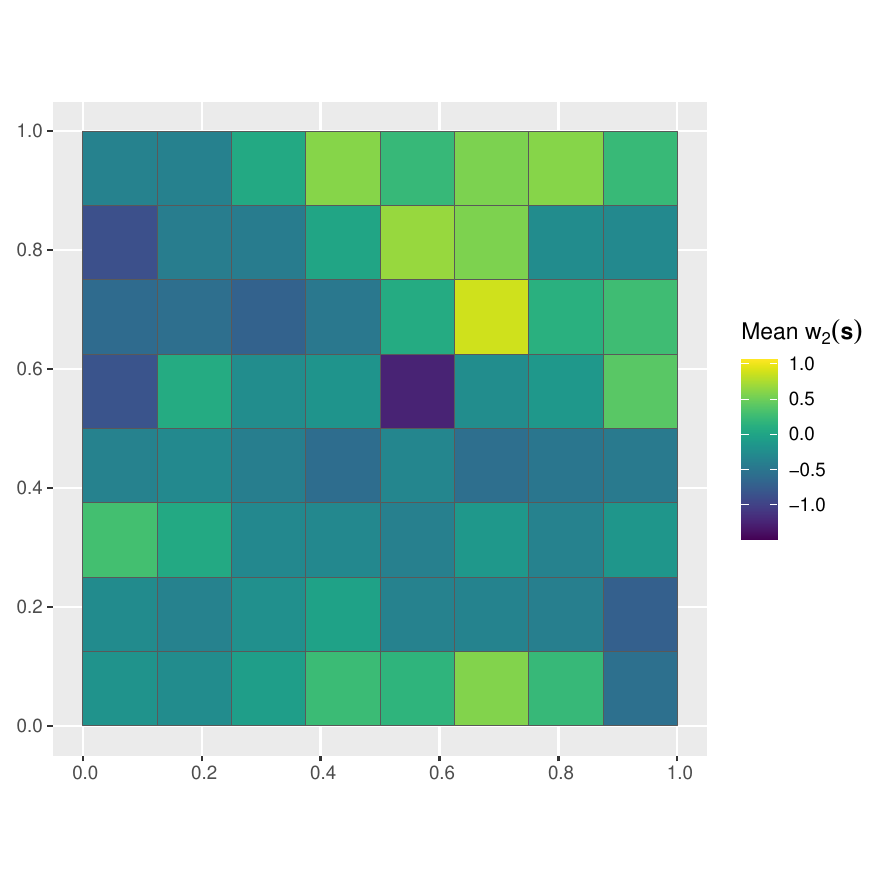}
    \end{minipage}
    \begin{minipage}{0.32\textwidth}
        \includegraphics[width=\linewidth]{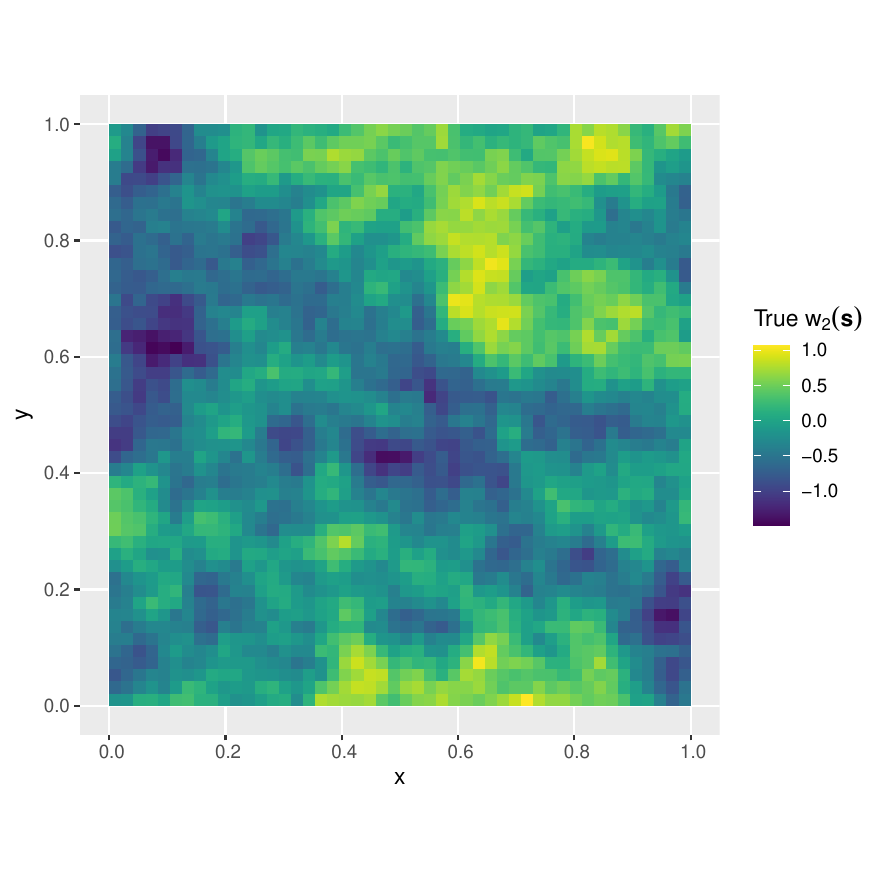}
    \end{minipage}
    \caption{$W_2(\textbf{s})$}
\end{subfigure}
\hfill
\begin{subfigure}{1\textwidth}
    \begin{minipage}{0.32\textwidth}
        \includegraphics[width=\linewidth]{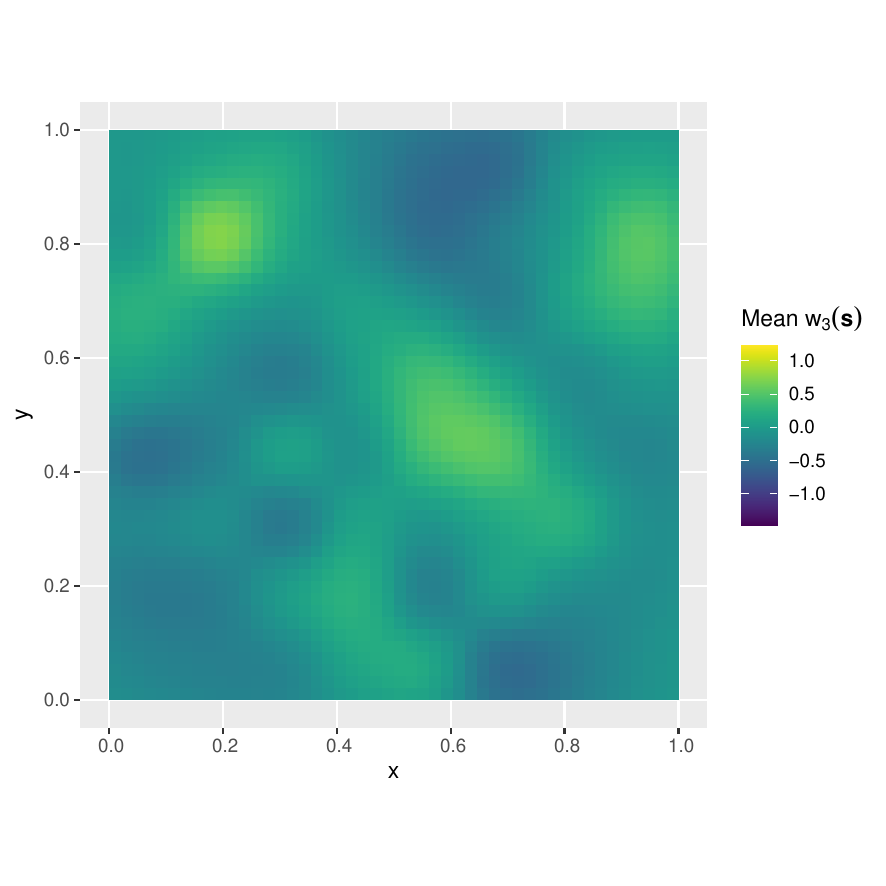}
    \end{minipage}
    \begin{minipage}{0.32\textwidth}
        \includegraphics[width=\linewidth]{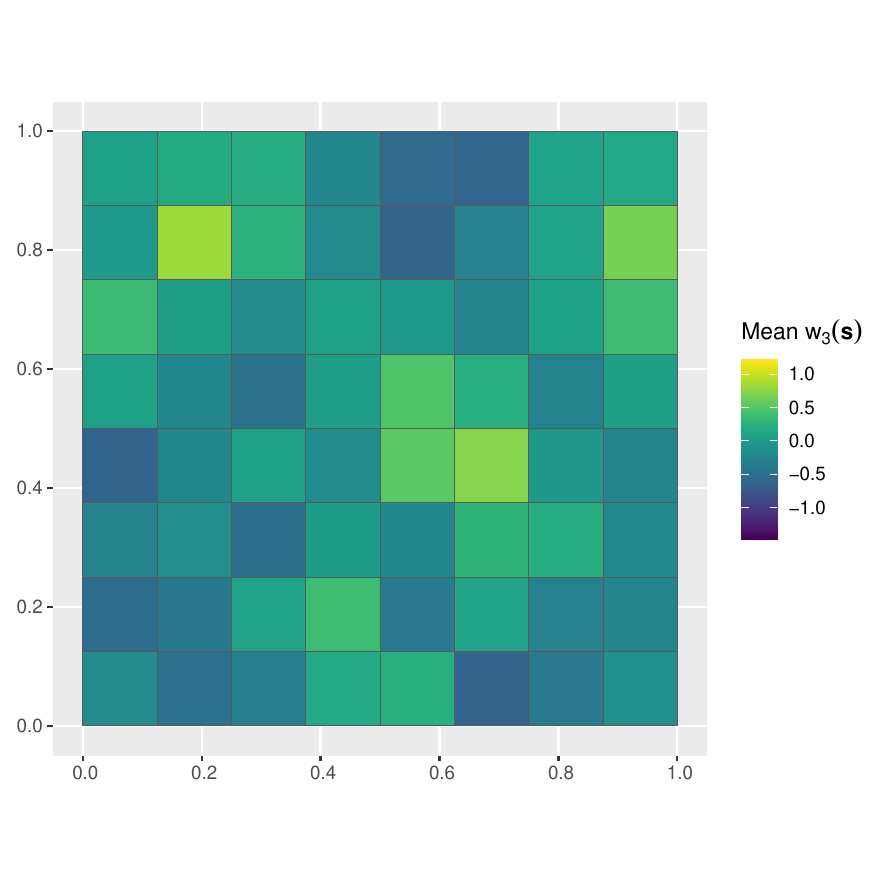}
    \end{minipage}
    \begin{minipage}{0.32\textwidth}
        \includegraphics[width=\linewidth]{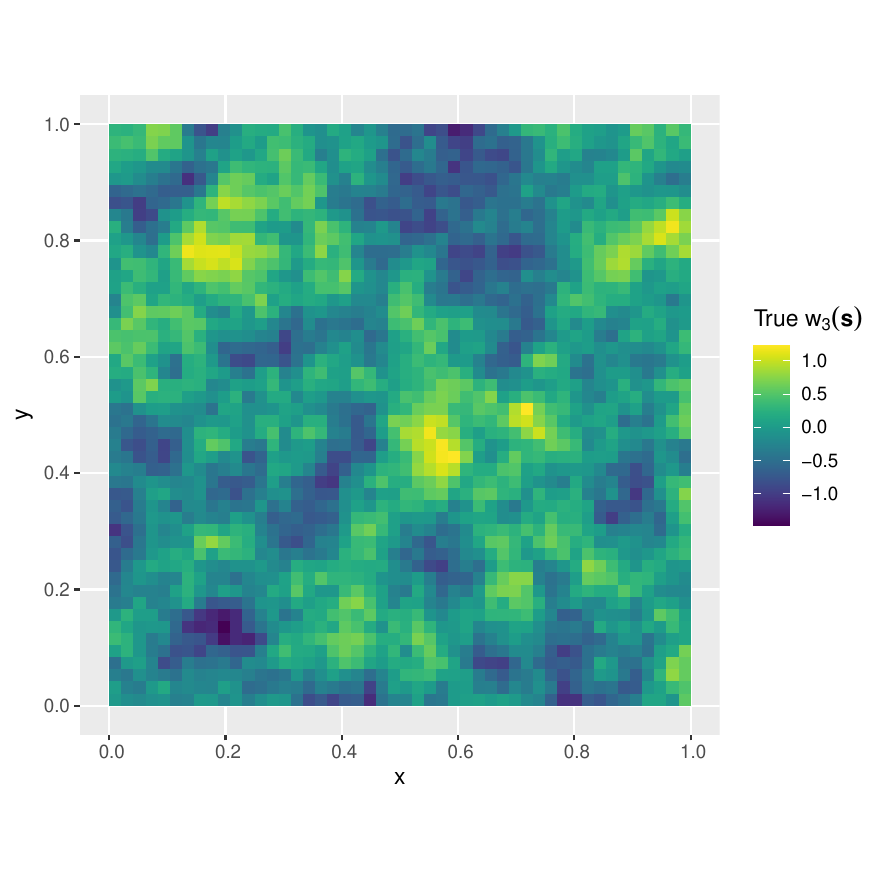}
    \end{minipage}
    \caption{$W_3(\textbf{s})$}
\end{subfigure}
\caption{The left column displays the predictions from the disaggregation model, the central column shows the predictions from the areal model, and the right column presents the realization of the latent space. Each row corresponds to one of the three components of the latent space.}
\label{fig:ex1_s_5}
\end{figure}


\begin{figure}[htp]
\centering
\large{\textbf{Scenario 6}\par\medskip}
\begin{subfigure}{1\textwidth}
    \begin{minipage}{0.32\textwidth}
        \includegraphics[width=\linewidth]{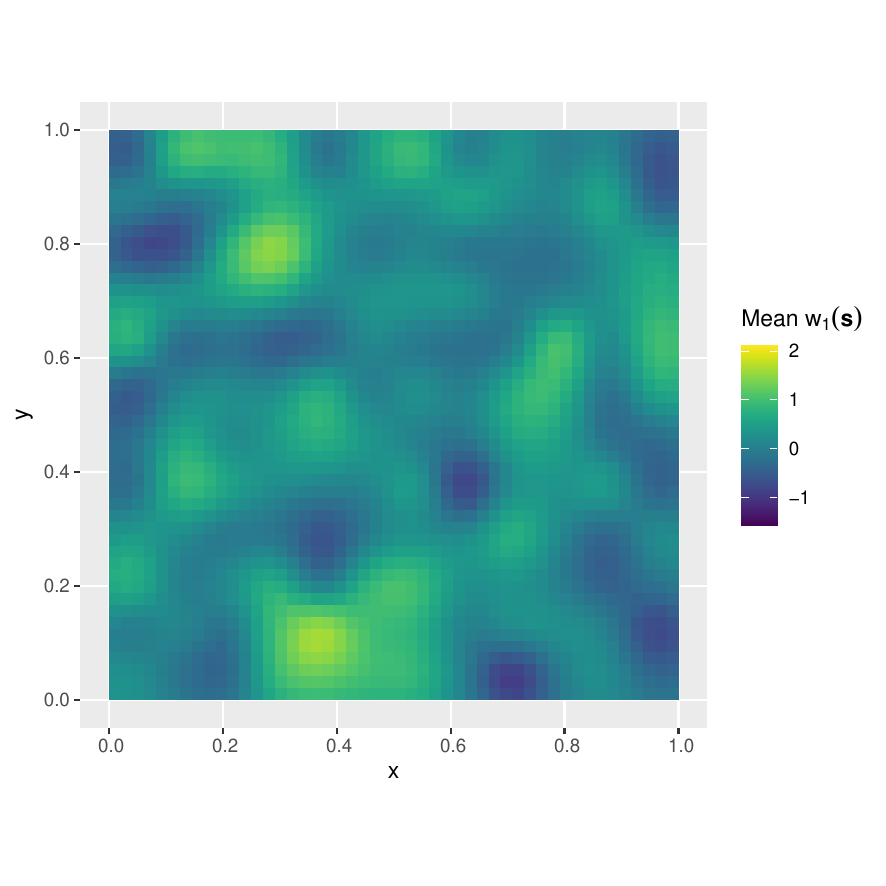}
    \end{minipage}
    \begin{minipage}{0.32\textwidth}
        \includegraphics[width=\linewidth]{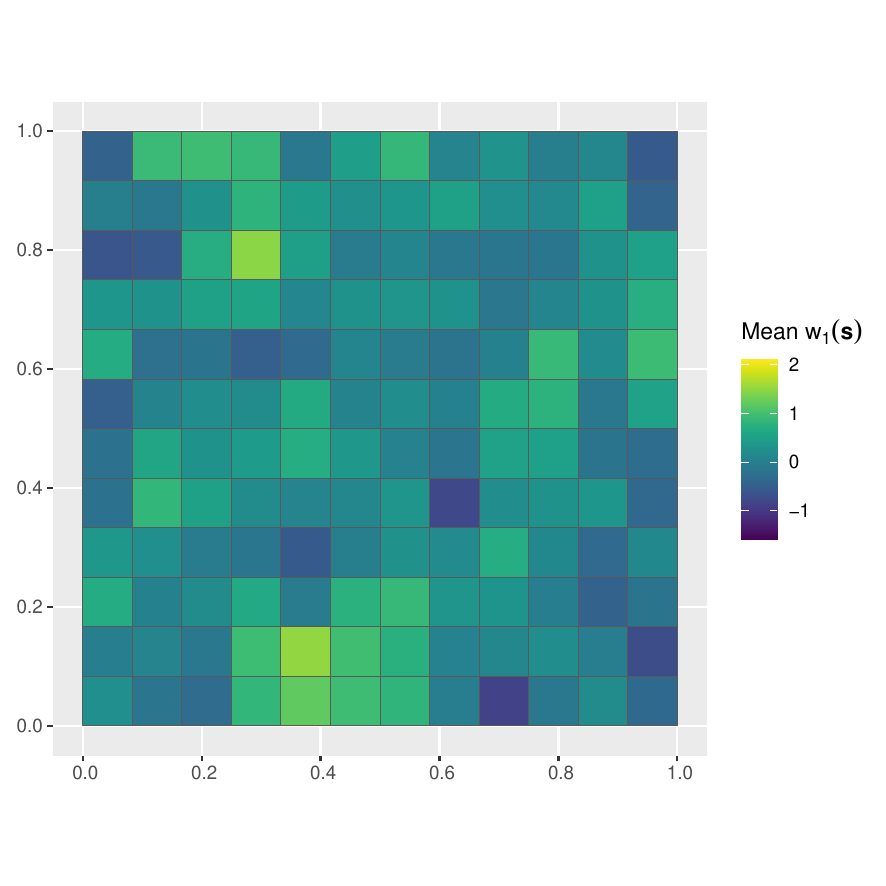}
    \end{minipage}
    \begin{minipage}{0.32\textwidth}
        \includegraphics[width=\linewidth]{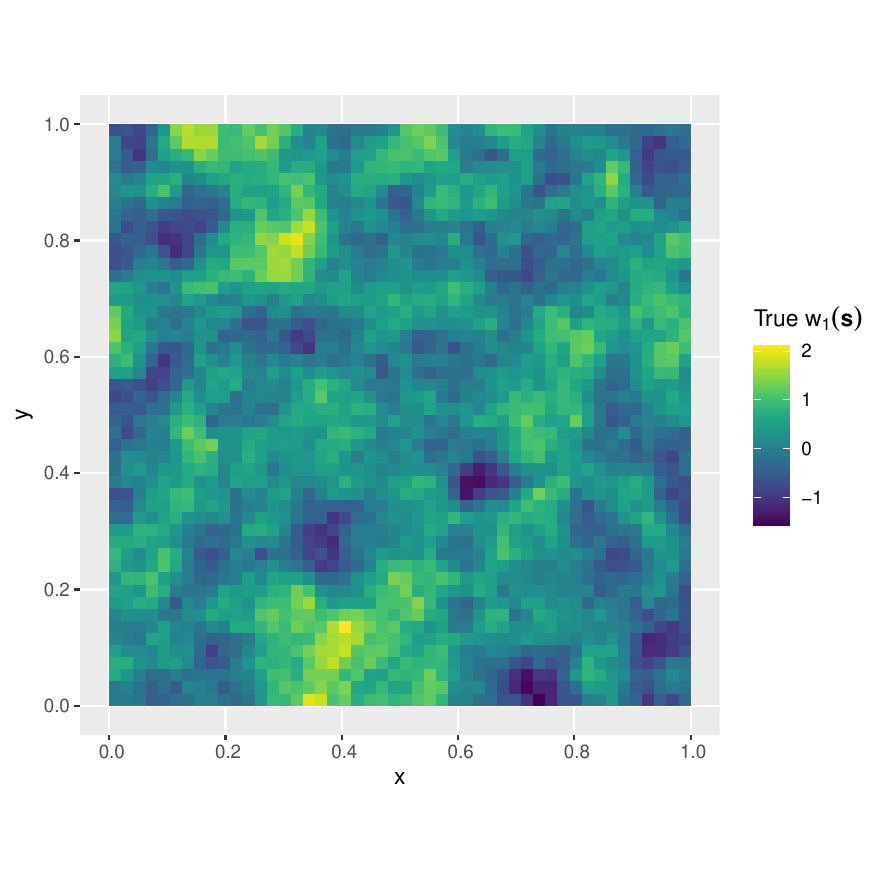}
    \end{minipage}
    \caption{$W_1(\textbf{s})$}
\end{subfigure}
\hfill
\begin{subfigure}{1\textwidth}
    \begin{minipage}{0.32\textwidth}
        \includegraphics[width=\linewidth]{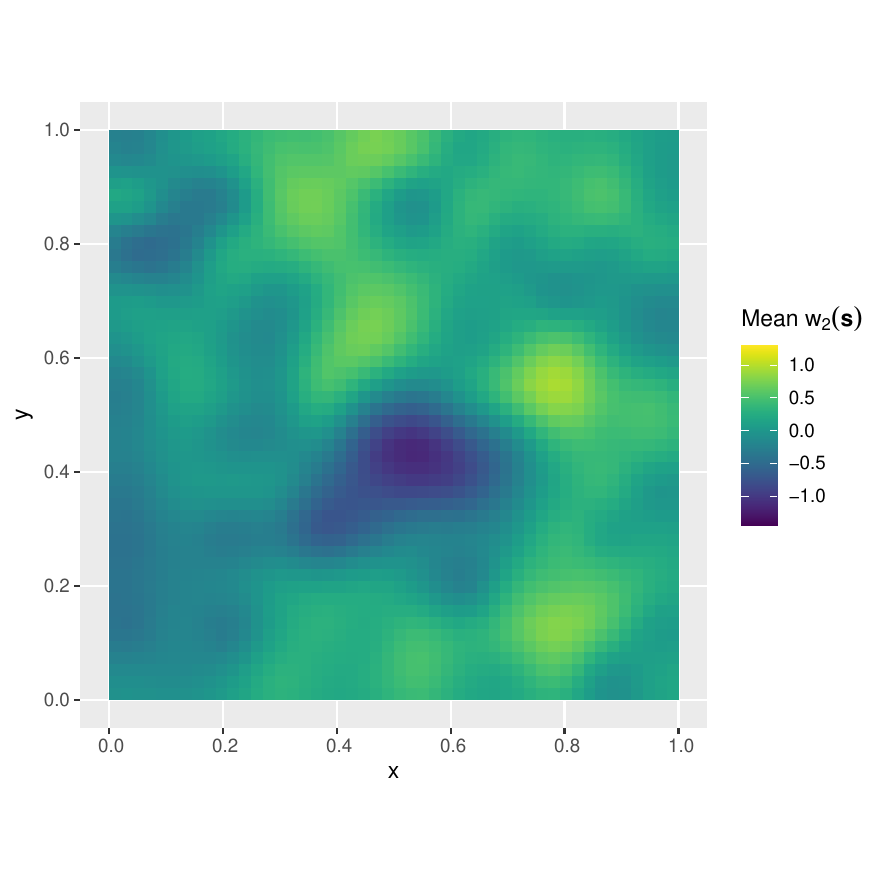}
    \end{minipage}
    \begin{minipage}{0.32\textwidth}
        \includegraphics[width=\linewidth]{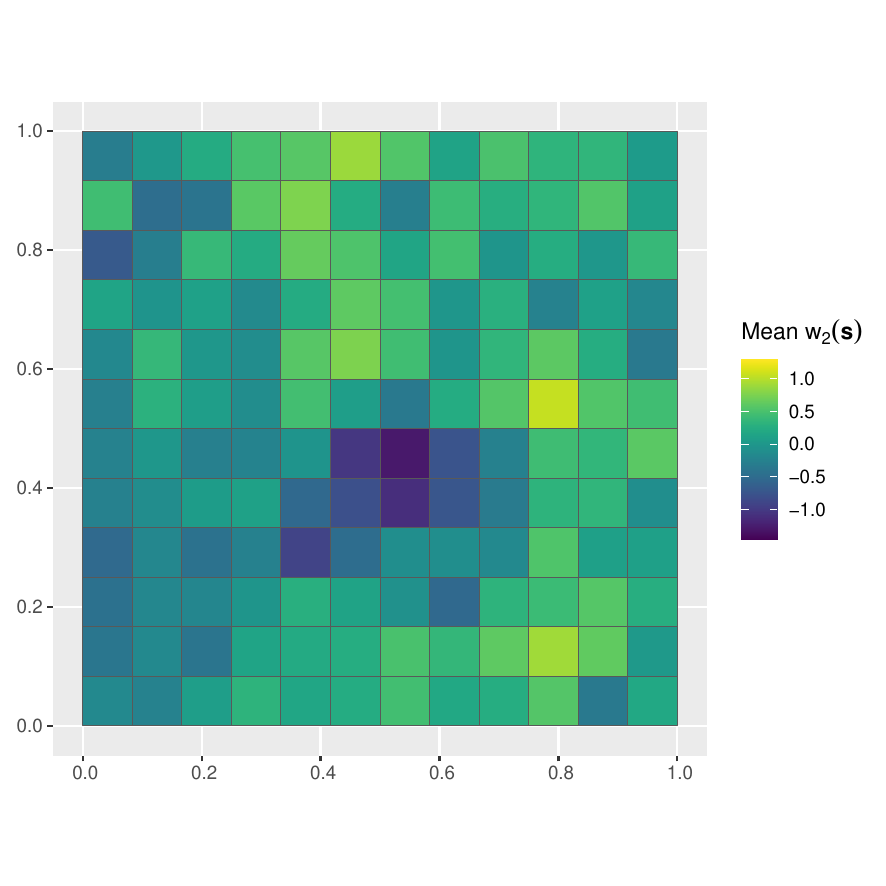}
    \end{minipage}
    \begin{minipage}{0.32\textwidth}
        \includegraphics[width=\linewidth]{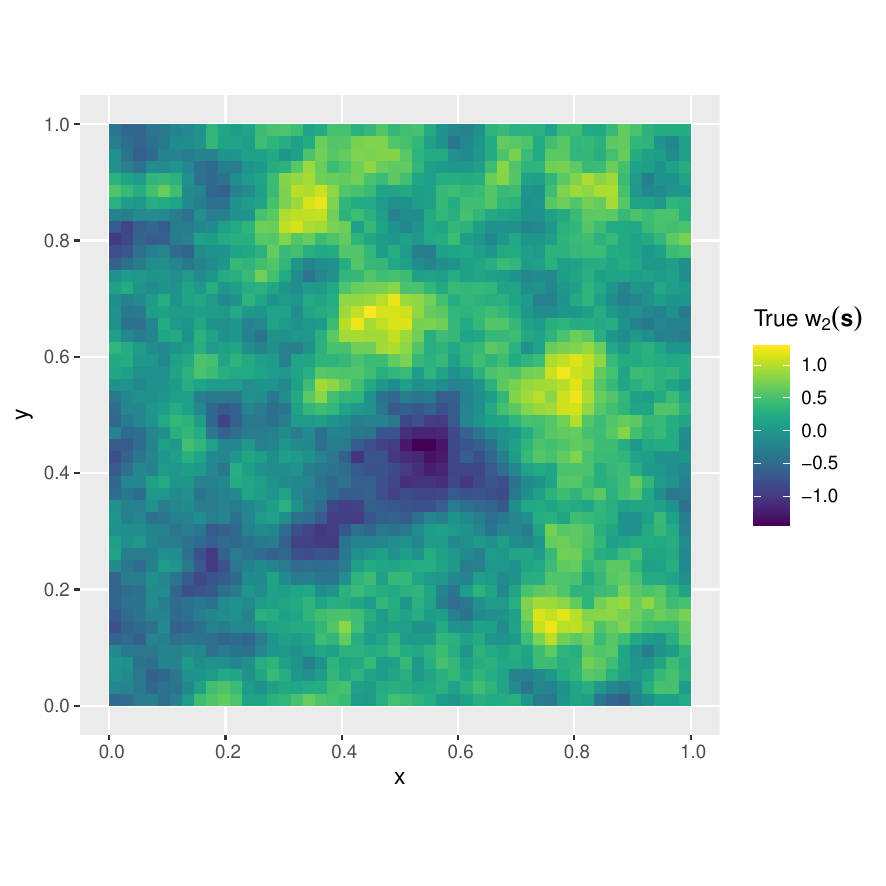}
    \end{minipage}
    \caption{$W_2(\textbf{s})$}
\end{subfigure}
\hfill
\begin{subfigure}{1\textwidth}
    \begin{minipage}{0.32\textwidth}
        \includegraphics[width=\linewidth]{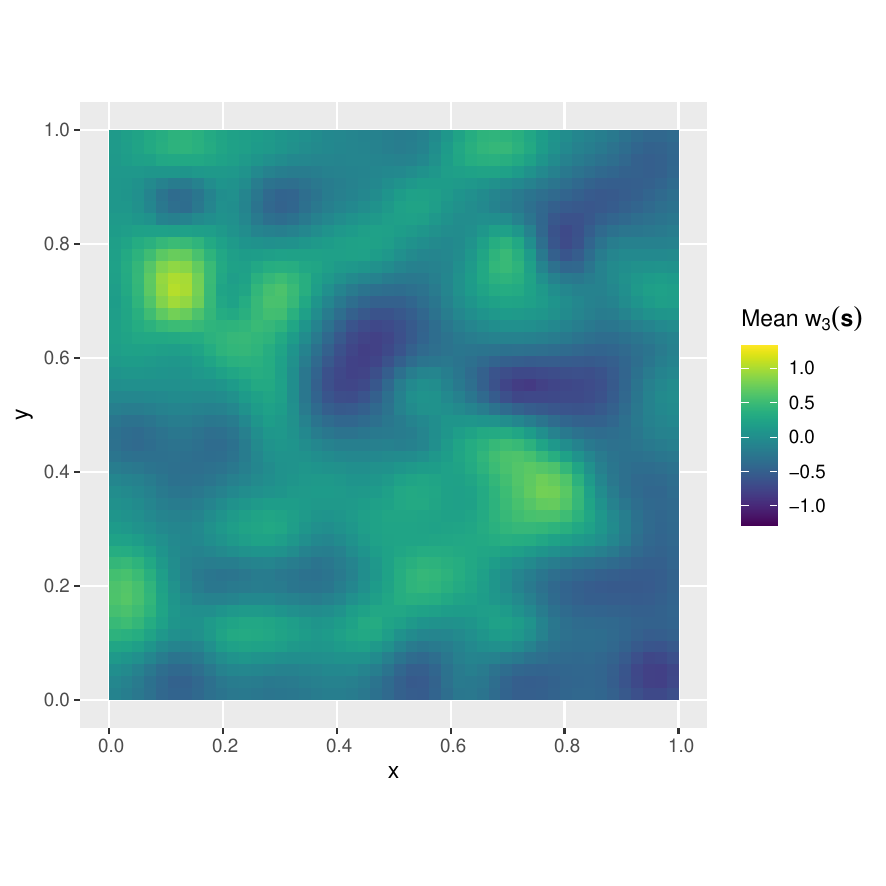}
    \end{minipage}
    \begin{minipage}{0.32\textwidth}
        \includegraphics[width=\linewidth]{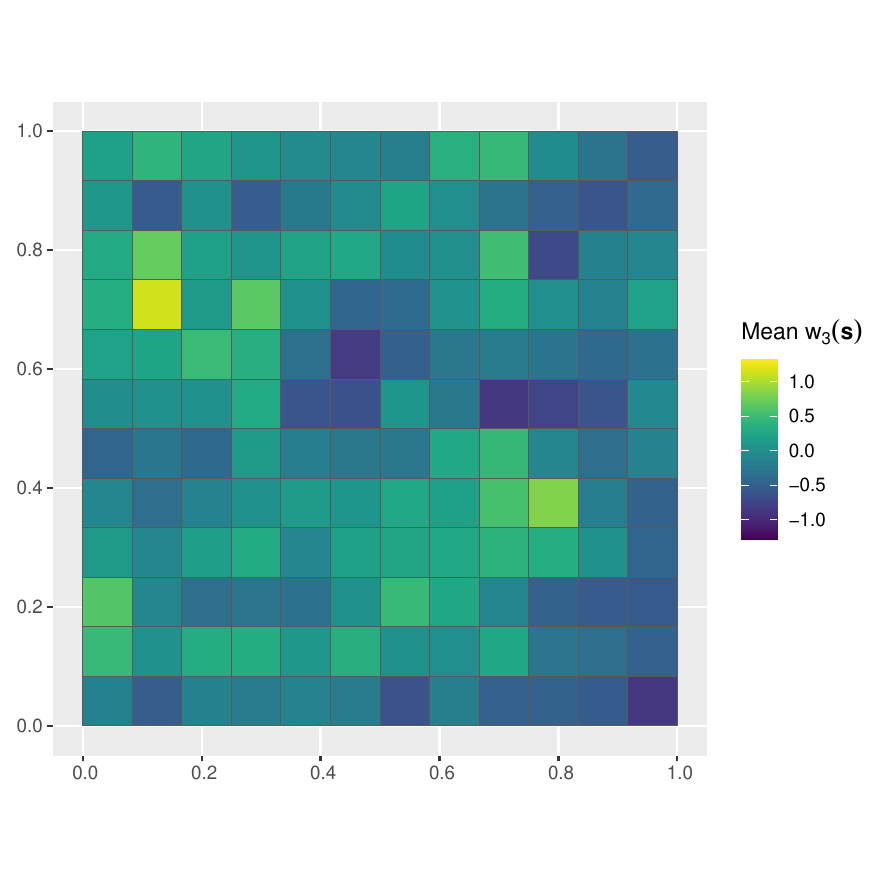}
    \end{minipage}
    \begin{minipage}{0.32\textwidth}
        \includegraphics[width=\linewidth]{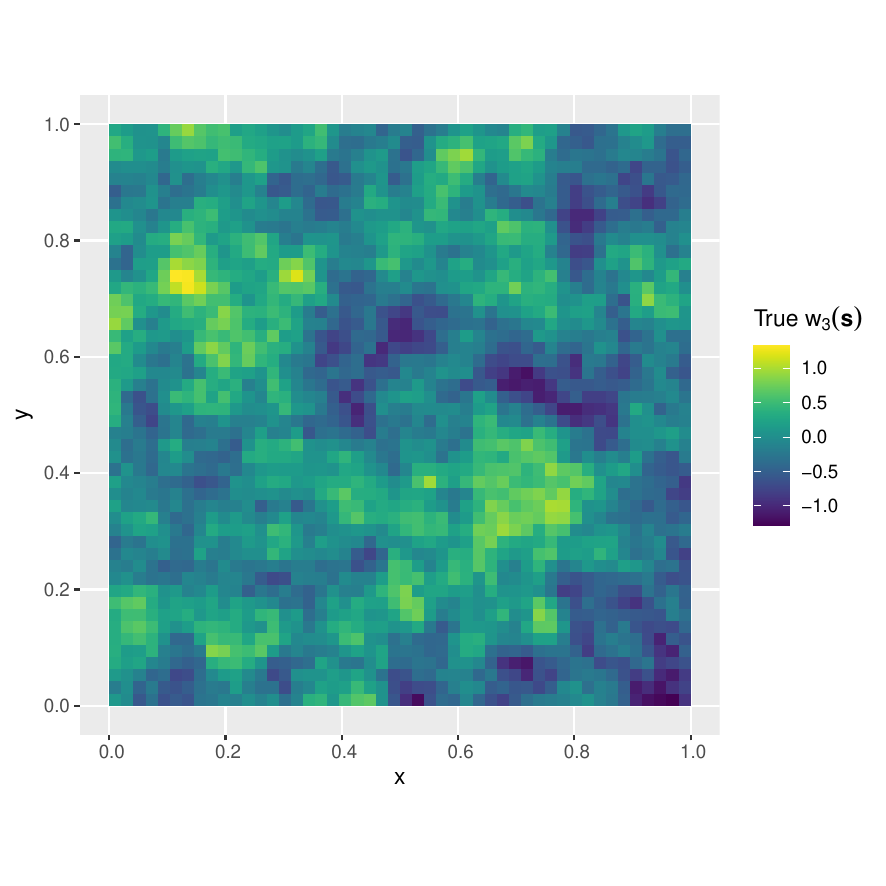}
    \end{minipage}
    \caption{$W_3(\textbf{s})$}
\end{subfigure}
\caption{The left column displays the predictions from the disaggregation model, the central column shows the predictions from the areal model, and the right column presents the realization of the latent space. Each row corresponds to one of the three components of the latent space.}
\label{fig:ex1_s_6}
\end{figure}

\end{document}